\documentclass[pra,twocolumn,superscriptaddress,amssymb]{revtex4}

\usepackage[colorlinks=true,linkcolor=magenta,citecolor=magenta,urlcolor=magenta,linktocpage=true]{hyperref}
\usepackage[utf8]{inputenc}
\usepackage[english]{babel}
\usepackage[T1]{fontenc}
\usepackage{amsmath}
\usepackage{tabularx}
\usepackage{array, makecell}

\usepackage{xcolor}
\definecolor{codegreen}{rgb}{0,0.6,0}
\definecolor{codegray}{rgb}{0.5,0.5,0.5}
\definecolor{codepurple}{rgb}{0.58,0,0.82}
\definecolor{backcolour}{rgb}{0.95,0.95,0.92}

\usepackage{psfrag,graphicx}
\usepackage{dcolumn}
\usepackage{bm}
\usepackage{amsfonts,amssymb,amsmath}        
\usepackage{slashed}
\usepackage[utf8]{inputenc}
\usepackage{graphicx}
\usepackage[caption=false]{subfig}
\usepackage{cancel}
\usepackage{xcolor}
\usepackage{amsmath}
\setcitestyle{numbers,square}

\newcommand{\figref}[1]{\mbox{Fig.~\ref{#1}}}

\newcommand{\be}{\begin{equation}}
\newcommand{\ee}{\end{equation}}
\newcommand{\bq}{\begin{eqnarray}}
\newcommand{\eq}{\end{eqnarray}}

\newcommand{\ket}[1]{\left |#1 \right\rangle}
\newcommand{\bra}[1]{\left \langle #1 \right |}
\newcommand{\ketbra}[2]{\left|#1\right\rangle\left\langle#2\right|}

\newcommand{\hhat}[1]{\hat{\hat{#1}}}

\makeatletter

\renewenvironment{widetext@grid}{
  \par\ignorespaces
  \setbox\widetext@top\vbox{
   \vskip15\p@
   \hb@xt@\hsize{
    \leaders\hrule\hfil
    \vrule\@height6\p@
   }
   \vskip6\p@
  }
  \setbox\widetext@bot\hb@xt@\hsize{
    \vrule\@depth6\p@
    \leaders\hrule\hfil
  }
  \onecolumngrid

  \let\set@footnotewidth\set@footnotewidth@ii
}{
  \par

  \twocolumngrid\global\@ignoretrue
  \@endpetrue
}

\makeatother

\begin{document}
\title{A quantum-classical decomposition of Gaussian quantum environments: a stochastic pseudomode model}
\author{Si Luo}
\affiliation{Graduate School of China Academy of Engineering Physics, Haidian District, Beijing, 100193, China}
\author{Neill Lambert}
\email{nwlambert@gmail.com}
\affiliation{Theoretical Quantum Physics Laboratory, RIKEN Cluster for Pioneering Research, Wako-shi, Saitama 351-0198, Japan}
\author{Pengfei Liang}
\affiliation{Graduate School of China Academy of Engineering Physics, Haidian District, Beijing, 100193, China}
\author{Mauro Cirio}
\email{cirio.mauro@gmail.com}
\affiliation{Graduate School of China Academy of Engineering Physics, Haidian District, Beijing, 100193, China}

\date{\today}

\begin{abstract}
We show that the effect of a Gaussian Bosonic environment linearly coupled to a quantum system can be simulated by a stochastic Lindblad master equation characterized by a set of ancillary Bosonic modes initially at zero temperature and classical stochastic fields. {  We test the method for Ohmic environments with exponential and polynomial cut-offs against, respectively, the Hierarchical Equations of Motion and the deterministic  pseudomode model with respect to which the number of ancillary quantum degrees of freedom is reduced.}
For a subset of rational spectral densities, all parameters are explicitly specified without the need of any fitting procedure, thereby simplifying the modeling strategy.  {  Interestingly, the classical fields  in this decomposition must sometimes be imaginary-valued, which can have counter-intuitive effects on the system properties which we demonstrate by showing that they can decrease the entropy of the system, in contrast to real-valued fields.} 
\end{abstract}
\maketitle
\vspace{5mm}
\newpage
Most quantum systems are inherently open \cite{Petruccione,Gardiner}, i.e., their dynamics depends on the interaction with an external environment. A typical example is when the system is in contact with a continuum of degrees of freedom constituting a thermal bath. 
The system dynamics can take qualitatively different forms depending on the characteristics of the bath and on the system-bath interaction. For example, under the Born-Markov approximation, the reduced system dynamics can be described by a Lindblad equation \cite{Lindblad, Gorini} or by a more general non-secular Redfield equation \cite{Redfield}, depending on the presence of near-degeneracies (compared to the broadening) in the system's spectrum \cite{Ishizaki_Fleming,Dodin,PhysRevA.94.012110,PhysRevA.103.062226}. In regimes where a coherent exchange of information with the bath occurs, more general master equations (possibly non-local in time) are required to correctly account for these memory effects \cite{Petruccione,Fruchtman,RevModPhys.89.015001}.

Within this general landscape, a wide range of physical environments can be represented by Gaussian baths (such as those made out of Bosonic and Fermionic degrees of freedom at thermal equilibrium) linearly interacting with the system. In this case, the effects of the bath on the system are fully characterized by the properties of correlation functions involving the bath operators which mediate the interaction. As a consequence, the regimes mentioned above can now all be described by properties of these correlation functions which also bear information about the non-classicality of the system dynamics \cite{Clemens1,Clemens2}. 

Intuitively, when the correlations decay faster than the characteristic time associated with the system, the regime is Markovian and the effects of the bath on the system can be modeled by master equations local in time and written in Lindblad form. In the opposite case, the bath response-time is slow enough to allow the system and the bath to coherently exchange energy before its eventual dissipation  into the continuum. Several approaches have been developed to analyze this non-Markovian regime \cite{Petruccione,Gardiner,RevModPhys.94.045006} such as  generalized master equations derived from the Feynmann-Vernon influence functional formalism \cite{Vernon,Caldeira_Leggett_1,Caldeira_Leggett_2,Caldeira_0,Bonig,PhysRevB.78.235311,Jin_Matisse,PhysRevLett.109.170402,Xiong2015}, the Hierarchical Equation of Motion (HEOM) \cite{Tanimura_3,Tanimura_1,Ishizaki_1,Tanimura_2,Ishizaki_2,PhysRevLett.104.250401,doi:10.1143/JPSJ.81.063301,Tanimura_2014,Lambert_Bofin,Tanimura_2020,Tanimura_2021}, the reaction coordinate method \cite{Garg,Martinazzo,iles2014environmental,Woods,PhysRevB.97.205405,Melina}, the polaron transformation \cite{Holstein1,Holstein2,Jackson,Silbey,Silbey2,Weiss,Jang,Jang2,PhysRevLett.103.146404,McCutcheon,Jang3,PhysRevB.83.165101,Kolli,PollockThesis,PollockNazir,Xu}, { collisional methods} \cite{PhysRev.129.1880,PhysRevLett.88.097905,PhysRevA.65.042105,PhysRevB.104.045417}, { cascaded networks \cite{Hudson,PhysRevA.31.3761,PhysRevLett.70.2273,PhysRevA.87.032117,Gough,5286277,Mabuchi,ZHANG20171,doi:10.1080/23746149.2017.1343097}} and the pseudomode model \cite{PhysRevA.50.3650,PhysRevA.55.2290,PhysRevLett.110.086403,PhysRevB.89.165105,PhysRevB.92.245125,Schwarz,Dorda,Mascherpa,Lemmer_2018,Tamascelli,Lambert,PhysRevLett.123.090402,PhysRevResearch.2.043058,PhysRevA.101.052108,Cirio2022}. These methods are also non-perturbative in the interaction strength thereby allowing to model the correct system-bath hybridization properties of the steady state.

Among the methods listed above, the pseudomode model consists in replacing the continuum of environmental degrees of freedom with a discrete set of effective {  dissipative} quantum {  harmonic} modes. These modes are designed to approximate the original bath correlation function as a discrete sum of decaying exponentials, similarly to the HEOM method. As shown in \cite{Lambert, Cirio2022}, the domain of this ansatz can be further enlarged by allowing some of the model parameters to take  unphysical values. Depending on the correlation function, different regimes might allow a more efficient representation in terms of pseudomodes. For example,  in certain bath models at zero temperature \cite{Lambert}, the continuum of Matsubara frequencies implies correlations decaying polynomially in time, thereby posing a possible challenge from a simulation standpoint \cite{Fruchtman,Tang,PhysRevB.95.214308} as the number of pseudomodes required necessarily ends up scaling with the simulation time.  While this issue is often not evident in practical applications, it emerges when vanishing spectral properties of the system need to be resolved, such as in the low-temperature Kondo regime.

{ Alongside these fully deterministic methods, stochastic techniques have also been used to describe the effects of a quantum environment by driving the dynamics with classical noise. 
This was originally implemented in the context of Markovian \cite{PhysRevLett.68.580,PhysRevA.45.4879,PhysRevE.52.428,Gisin,Percival,PhysRevA.104.062212,PhysRevResearch.4.023036} and non-Markovian \cite{PhysRevLett.82.1801,PhysRevA.58.1699,Diosi_1,Diosi_2} quantum state diffusion. To simplify the presence of memory kernels, this formalism was lifted to Liouville space \cite{PhysRevLett.88.170407,Shao,STOCKBURGER2004159,Stockburger_JCP1999,Stockburger_CP2001} which has been further extended to also include Fermionic baths \cite{Chernyak1,Chernyak2,PhysRevLett.123.050601}. The difficulty to average over correlated stochastic processes led to the development of hybrid techniques involving the HEOM \cite{Hsieh_1,Hsieh_2,Yan_Shao,Tanimura_2020,Moix,YAN2004216}, classical and semi-classical optimizations \cite{PhysRevLett.100.230402,Stockburger_2016,PhysRevE.102.062134}, and piece-wise ensemble averaging \cite{Yun-AnShao}.}

{ Here, we introduce a hybrid approach to reproduce non-Markovian effects in open quantum systems in the context of the pseudomode model, i.e., using ancillary quantum harmonic modes alongside a single stochastic classical drive. The system reduced dynamics can be computed by solving a Lindblad-like master equation which  does not involve memory kernels (as in the pure quantum state diffusion) and it does not require multiple cross-correlated stochastic processes (as in the purely stochastic Liouville space approaches). }

{  In contrast to the mathematical auxiliary degrees of freedom present in HEOM-related works, this effective master equation {  provides a rather intuitive interpretation where system is directly coupled to one or more leaky modes  and to a classical stochastic field}. The effective nature of this model does not grant a physical meaning for this interpretation {  and, depending on the decomposition, either the leaky modes or the classical field can be unphysical in the sense they depend on imaginary-value non-Hermitian parameters.} However, at the same time, it allows us explore wider (unphysical) regimes for the ancillary degrees of freedom which can improve the optimization of the model, 

We show that this hybrid approach can be introduced as a consequence of a classical-quantum decomposition of the environmental \emph{effects on the system} (i.e., not directly as a physical description of the properties of the bath) which, in the Gaussian case, translates into defining {  the corresponding (classical or quantum)} contributions to the bath correlation function. Interestingly, this decomposition is not unique but it can be chosen to optimize the model. For example, this feature leads to the possibility to initialize the  ancillary harmonic quantum degrees of freedom in their vacuum state, \emph{for any temperature of the original bath}. This directly allows for a further, practical, optimization of the Hilbert space {  dimension due to a smaller Fock-space construction}.

}

In addition, in the case where  the environment can be described as a sum of underdamped spectral densities, the corresponding stochastic pseudomode model only requires a single, physical, zero temperature quantum degree of freedom per spectral density. {  This is in stark contrast with the fully deterministic pseudomode model where further quantum degrees of freedom are needed by fitting the Matsubara part of the correlation.  This leads to advantages in modeling certain open quantum systems such as when the system is coupled to multiple baths at different temperature. In this case, we show that the possibility to encode all collective temperature effects into a single field substantially optimizes the simulation resources with respect to the fully deterministic pseudomode model. {  We demonstrate this with a practical example where a system is coupled to multiple baths at different temperatures.}

{  Importantly, for the general case} we analytically determine} an expression for the number of pseudomodes for rational spectral densities  { for which the model is defined without free parameters (i.e., without requiring any fitting) for all temperatures}. For non-rational spectral densities, a fitting procedure is still required to optimize the number of pseudomodes.

 {  Finally, we also demonstrate the unusual feature that, for a zero-temperature bath, the imaginary nature of the classical fields can increase the order in the system, as described by it's von Neumann entropy (which can be understood as the system-environment entanglement entropy at zero temperature). This is contrast to real-valued classical fields which are expected to increase the disorder of the system state.}
 
This article is organized as follows. In section \ref{sec:pseudomode_model} we introduce the pseudomode model in its fully quantized, { deterministic} version. In section \ref{sec:Quantum-Classical} we introduce a quantum-classical decomposition of the environmental correlation function which we then use in section \ref{sec:stochastic} to define a stochastic version of the pseudomode model. 
In subsection \ref{sec:bias_stoch}, we further analyze the bias and stochastic sources of errors in the model and, in subsection \ref{sec:ZeroT_main}, we present a variant of the model with the advantage of allowing all quantum modes to be initially at zero temperature. In section \ref{sec:rational} we show that all parameters of the model can be computed analytically for spectral densities which can be written as a specific class of rational functions. In section \ref{sec:examples} we consider two specific examples of spectral densities. In subsection \ref{sec:underdamped} we test the stochastic method against the fully quantized, { deterministic} pseudomode model for the case of underdamped Brownian spectral densities. In subsection \ref{sec:Ohmic}, we consider Ohmic spectral densities with exponential cut-off which are outside of the rational domain thereby not allowing for a full analytical analysis. 
{  In Section \ref{sec:MultipleBaths}, we apply the hybrid formalism to describe a system coupled to multiple baths at different initial temperatures. We finish in section \ref{sec:Entropy}, where we analyze the effects of unphysical stochastic fields on the von Neumann entropy of the system.}
Each section in the main text is further accompanied by a corresponding one in the Appendix where we describe all technical details.

\section{Pseudomode model}
\label{sec:pseudomode_model}
In this section we review the pseudomode model and introduce the main notation used throughout this article. We consider a system $S$ coupled to a bosonic environment $B$ as described by the Hamiltonian ($\hbar=1$)
\begin{equation}
\label{eq:full_Hamiltonian}
H=H_S+H_I+H_B\;,
\end{equation}
where $H_S$ is the system Hamiltonian and $H_B=\sum_k \omega_k b_k^\dagger b_k$ characterizes the energy $\omega_k$ of each environmental mode $b_k$. We suppose the system-environment coupling to be in the form $H_I= {s}X$, where ${s}$ is a system operator and $X=\sum_k g_k (b_k+b_k^\dagger)$ is a linear interaction operator (in terms of the real coupling strengths $g_k\in\mathbb{R}$). When the environment is initially in thermal equilibrium at inverse temperature $\beta$, the dynamics of the reduced system density matrix $\rho_S(t)$ can be written  as
\begin{equation}
\label{eq:rhoS}
\rho_S(t)=\mathcal{T}e^{-\mathcal{F}(t,{s},C(t))}\rho_S(0)\;.
\end{equation}
Here, the super-operator $\mathcal{F}(t,{s},C(t))$, whose explicit form is shown in Appendix \ref{app:influenceSuperoperator}, only depends on the system coupling operator ${s}$ and the free correlation function of the environmental interaction operator 
\begin{equation}
\label{eq:correlation}
\begin{array}{lll}
C(t)&=&\langle X(t_2)X(t_1)\rangle\\
&=&\displaystyle\int_0^\infty d\omega~\frac{J(\omega)}{\pi}\left[\coth(\beta\omega/2)\cos(\omega t)-i\sin(\omega t)\right],
\end{array}
\end{equation}
where $t=t_2-t_1$, and where we introduced the spectral density $J(\omega)=\pi\sum_k  g_k^2\delta(\omega-\omega_k)$. Here $X(t)=U_B^\dagger (t) X U_B(t)$ is the bath coupling operator in the Heisenberg picture of the free bath  (in terms of $U_B(t)=e^{-iH_Bt}$), and the expectation value is taken with respect to a thermal state  $\rho_B=\text{exp}[-\beta\sum_k\omega_k b_k^\dagger b_k]/Z_\beta$ at inverse temperature $\beta$, with $Z_\beta$ enforcing unit trace.

The formal expression in Eq.~(\ref{eq:rhoS}) is not easily evaluated as it includes environmental memory effects encoded in the time ordering of the double time-integral in the exponential. However, the expression of the superoperator $\mathcal{F}(t,{s},C(t))$ suggests the possibility to define approximation schemes in which the environment is replaced by a fictitious one whose two-point free correlation function and system coupling operator match the original ones. In particular, the pseudomode model introduces an environment made out of lossy harmonic modes which are purely mathematical entities as they have no direct relation to the original environmental modes. Even more, as noted in \cite{Lambert}, since we only focus on the reduced system dynamics, the domain of the parameters in the model can be further enlarged to unphysical values, thereby allowing more flexibility in  optimizing the approximation.

Specifically, the pseudomode model consists of approximating the reduced system dynamics of the original open quantum system in Eq.~(\ref{eq:full_Hamiltonian}) as
\begin{equation}
\label{eq:PM_dynamics}
\rho_S(t)\simeq\text{Tr}_\text{PM}\rho(t)\;,
\end{equation}
where $\rho(t)$ is the density matrix in an enlarged Hilbert space $\mathcal{H}=\mathcal{H_S}\otimes \mathcal{H}_\text{PM}$ where $\mathcal{H}_S$ and $\mathcal{H}_\text{PM}= \otimes_{j=1}^{N_\text{PM}} \mathcal{H}_j$ represent the system and the pseudomodes Hilbert spaces respectively (in which $\mathcal{H}_j$  is the space of the $j$th pseuomode). The dynamics in this enlarged space is computed through the Lindblad equation
\begin{equation}
\label{eq:pseudoShr\"{o}dinger}
\dot{\rho}(t)=-i[H_\text{PM},\rho(t)]+\sum_{j=1}^{N_\text{PM}} D_j[\rho(t)]\;,
\end{equation}
where
\begin{equation}\label{eq:pmhamiltonian}
H_\text{PM}=H_S+\sum_{j=1}^{N_\text{PM}} \left(\lambda_j X_j{s}+\Omega_j a^\dagger_j a_j\right)\;,
\end{equation}
\begin{figure}[t!]
\includegraphics[width = \columnwidth]{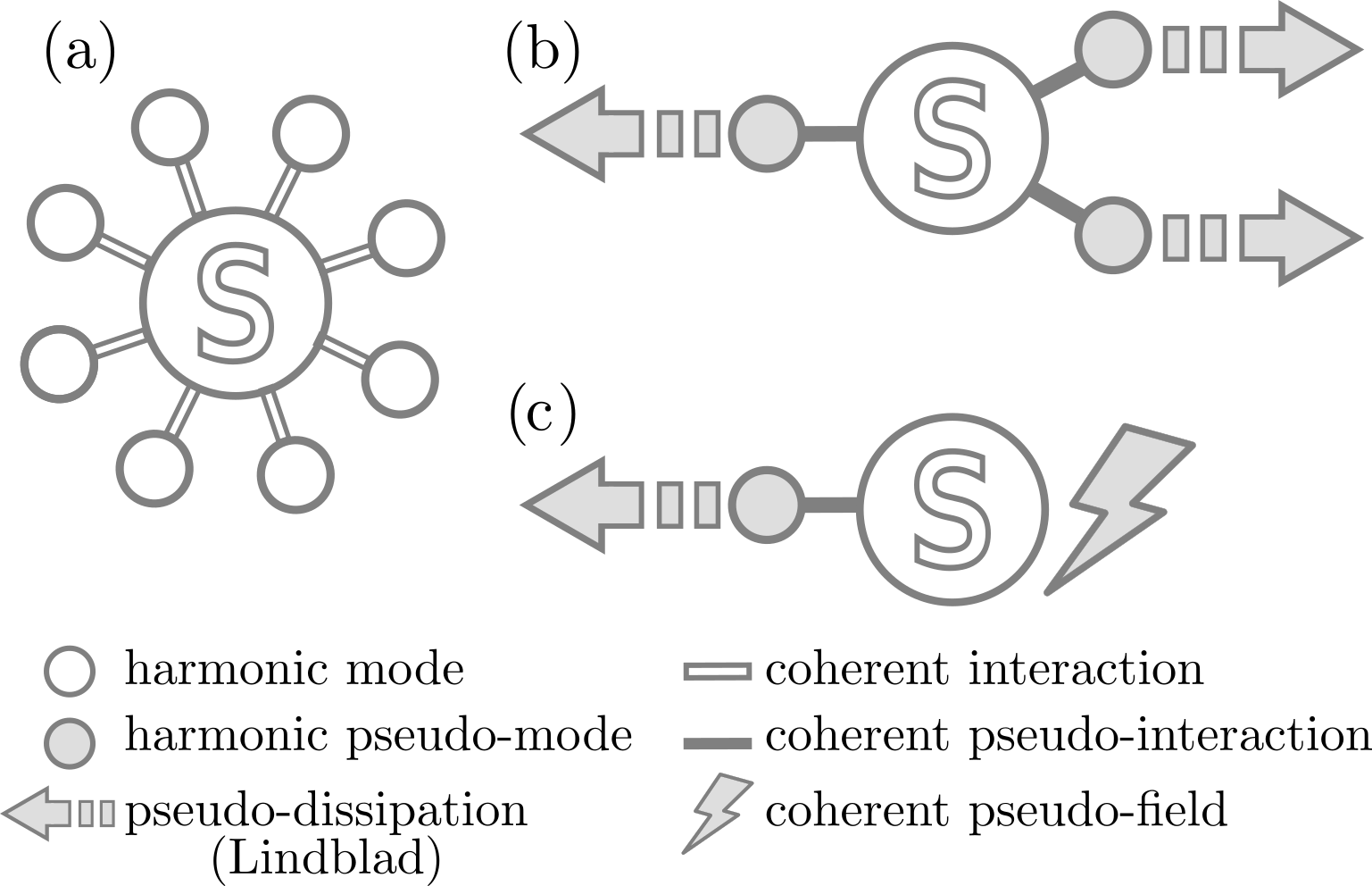} 
\caption{
 \label{fig:cartoon}Schematic representation of the stochastic pseudomode model. (a) A system $S$ interacts with { a physical bath made out of} a continuum of environmental modes. (b) The continuum can be approximated by a discrete set of leaky quantum  pseudomodes, see Eq.~(\ref{eq:PM_dynamics}). (c) The subset of the quantum pseudomodes which generate classical correlations can be replaced [see Eq.~(\ref{eq:stochLindblad})] by the stochastic classical field defined in Eq.~(\ref{eq:xi_spectral_representation_2_main}).}
\end{figure}
with $X_j=a_j+a^\dagger_j$ in terms of the bosonic pseudomode $a_j$ and where $\lambda_j,\Omega_j\in\mathbb{C}$ are formal interaction strengths and energies. We note that when these parameters are allowed to be complex, the original {  Lindblad} equation does not acquire any complex conjugation as one would expect in non-Hermitian systems \cite{PhysRevLett.80.5243,Bender2,Ali,Ozdemir}. For this reason, this equation was dubbed \emph{pseudoShr\"{o}dinger} equation in \cite{Lambert}. Each pseudomode undergoes non-unitary dynamics as described by the dissipators
\begin{equation}
\label{eq:dissipator}
\begin{array}{lll}
D_j[\cdot]=&\Gamma_j[(1+n_j)(2a_j\cdot a_j^\dagger-a_j^\dagger a_j\cdot -\cdot a^\dagger_j a_j)\\
&+n_j(2a^\dagger_j\cdot a_j-a_j a^\dagger_j\cdot -\cdot a_j a^\dagger_j)]\;.
\end{array}
\end{equation}
The initial condition for Eq.~(\ref{eq:pseudoShr\"{o}dinger}) is given by $\rho(0)=\rho_S(0)\prod_j\text{exp}\left[-\beta_j\Omega_j a^\dagger_j a_j\right]/Z_j$, where $Z_j=1/(1-\text{exp}[-\beta_j\Omega_j])$, in terms of the temperature-like parameters $\beta_j\in\mathbb{C}$.
Importantly, the non-unitary dynamics in the system+pseudomodes space described in Eq.~(\ref{eq:pseudoShr\"{o}dinger}) is equivalent to that of an open quantum system in which each pseudomode interacts with quantum-white noise residual environments, see Appendix \ref{app:pseudomode_model}. The Gaussianity of this formal open quantum system implies that the quality of the approximation in Eq.~(\ref{eq:PM_dynamics}) depends on the accuracy to which the correlation of the  pseudomodes coupling operators match the original one. More specifically, the pseudomode model parameters are optimized to minimize the following approximation
\begin{equation}
\label{eq:PM_Corr}
\begin{array}{lll}
C(t)&\simeq&\displaystyle\sum_{j=1}^{N_\text{PM}}\lambda_j^2\langle X_j(t) X_j(0)\rangle \\
&=&\displaystyle\sum_{j=0}^{N_\text{PM}}\lambda_j^2 e^{-\Gamma_j|t|}[(1+n_j)e^{-i\Omega_j t}+n_j e^{i\Omega_j t}]\;,
\end{array}
\end{equation}
for all $t\in\mathbb{R}$ \footnote{ We explicitly note that the constraint in Eq.~(\ref{eq:PM_Corr}) needs to hold for both positive and negative times, see \cite{Paul} for more details. We further comment on the fact that proving Eq.~(\ref{eq:PM_Corr}) for positive times only, and extending the result using the general identity  $C(t)=\bar{C}(-t)$ might require some extra care. In fact, when the parameters on the right hand-side are analytically continued, complex conjugation is, in general, no longer associated with the sign of time. In fact, the parameters of the pseudomode model have specific even (for $\lambda_j, \Omega_j, n_j$) and odd (for $\Gamma_j$) symmetries under time reversal irrespective of their complex nature.} and where $n_j=1/(\exp[\beta\Omega_j]-1)$.  In general, physical correlations of the original model include non-exponential decay at large times which  imply the above approximation can become an exact equality only in the limit $N_{\text{PM}}\rightarrow\infty$.\\

In summary, the pseudomode model allows to replace the original continuum of environmental modes with a discrete set of unphysical modes. This allows to compute the reduced system dynamics non-perturbatively in the system-environment coupling by solving a Lindblad-like differential equation. {  We stress that while Lindblad master equations are usually derived under some perturbative approximation, Eq.~(\ref{eq:pseudoShr\"{o}dinger}) does not rely on such assumption. In fact, as shown in Appenidix \ref{app:pseudomode_model} (see also \cite{Tamascelli,Lambert}), Lindblad dissipation exactly corresponds to an open quantum system whose correlation function is a sum of decaying exponentials, i.e., to the ansatz on the right hand-side of Eq.~(\ref{eq:PM_Corr}). As a consequence, the only approximation made by the pseudomode model is in the bias-error between this ansatz and the original correlation [left hand-side of Eq.~(\ref{eq:PM_Corr})].

In this context, we now introduce a hybrid model in which classical stochastic fields are used alongside quantum pseudomodes. This allows to lift the constraint on the ansatz given in Eq.~(\ref{eq:PM_Corr}) to further reduce, on average, the bias-error. For spectral densities which can be written as a rational function, this error becomes negligible. This comes at the cost of introducing stochastic errors whose magnitude scale inversely proportional to the square-root of the number of realizations used in the averaging. This highlights a qualitative difference in the requirements of the computational resources. In fact, by replacing some of the original quantum modes with classical fields, the dimension of the Hilbert space of the effective model is reduced at the cost of the additional time required for the averaging.
}

\begin{figure*}[t!]
\includegraphics[width = \textwidth]{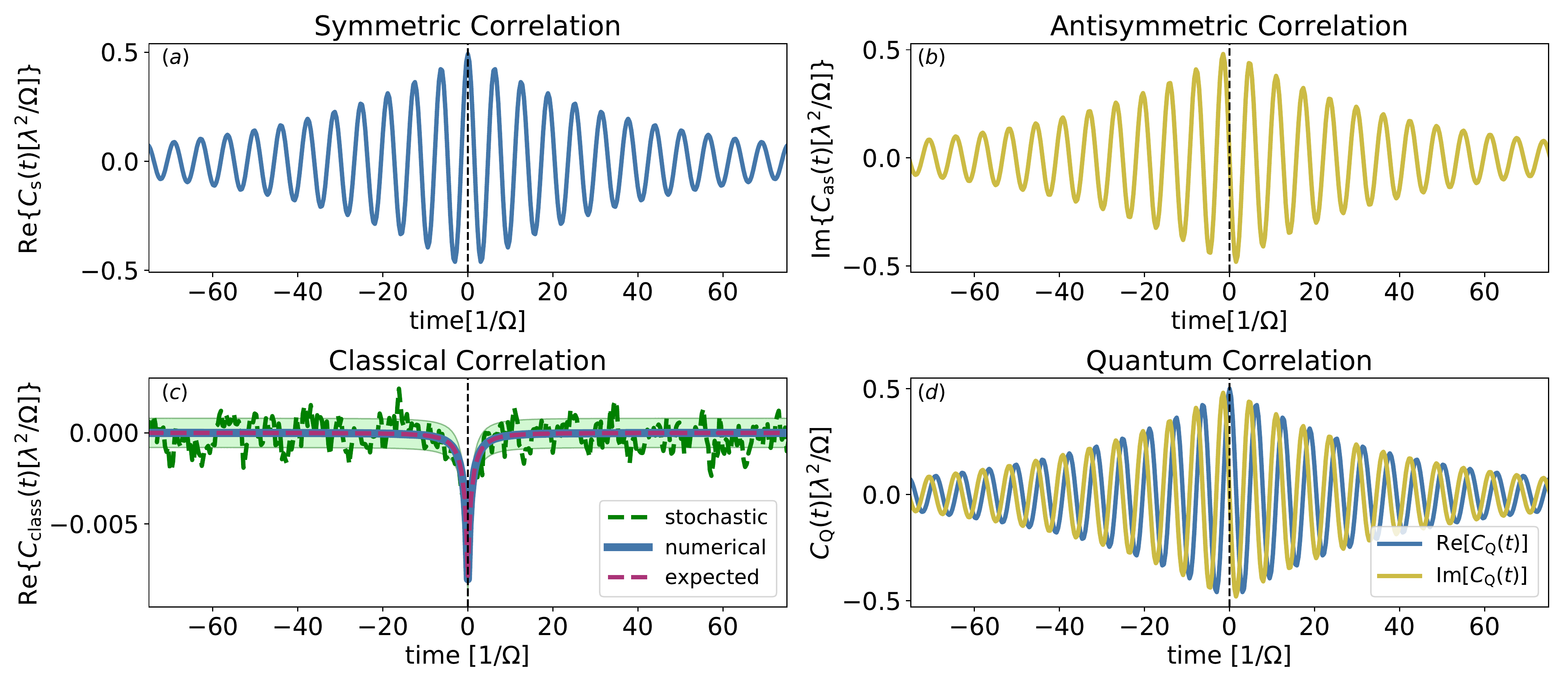} 
\caption{Correlations for the Brownian spectral density at zero temperature. 
 In $(a)$ and $(b)$ we plot the symmetric [Eq.~(\ref{eq:C_s_brownian_underdamped})] and antisymmetric [Eq.~(\ref{eq:C_as_brownian_underdamped})] contribution to the correlation. In $(c)$/$(d)$ we plot the classical [Eq.~(\ref{eq:C_class_brownian_underdamped})] and quantum [Eq.~(\ref{eq:C_Q_brownian_underdamped})] contributions which are obtained by subtracting/adding the symmetric function given in Eq.~(\ref{eq:fas})). At zero temperature, the classical part of the correlation $C_\text{class}(t)$ shown in $(c)$ consists of the Matsubara contribution, see Eq.~(\ref{eq:C_mats_brownian_underdamped}).  In $(c)$, we further show  the empirical average $C^\text{emp}_\text{class}(t)$  [in dashed green, see Eq.~(\ref{eq:C_emp})] of the correlation function, its expected value $C^\mathbb{E}_\text{class}(t)$ [in dashed red, see Eq.~(\ref{eq:E[xixi]})], and its variance [in light green, see Eq.~(\ref{E_emp_sigma})]. As the number $N_\text{stoch}$ of samples  increases, the stochastic reconstruction (green) averages to its expectation value (red) which, in turn, tends to the { exact} numerical expression as the number $N_\xi$ of spectral components in the fields in Eq.~(\ref{eq:xi_spectral_representation_2_main}) approaches infinity. In $(d)$, the quantum contribution $C_\text{Q}(t)$ is modeled by a single exponential function and it does not have any definite symmetry under time reversal. Here, the parameters are the same as those used in Fig.~\ref{fig:brownian_dynamics_T0}(a). Specifically, $N_\xi=1000$ and $N_\text{stoch} = 100$.\label{fig:brownian_corr}}
\end{figure*}
\section{Quantum-Classical decomposition of the correlation function}
\label{sec:Quantum-Classical}
In this section we introduce a non-unique decomposition of the bath correlation function in terms of a quantum and a classical part. By using a spectral expansion, the latter contribution can be reproduced using classical stochastic noise.

We start by noting that the correlation in Eq.~(\ref{eq:correlation}) can be decomposed into two contributions with different symmetries under time reversal, i.e.,  
\begin{equation}
\label{eq:C_symm_antisymm}
\begin{array}{lll}
C(t)&=&C_\text{s}(t)+C_\text{as}(t)\;,
\end{array}
\end{equation}
where
\begin{equation}
\label{eq:C_symm_antisymm_def}
\begin{array}{lll}
C_\text{s}(t)&=&\displaystyle\frac{1}{\pi}\int_0^\infty d\omega~J(\omega)\coth(\beta\omega/2)\cos(\omega t))\\
C_\text{as}(t)&=&-\displaystyle\frac{i}{\pi}\int_0^\infty d\omega~J(\omega)\sin(\omega t)\;.
\end{array}
\end{equation}
are the symmetric and antisymmetric contribution, respectively. In order to provide a more general setup, we are going to consider the slightly different decomposition
\begin{equation}
\label{eq:C_Q_class}
\begin{array}{lll}
C(t)&=&C_\text{class}(t)+C_\text{Q}(t)\;,
\end{array}
\end{equation}
in terms of a ``classical'' and ``quantum'' component defined as $C_\text{class}(t)=C_\text{s}(t)+f_\text{s}(t)$ and $C_\text{Q}(t)=C_\text{as}(t)-f_\text{s}(t)$ respectively. Here, $f_s(t)$ is a generic symmetric function which makes this decomposition non-unique. As we will show, it is possible to take advantage of this feature in order to simplify the description of the quantum degrees of freedom of the pseudomode model. Importantly, the quantum component of the correlation does not have, in general, any  particular symmetry under time reversal (i.e., it can be any linear combination of symmetric and antisymmetric functions) while the classical part is symmetric. This feature allows to interpret the classical contribution to the correlation as originating from stochastic noise.  This can be seen by explicitly writing the following spectral representation of the classical correlation (interpreted as a function in $\mathcal{L}^2([-T,T])$, with $T\in\mathbb{R}$)
\begin{equation}
\label{L2_expansion}
    C_\text{class}(t)=c_0+2\sum_{n=1}^{\infty}c_n \cos[n\pi t/T]\;,
\end{equation}
where
\begin{equation}
\label{eq:xi_cn}
    c_n=\frac{1}{2T}\int_{-T}^Td\tau\cos(n\pi \tau/T)C_\text{class}(t)\;,
\end{equation}
which denotes the inner product between the classical correlation and the sinusoidal functional basis in $\mathcal{L}^2([-T,T])$.
Given this expansion, we can further define a stochastic field as
\begin{equation}
\label{eq:xi_spectral_representation_2_main}
    \xi(t)=\sqrt{c_0} \xi_0+\sum_{n=1}^{N_\xi}\sqrt{2c_n}[\xi_n \cos(n\pi t/T)+\xi_{-n}\sin(n\pi t/T)].
\end{equation}
where $N_\xi\in\mathbb{N}$ is  a truncation parameter $N_\xi\in\mathbb{N}$ and $\xi_n$, $n=-N_\xi,\cdots,N_\xi$ are independent Gaussian random variables with zero mean and unit variance. This spectral representation is designed to allow the correlation of the stochastic field to be equivalent to the classical contribution $C_\text{class}(t)$, i.e., 
\begin{equation}
\label{eq:E[xixi]}
\begin{array}{lll}
    C_\text{class}^\mathbb{E}(t)&\equiv&\displaystyle\mathbb{E}[\xi(t_2)\xi(t_1)] =c_0+2\sum_{n=1}^{N_\xi}c_n \cos[n\pi t/T]\\
    &\rightarrow & C_\text{class}(t)\;,
    \end{array}
\end{equation}
in the limit $N_\xi\rightarrow\infty$, and
where $\mathbb{E}[\cdot]$ denotes the average over the Gaussian random variables $\xi_n$. It is interesting to note that the presence of the (antisymmetric) sine function in the expression for $\xi(t)$ is essential to reproduce a stationary correlation, see Appendix \ref{app:spectral_representation} for more details. \\
{ In summary, the procedure to define the stochastic process can be outlined as follows.
\begin{itemize}
    \item First, we decompose the classical part of the correlation function into Fourier components as described in Eq.~(\ref{L2_expansion}). This involves the computation of $N_\xi+1$ coefficients given explicitly in Eq.~(\ref{eq:xi_cn}). 
    \item Second, we proceed to extract $2N_\xi+1$ Gaussian random numbers $\xi_n$ which can, together with the coefficients computed above, directly used in Eq.~(\ref{eq:xi_spectral_representation_2_main}) to compute the stochastic field. 
    \item The field in Eq.~(\ref{eq:xi_spectral_representation_2_main})  can then be used to solve the stochastic Shr\"{o}dinger equation in Eq.~(\ref{eq:stochLindblad}) presented in the next section.
\end{itemize}}

{ The above mentioned Eq.~(\ref{eq:stochLindblad}) defines} a stochastic pseudomode model in which classical stochastic noise {  is used alongside quantum pseudomodes to reduce both the  dimension of the effective Hilbert space and the bias errors in the approximation of the original correlation function.} Furthermore, it is possible to take advantage of the non-uniqueness of the representation in Eq.~(\ref{eq:C_Q_class}) to allow all quantum degrees of freedom to be initially at zero temperature. We show this in the following section.

\section{A Stochastic Pseudomode model}
\label{sec:stochastic}
In this section we introduce a { hybrid pseudomode model in which classical stochastic fields and quantum pseudomodes are used alongside} to simulate different contributions to the original bath.

We  define  the following stochastic pseudomode model
\begin{equation}
\label{eq:stochLindblad}
\dot{\rho}^\xi(t)=-i[H^\xi_\text{PM},\rho^\xi(t)]+\sum_{j=0}^{N_\text{PM}} D_j[\rho^\xi(t)]\rho(t)\;,
\end{equation}
defined for times $t\in[0,T_\text{dynamics}]$, with $T_\text{dynamics}\leq T$ and where the  Hamiltonian
\begin{equation}
\label{eq:Hybrid model}
H^\xi_{\text{PM}}=H_S+\sum_{j=1}^{N_\text{PM}} \left(\lambda_j X_j{s}+\Omega_j a^\dagger_j a_j\right)+{s}\xi(t)\;,
\end{equation}
 depends on $N_\text{PM}$ pseudomodes $a_j$ [having frequency $\Omega_j$ and interaction operator $\lambda_j X_j=\lambda_j(a_j+a_j^\dagger)$] and an extra classical Gaussian stationary stochastic field $\xi(t)$ defined by Eq.~(\ref{eq:xi_spectral_representation_2_main}). Both the pseudomodes and classical field interact to the system through the system operator ${s}$. The pseudomodes are initially in the ``thermal'' state $\rho(0)=\rho_S(0)\text{exp}[-\sum_j\beta_j\Omega_j a^\dagger_j a_j]/Z_j$, where $Z_j=1/(1-\text{exp}[-\beta_j\Omega_j])$, where $\beta_j\in\mathbb{C}$. The classical and quantum degrees of freedom are designed to respectively reproduce the classical and quantum part of the original correlation function, i.e., 
\begin{equation}
\label{eq:CCCQ}
\begin{array}{lll}
C^\mathbb{E}_\text{class}(t)&=&\displaystyle\mathbb{E}[\xi(t_2)\xi(t_1)]\simeq C_\text{class}(t)\\
C_\text{Q}(t)&\simeq&\displaystyle\sum_{j=0}^{N_\text{PM}}\lambda_j^2\langle X_j(t_2)X_j(t_1)\rangle\\
&=&\displaystyle\sum_{j=0}^{N_\text{PM}}\lambda_j^2 e^{-\Gamma_j|t|}[(1+n_j)e^{-i\Omega_j t}+n_j e^{i\Omega_j t}]\;,
\end{array}
\end{equation}
where $t=t_2-t_1$ and $n_j=1/(\exp[\beta\Omega_j]-1)$.
Here, the first equation  is a version of Eq.~(\ref{eq:E[xixi]}) for finite $N_\xi$, while the second one corresponds to the pseudomode representation  of the correlation function as given in Eq.~(\ref{eq:PM_Corr}). We note that, { within the pseudomode model defined by Eq.~(\ref{eq:PM_dynamics})},  it is not possible to reproduce this quantum part without either introducing some complex-conjugation (which, in turn, would lead to different higher order statistics) or, more generally, non-commuting variables (i.e., quantum degrees of freedom, as done in here). 
 Under the approximations in Eq.~(\ref{eq:CCCQ}), it is possible to write the following stochastic version of the main pseudomode equation as
\begin{equation}
\label{eq:Erho}
    \rho_S(t)\simeq \mathbb{E}\left[\rho_S^\xi(t)\right]\;,
\end{equation}
where we defined $\rho_S^\xi(t)=\text{Tr}_\text{PM}\rho^\xi(t)$.
An exact evaluation of  this expectation value requires the knowledge of $\rho_S^\xi(t)$ for all $\xi(t)$ which is, in general, not available. However, for any system observable $O$, we can define the following empirical average
\begin{equation}
\label{eq:empirical_average_main}
\mathcal{O}_\text{emp}^\xi(t;N_\text{stoch})=\frac{1}{N_\text{stoch}}\sum_{j=1}^{N_\text{stoch}}\mathcal{O}^{\xi_j}(t)\;,
\end{equation}
in terms of the quantity
\begin{equation}
\mathcal{O}^{\xi_j}(t)=\text{Tr}_{S} \left[O\rho^{\xi_j}_S(t)\right], 
\end{equation}
where the label $j=1,\cdots,N_\text{stoch}$ in the field $\xi_j$ characterizes one of 
 the different $N_\text{stoch}$ realizations of the noise. Using Eq.~(\ref{eq:Erho}), it is clear that this random variable has the correct expectation value, i.e.,
\begin{equation}
\label{eq:biasFrommean}
\mathcal{O}(t)\equiv\mathbb{E}\left[\mathcal{O}^\xi(t;N_\text{stoch})\right]\simeq\text{Tr}_{S} \left[{O}\rho_S(t)\right]\equiv \mathcal{O}_\text{true}\;.
\end{equation}
{  The model described by Eq.~(\ref{eq:stochLindblad}) and Eq.~(\ref{eq:Erho}) constitutes the main result of this work. It defines the dynamics of the reduced density matrix of an open quantum system in terms of a set of $N_\text{PM}$ quantum pseudomodes and a stochastic field $\xi(t)$ [explicitly defined in Eq.~(\ref{eq:xi_spectral_representation_2_main})] which reproduce the physical effects of the original bath related to, respectively, the quantum and classical contributions of the full bath correlation [see Eq.~(\ref{eq:CCCQ})]. 
As the statistics of the classical field can encode any time-symmetric functional dependence [see Eq.(\ref{eq:E[xixi]})], this method can model correlations which are not limited by the decaying-exponential ansatz in Eq.~(\ref{eq:PM_Corr}). In parallel, this model also allows to improve the accuracy of a deterministic model whenever this is prevented by the impossibility to further increase the Hilbert space dimension. Ultimately, the introduction of randomness in the model requires a characterization of the variance of the result which we are going to analyze in the next section. }

\subsection{Bias and Stochastic Errors}
\label{sec:bias_stoch}
We start this section by analyzing Eq.~(\ref{eq:CCCQ}) in some more detail. 
In fact, there is a conceptual difference between the approximations made in the two lines of Eq.~(\ref{eq:CCCQ}). On the quantum side [second line in Eq.~(\ref{eq:CCCQ})], the approximation consists in writing the correlation as a finite sum of decaying exponentials and its accuracy can only be increased by considering more pseudomodes or, in other words, by increasing the dimension of the Hilbert space.
On the classical side [first line in Eq.~(\ref{eq:CCCQ})], the approximation error  is gauged by the cut off parameter $N_\xi$ (which can, in principle, be tuned up to increase the accuracy). At the same time, the presence of stochastic noise in the Shoredinger equation in Eq.~(\ref{eq:stochLindblad}) means that all observables are now random variables whose statistics should be further analyzed. 

To quantify these sources of inaccuracy in more detail, we can define the following measure for the total error on the observable $O$
\begin{equation}
    \Delta \mathcal{O}(t)=\mathbb{E}[|\mathcal{O}^\xi(t;N_\text{stoch})-\mathcal{O}_\text{true}(t)|]\;,
\end{equation}
Using the triangle inequality, we can obtain an upper bound on this error as
\begin{equation}
    \Delta \mathcal{O}(t)\leq \Delta_\text{stoch} \mathcal{O}(t)+\Delta_\text{bias} \mathcal{O}(t)\;,
\end{equation}
where 
\begin{equation}
\begin{array}{lll}
      \Delta_\text{stoch} \mathcal{O}(t)&=&\displaystyle\mathbb{E}[|\mathcal{O}^\xi(t;N_\text{stoch})-\mathcal{O}(t)|]\\
     \Delta_\text{bias} \mathcal{O}(t)&=&\displaystyle|\mathcal{O}(t)-\mathcal{O}_\text{true}(t)|\;.
\end{array}
\end{equation}
are the stochastic and bias errors respectively. As shown in \cite{Mascherpa} and in Appendix \ref{app:bias}, the bias error can be upper bounded by a quantity which scales asymptotically as an exponential in 
\begin{equation}
    \Delta C=\int_0^T d t_2 \int_0^{t_2}dt_1 |C_\text{err}(t_2,t_1)|\;,
\end{equation}
where 
\begin{equation}
    C_\text{err}(t_2,t_1)=C_\text{class}+C_\text{Q}(t_2-t_1)-C(t_2-t_1)\;,
\end{equation}
is the approximation error of the original correlation function. As mentioned above, in general, the quality of this approximation can be improved by increasing the cut off $N_{\xi}$ (for the classical contribution), and the number $N_\text{PM}$ of pseudomodes (for the quantum contribution).  For the models in which the quantum part can be modeled exactly with a finite number of pseudomodes (see section \ref{sec:rational}), the bias error in the correlation is only due to the finiteness of the spectral expansion of the classical field and can always be improved. Specifically, the bias is simply given
by the difference between Eq.~(\ref{L2_expansion}) and Eq.~(\ref{eq:E[xixi]}) and corresponds to $2\sum_{n=N_\xi+1}^\infty c_n \cos[n\pi t/T]$ so that $ \Delta_\text{bias}\rightarrow 0$ in the $N_{\xi}\rightarrow\infty$ limit.

 By analyzing the properties of the random variables involved (see  Appendix \ref{app:stochastic_error} for specific details), it is also possible to bound the stochastic error by a quantity which scales as $O({1}/{\sqrt{N_\text{stoch}}})$. However, as with the bias, the explicit form of this bound asymptotically scales exponentially in time (see Appendix \ref{app:stochastic_error}), effectively limiting its practicality. However, it is  possible to further provide an alternative empirical bound in the form
\begin{equation}
    \Delta^2_\text{stoch}\mathcal{O}(t)\leq \sigma^2_{\mathcal{O}^\xi(t;N_\text{stoch})}
    \;,
\end{equation}
in terms of the variance of the empirical average $\mathcal{O}^\xi(t;N_\text{stoch})$. This variance can  be further estimated as
\begin{equation}
\label{eq:empirical_sigma}
\sigma^2_{\mathcal{O}^\xi(t;N_\text{stoch})}=\displaystyle\frac{\displaystyle\frac{\sum_j [\mathcal{O}^{\xi_j}(t)]^2}{N_\text{stoch}}-\left[\frac{\sum_j\mathcal{O}^{\xi_j}(t)}{N_\text{stoch}}\right]^2}{N_\text{stoch}}\;,
\end{equation}
up to terms of order $O(1/N^{3/2}_\text{noise})$, thereby allowing a numerical estimate.
Similarly, it is possible to define an empirical average to estimate the expected correlation in Eq.~(\ref{eq:E[xixi]}) as
\begin{equation}
\label{eq:C_emp}
    C^\text{emp}_\text{class}(t_2,t_1)=\frac{1}{N_\text{stoch}}\sum_{j=1}^{N_\text{stoch}}\xi_j(t_2)\xi_j(t_1)\;,
\end{equation}
whose expectation value is $\mathbb{E}[C^\text{emp}_\text{class}(t_2,t_1)]=C^\mathbb{E}_\text{class}(t_2,t_1)$, and whose variance satisfies the bound $\sigma^2_\text{emp}\leq[|{C}(0)|^2+|{C}(t_2-t_1)|^2]/{N_\text{stoch}}$.

 In summary, for all models in which the bias error in reproducing the correlation function is negligible, we can estimate the value of an observable $\mathcal{O}(t)$ by using a single realization of the empirical average in Eq.~(\ref{eq:empirical_average_main}) which requires to solve the dynamics of the system $N_\text{stoch}$ times and whose variance can be estimated using Eq.~(\ref{eq:empirical_sigma}). We note that, by using the stochastic formalism presented here, reaching a regime with negligible bias error is possible. In fact, for all models (with rational spectral densities) presented in section \ref{sec:rational} in which  the quantum part of the correlation can be exactly reproduced using a finite number of pseudomodes, the bias error can be made arbitrary small by simply increasing the number $N_\xi$ of classical spectral components of the field in Eq.~(\ref{eq:E[xixi]}).

\subsection{A Zero-Temperature Stochastic Pseudomode}
\label{sec:ZeroT_main}

The stochastic pseudomode model presented in the previous section is valid for any decomposition of the correlation function into a classical and quantum part as described in Eq.~(\ref{eq:C_Q_class}). In this section, we show that it is possible to take advantage of the non-uniqueness of this decomposition in order to define all the quantum degrees of freedom initially at zero temperature.  This is consistent  with the general intuition that ``quantumness'' is required to generate zero-temperature effects (or, more precisely, to generate the asymmetric tuning of absorption and emission rates required to model detailed balance at finite temperatures) while stochastic  noise can be used to reproduce classical statistical uncertainty.

To achieve this, we follow the straightforward strategy of adapting the pseudomode ansatz to the antisymmetric part of the correlation. We then use this ansatz to characterize a symmetric function $f_s(t)$ leading to a quantum correlation compatible with a zero temperature model. Explicitly, our starting point is the (unique) decomposition in Eq.~(\ref{eq:C_symm_antisymm}) of the correlation function into a symmetric and antisymmetric part. Given its symmetries, and compatibly with the pseudomode ansatz, we assume the following expansion for the antisymmetric contribution
\begin{equation}
\label{eq:as_ansatz}
C_\text{as}(t)=\sum_{j=1}^{N_\text{PM}}a_j \sin(b_j t)e^{-c_j |t|}\;,
\end{equation}
in terms of the parameters $a_j, b_j, c_j\in\mathbb{C}$, $j=1,\cdots,N_\text{PM}$. We note that this is assumption does not imply more restrictions than those required by the pseudomode model as it writes the antisymmetric contribution as a sum of decaying sines. The ansatz is also compatible with a wide range of spectral densities, see section \ref{sec:rational}. We can now define
\begin{equation}
\label{eq:fas}
f_\text{s}(t)=-i\sum_{j=1}^{N_\text{PM}}a_j\cos(b_j t)e^{-c|t|}\;,
\end{equation}
which, inserted in Eq.~(\ref{eq:C_Q_class}), leads to the following classical and quantum contributions
\begin{equation}
\label{eq:class_Q}
    \begin{array}{lll}
C_\text{class}(t)&=&C_\text{s}(t)+f_\text{s}(t)\\
C_\text{Q}(t)&=&C_\text{as}(t)-f_\text{s}(t)=\displaystyle\sum_{j=1}^{N_\text{PM}}ia_j e^{-ib_j t}e^{-c_j |t|}\;,
    \end{array}
\end{equation}
where we used Eq.~(\ref{eq:as_ansatz}) and Eq.~(\ref{eq:fas}). Comparing this expression with Eq.~(\ref{eq:CCCQ}), we can appreciate the presence of only zero-temperature pseudomodes, i.e., $n_j=0$ for all $j$. More specifically, the parameters of the pseudomode model are $\lambda^2_j=ia_j$, $\Omega_j = b_j$, $\Gamma_j=c_j$. It is interesting to explicitly note that the pseudomodes $j$ is physical for $b_j,c_j\in\mathbb{R}$ and for imaginary $a_j$ (compatibly with the imaginary nature of the  antisymmetric correlation). We also explicitly note that Eq.~(\ref{eq:fas}) is not uniquely defined as, for example, an overall sign in Eq.~(\ref{eq:as_ansatz}) can be encoded in either the parameters $a_j$ or $b_j$.

We conclude mentioning that the procedure presented in this section is not the only possible one as any of the quantum-classical decomposition described in section \ref{sec:Quantum-Classical} can be more appropriate to model a specific environment. We show this in Appendix \ref{app:ohmic} where, in the case of a pure dephasing environment, the decomposition is chosen imposing $f_s(t)=0$ [instead of Eq.~(\ref{eq:fas})] in Eq.~(\ref{eq:C_symm_antisymm}).

\section{Stochastic Pseudomode model for Rational Spectral densities}
\label{sec:rational}
In this section, we analyze the stochastic pseudomode method to model Gaussian environments characterized by rational spectral densities. More specifically, we consider spectral densities which can be written as
\begin{equation}
J_\text{r}(\omega)=\frac{p(\omega)}{q(\omega)}\;,
\end{equation}
where $p$ and $q$ are polynomials such that the order of the denominator is strictly higher than the numerator and such that  the zeros of the denominator are different than the zeros of the numerator. We further assume that $J_\text{r}(\omega)\in\mathbb{R}$ for $\omega\in\mathbb{R}$, that $J_\text{r}(-\omega)=-J_\text{r}(\omega)$, and that all poles of $J_\text{r}(\omega)$ are simple and not located on the real axis. Since these spectral densities are real on the real axis, their poles come in conjugate pairs. Furthermore, antisymmetry implies that the poles come either in imaginary conjugate pairs (labeled by $k_\text{i}=1,\cdots, N_\text{i}$) such as $(\omega_{k_\text{i}},\omega^{\mathcal{R}}_{-k_\text{i}})$ for $\text{Re}[\omega_{k_\text{i}}]=0$ and $\text{Im}[\omega_{k_\text{i}}]>0$ or in quadruples  (labeled with $k_\text{q}=1,\cdots, N_\text{q}$) as $(\omega_{k_\text{q}},\omega_{-k_\text{q}},-\omega_{k_\text{q}},-\omega_{-k_\text{q}})$ for $\text{Re}[\omega_{k_\text{q}}]>0$. Here, we used the notation $\omega_{k}=\bar{\omega}_{-k}$. We can explicitly note that the total number of poles is $N_\text{poles}=2N_\text{i}+4N_\text{q}$.

Using this notation, the antisymmetric  part of the correlation  in Eq.~(\ref{eq:C_symm_antisymm_def}) reads
\begin{equation}
\label{eq:Cas_sign}
     \begin{array}{lll}
    C_\text{as}(t)
   &=&-2\displaystyle i\text{sg}(t)\sum_{{k_\text{q}}=1}^{N_\text{q}}R^{\mathcal{R}}_{k_\text{q}}\cos(\omega_{k_\text{q}}^{\mathcal{R}}t)e^{-\omega^\mathcal{I}_{k_\text{q}}|t|}\\
   &&+2\displaystyle i\sum_{{k_\text{q}}=1}^{N_\text{q}}R^\mathcal{I}_{k_\text{q}}\sin(\omega_{k_\text{q}}^{\mathcal{R}}t)e^{-\omega^\mathcal{I}_{k_\text{q}}|t|}\\

   &&-\displaystyle i\text{sg}(t)\sum_{{k_\text{i}}=1}^{N_\text{i}}R^\mathcal{R}_{k_\text{i}}\cos(\omega_{k_\text{i}}^\mathcal{R}t)e^{-\omega^\mathcal{I}_{k_\text{i}}|t|}\;,
    \end{array}
 \end{equation}
where $R_k=\text{Res}[J_\text{r}(\omega);\omega_k]$ denotes the residue of the spectral density at the simple  pole $\omega_k\in\mathbb{C}$. Furthermore, we defined $\text{sg}(t)=t/|t|$ for $t\neq 0$, and we used the labels $\mathcal{R}$ and $\mathcal{I}$ to denote the real and imaginary part of a parameter, respectively. We can note that the sign function appearing in this expression makes a direct comparison  with the ansatz in Eq.~(\ref{eq:as_ansatz}) not immediately available. However, it is possible to make progress by weakening the meaning of correlation function to a distribution. In fact, since we are ultimately interested in computing the reduced density matrix of the system, the correlation in Eq.~(\ref{eq:correlation}) always appear inside definite integrals in time thereby effectively defining it as a functional over system's observables, i.e., a distribution. This conceptual observation is not just abstract as it allows to concretely modify the expression on the right hand-side of Eq.~(\ref{eq:Cas_sign}) on any zero-measure subspace of the real axis (by simply reinterpreting the equal sign in the  distributions space). These considerations allow us to write (omitting the time dependence on the right hand-side)
  \begin{equation}
  \label{eq:C_as(t)_before_PM_main}
     \begin{array}{lll}
    C_\text{as}
       &=&\displaystyle 2\sum_{{k_\text{q}}=1}^{N_\text{q}}\left\{iR^{\mathcal{I}}_{k_\text{q}}\sin(\omega_{k_\text{q}}^{\mathcal{R}}t)e^{-\omega^{\mathcal{I}}_{k_\text{q}}|t|}\right.\\
   &+&\displaystyle R^{\mathcal{R}}_{k_\text{q}}\sin\left[\frac{-i(W-\omega^\mathcal{I}_{k_\text{q}})+2\omega_{k_\text{c}}^{\mathcal{R}}}{2}t\right]e^{-(W+\omega^{\mathcal{I}}_{k_\text{q}})|t|/2}\\
   &+&\left.\displaystyle R^{\mathcal{R}}_{k_\text{q}}\sin\left[\frac{-i(W-\omega^{\mathcal{I}}_{k_\text{q}})-2\omega_{k_\text{c}}^{\mathcal{R}}}{2}t\right]e^{-(W+\omega^{\mathcal{I}}_k)|t|/2}\right\}\\

   &+&\displaystyle 2\sum_{k_\text{i}=1}^{N_\text{i}}\displaystyle R^{\mathcal{R}}_{k_\text{i}}\sin\left[\frac{-i(W-\omega^{\mathcal{I}}_{k_\text{i}})}{2}t\right]e^{-(W+\omega^{\mathcal{I}}_{k_\text{i}})|t|/2}\;.
    \end{array}
 \end{equation}
in terms of an arbitrary energy-parameter $W$ which is supposed to be much bigger than any frequency which can be associated with the system. This expression allows a direct comparison to the ansatz in Eq.~(\ref{eq:as_ansatz}) and ultimately implies that $N_\text{PM}=3N_\text{q}+2N_\text{i}$ zero-temperature pseudomodes are needed to \emph{exactly} reproduce the antisymmetric part of the correlation for this class of spectral densities. We note that, in practice, a smaller number of pseudomodes can be considered at the price of downgrading Eq.~(\ref{eq:as_ansatz}) to a best-fit approximation.
\begin{figure*}[t!]
\includegraphics[width = \columnwidth]{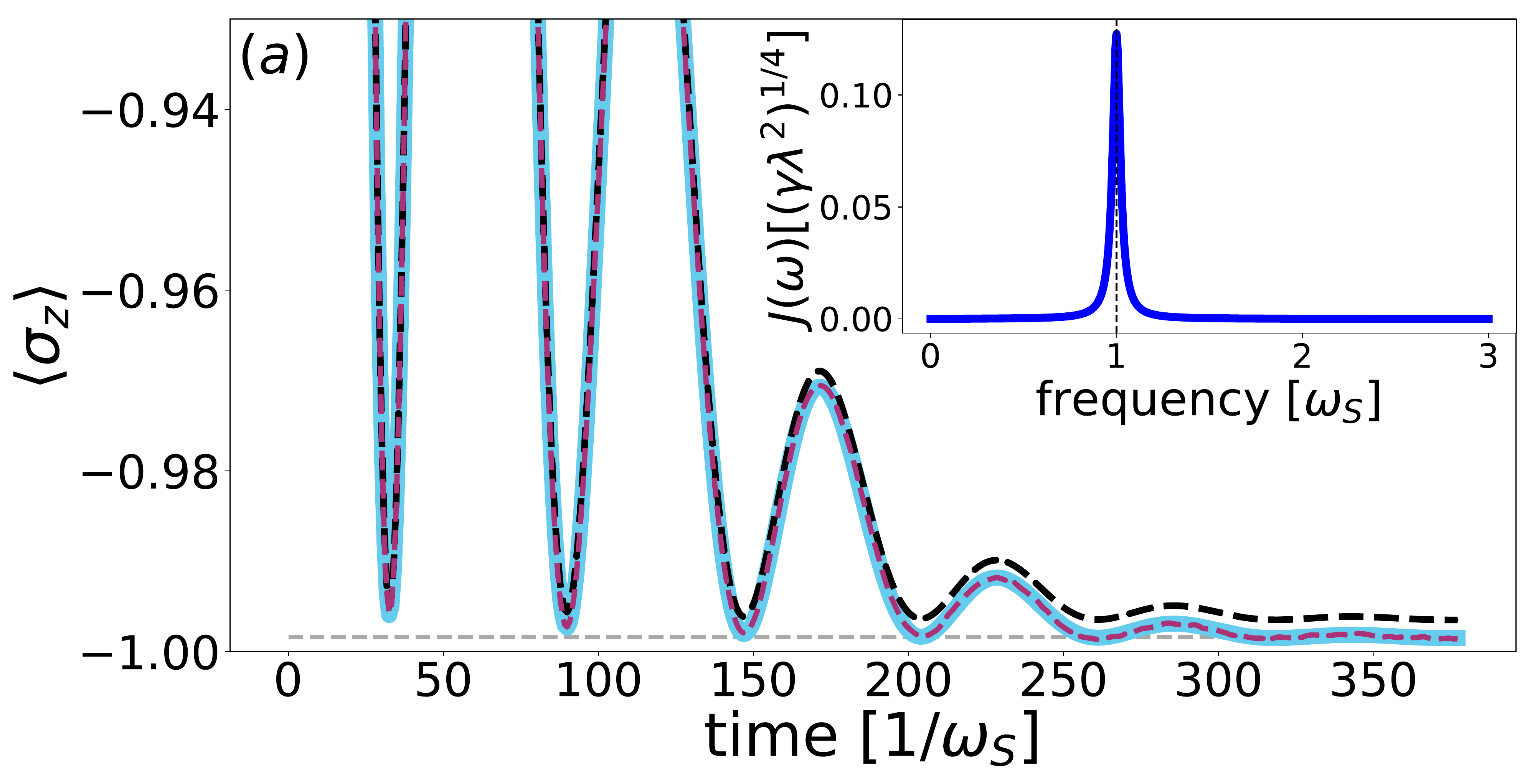} 
\includegraphics[width = \columnwidth]{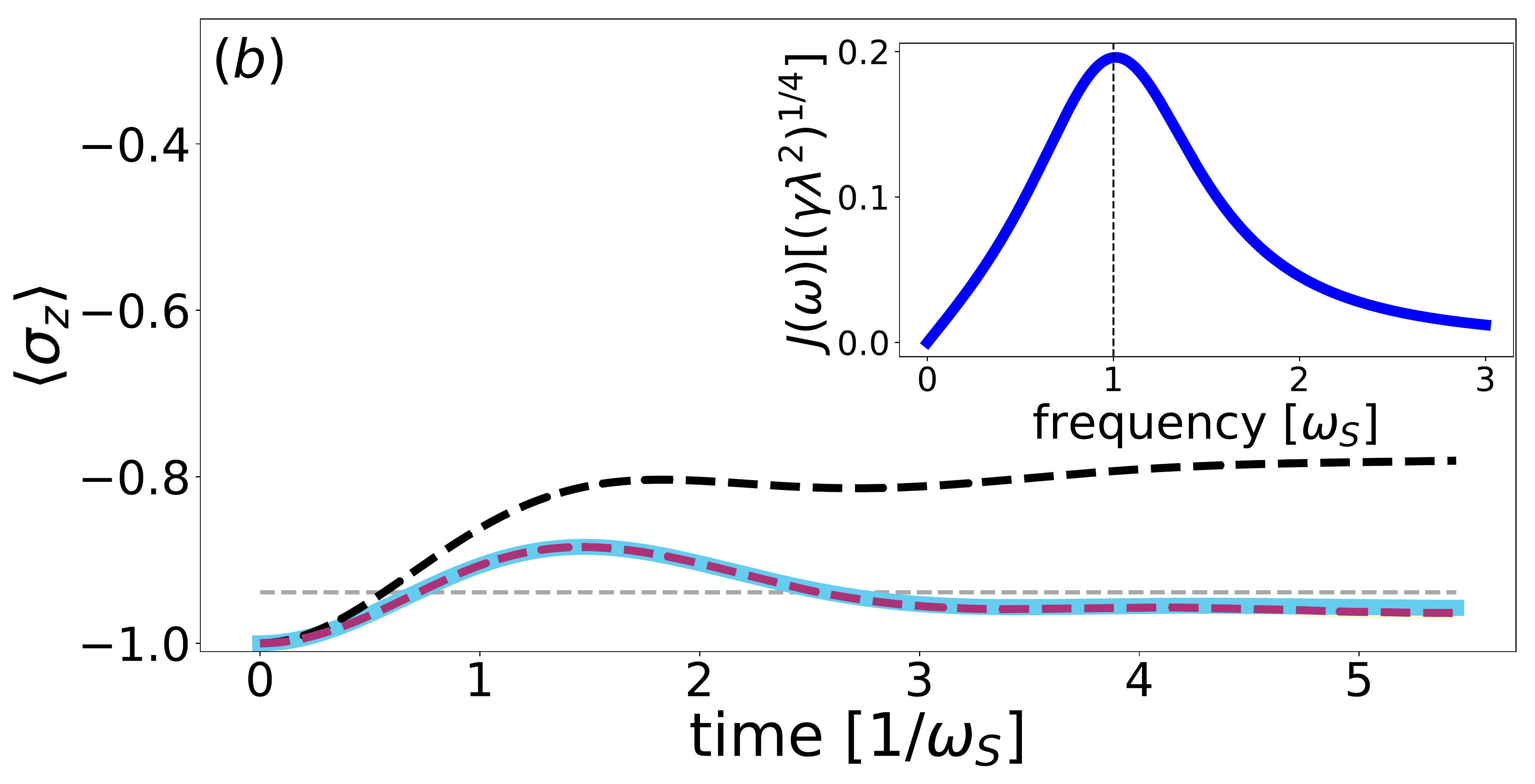}
\caption{Dynamics of a two level system with Hamiltonian $H_S=(\omega_s/2)\sigma_z$ and coupling operator ${s}=\sigma_x$ strongly coupled to a bath with a narrow and broad Brownian spectral density (insets) at zero temperature. In blue the evolution follows a standard pseudomode model in which all degrees of freedom are quantized (1 mode for the quantum contribution and 2 modes for the classical one). In dashed-black the evolution without the contribution from the Matsubara correlation. In dashed-red the evolution of the empirical average in Eq.~(\ref{eq:empirical_average_main}) for $N_\text{stoch}=100$
is evaluated with the stochastic pseudomode model  involving a single harmonic oscillator to model the quantum contribution. 
The horizontal grey dashed line represents the value of the observable for the ground state of a Rabi model with Hamiltonian $H_B$ and zero fields. In $(b)$, the failure to reach to this value can be interpreted { as due to} hybridization to residual modes in the environment (as a consequence of the broad resonance) which are correctly taken into account by both the standard and stochastic pseudomode model,  see \cite{Lambert} for more details about the effects of the Matsubara terms in the { deterministic} pseudomode model. In $(a)$, the parameters are $\omega_0=\omega_s$, $\gamma=0.05\omega_s$, $\lambda=(0.2/\sqrt{2\pi})\omega^{3/2}_s$, $N_\xi=1000$, $N_\text{stoch} = 100$. In $(b)$, the parameters are $\omega_0=\omega_s$, $\gamma=\omega_s$, $\lambda=(1/\sqrt{2\pi})\omega^{3/2}_s$, $N_\xi=100$, and $N_\text{stoch}=1000$. {  To further highlight the different hybridization properties of the two cases, we used different initial conditions for the two-level system (excited state in $(a)$ and ground state in $(b)$.}\label{fig:brownian_dynamics_T0}}
\end{figure*}
While the analytical expression for the antisymmetric contribution to the correlation function above allows us to give an upper bound on the number of pseudomodes, a similar analytical expression for the symmetric part is not as useful as the expansion in Eq.~(\ref{L2_expansion}) can always be computed numerically. Nonetheless, such an expression can be obtained and it explicitly reads (for $\beta\in\mathbb{R}$)
\begin{equation}
\begin{array}{lll}
    C_\text{s}
      &=&-2\displaystyle\sum_{{k_\text{q}}}[{R}^{\beta,R}_{k_\text{q}} \sin{(\omega^{R}_{k_\text{q}} |t|)}+{R}^{\beta,I}_{k_\text{q}} \cos{(\omega^{R}_{k_\text{q}} t)}] e^{-\omega_{k_\text{q}}^I|t|}\\

            &&-\displaystyle\sum_{{k_\text{i}=1}}^{N_\text{i}}{R}^{\beta,I}_{k_\text{i}}e^{-\omega_{k_\text{i}}^I|t|}+\displaystyle\frac{2i}{\beta}\sum_{k>0}J(\omega^\text{M}_{k})e^{-|\omega^\text{M}_{k}||t|}\;,
    \end{array}
\end{equation}
where $k_\text{q}=1,\dots,{N_\text{q}}$, $R^\beta_k=R_k \coth(\beta\omega_k/2)$, and in terms of the Matsubara poles $\omega^\text{M}_k=2\pi i k/\beta$ ($k>0$) of $\coth(\beta\omega/2)$.
By using the prescriptions described in section \ref{sec:ZeroT_main}, we can use Eq.~({\ref{eq:class_Q}}) to compute the quantum and classical contributions to the correlations $C_\text{Q}(t)=C_\text{as}(t)-f_\text{s}(t)$, as
 \begin{equation}
  \label{eq:C_as(t)_before_PM_3_main}
     \begin{array}{lll}
    C_\text{Q}(t)
       &=&\displaystyle -2\sum_{{k_\text{q}}=1}^{N_\text{q}}\left\{R^\mathcal{I}_{k_\text{q}}e^{-i\omega_{k_\text{q}}^\mathcal{R}t}e^{-\omega^\mathcal{I}_{k_\text{q}}|t|}\right.\\
   &+&i\displaystyle R^\mathcal{R}_{k_\text{q}}e^{-i[\omega_{k_\text{q}}^\mathcal{R}-i(W-\omega^\mathcal{I}_{k_\text{q}})/2]t}e^{-(W+\omega^\mathcal{I}_{k_\text{q}})|t|/2}\\
   &+&i\left.\displaystyle R^\mathcal{R}_{k_\text{q}}e^{-i[-\omega_{k_\text{q}}^\mathcal{R}-i(W-\omega^\mathcal{I}_{k_\text{q}})/2]t}e^{-(W+\omega^\mathcal{I}_{k_\text{q}})|t|/2}\right\}\\

   &+&\displaystyle 2i\sum_{k_\text{i}=1}^{N_\text{i}}\displaystyle R^\mathcal{R}_{k_\text{i}}e^{-i[-i(W-\omega^\mathcal{I}_{k_\text{i}})/2]t}e^{-(W+\omega^\mathcal{I}_{k_\text{i}})|t|/2}\;,
    \end{array}
 \end{equation}
 and 
 \begin{equation}
  \label{eq:C_as(t)_before_PM_33}
     \begin{array}{lll}
    C_\text{class}
          &=&i\displaystyle\sum_{{k_\text{q}=1}}^{N_\text{q}}[\Delta R_{k_\text{q}} ~e^{i\omega_{k_\text{q}}|t|}-\Delta'R_{k_\text{q}}~e^{-i\bar{\omega}_{k_\text{q}}|t|}]\\

            &+&i\displaystyle\sum_{{k_\text{i}=1}}^{N_\text{i}}\Delta R_{k_\text{i}}~e^{i\omega_{k_\text{i}}|t|}+\displaystyle\frac{2i}{\beta}\sum_{k>0}J(\omega^\text{M}_{k})e^{-|\omega^\text{M}_{k}||t|}.

    \end{array}
 \end{equation}
 in terms of the quantities $\Delta R_{k_\text{x}}=R^\beta_{k_\text{x}}-R_{k_\text{x}}$ and $\Delta' R_{k_\text{x}}=\bar{R}^\beta_{k_\text{x}}-R_{k_\text{x}}$, ($\text{x}=\text{q},\text{i}$).
 The quantum correlation  corresponds to a decomposition like the one presented in Eq.~(\ref{eq:CCCQ}) which can be modeled by pseudomodes initially at zero temperature.
On the other hand, the explicit expression for the classical correlation can further be used to explicitly compute the coefficients for the spectral decomposition of the field $\xi$ in Eq.~(\ref{eq:xi_spectral_representation_2_main}) as
\begin{equation}
\begin{array}{lll}
   c_n
          &=&i\displaystyle\sum_{{k_\text{q}=1}}^{N_\text{q}}[\Delta R_{k_\text{q}}~L({i\omega_{k_\text{q}}})-\Delta'R_{k_\text{q}}~L_n({-i\bar{\omega}_{k_\text{q}}})]\\

            &+&i\displaystyle\sum_{{k_\text{i}=1}}^{N_\text{i}}\Delta R_{k_\text{i}}~L_n({i\omega_{k_\text{i}}})+\displaystyle\frac{2i}{\beta}\sum_{k>0}J(\omega^\text{M}_{k})L_n({-|\omega^\text{M}_{k}|}),

    \end{array}
\end{equation}
where
\begin{equation}
L_n(\omega)=(e^{\omega T}e^{in\pi}-1)\frac{\omega T}{(\omega T)^2+(n\pi)^2}\;.
\end{equation}

\section{Examples}
\label{sec:examples}
In this section, we exemplify the model by considering two specific choices of spectral densities. First, we are going to test the stochastic method {  against} the fully quantized, { deterministic} pseudomode model using an underdamped Brownian spectral density. We then focus on a Ohmic spectral density with exponential cut-off which lies outside the domain of rational spectral densities analyzed analytically in the previous sections. 
\subsection{Underdamped Brownian Spectral Density}
\label{sec:underdamped}
The spectral density
\begin{equation}
\label{eq:spectral_density_main}
    J^{\text{B}}(\omega)=\frac{\gamma\lambda^{2}\omega}{(\omega^{2}-\omega_{0}^{2})^{2}+\gamma^{2}\omega^{2}}\;,
\end{equation}
defined in terms of the frequency parameters $\omega_0$ and $\gamma<2\omega_0$  and an overall strength $\lambda$ (having dimension of $\text{frequency}^{3/2}$) describes a structured environment characterized by an underdamped resonance having frequency $\Omega=\sqrt{\omega_0^2-\Gamma^2}$ and a broadening $\Gamma=\gamma/2$. It has the form described in section \ref{sec:rational} of a rational function with a Ohmic numerator (linear in $\omega$) and a polynomial cut off with poles located at $\pm\Omega\pm i\Gamma$.
For this spectral density, the symmetric and antisymmetric part of the correlation are
\begin{equation}
\label{eq:C_as_brownian_underdamped}
\begin{array}{lll}
    C^{\text{B}}_\text{as}(t)&=&\displaystyle-i\frac{\lambda^2}{2\Omega}\sin(\Omega t)e^{-\Gamma |t|}\;,
    \end{array}
\end{equation}
and
\begin{equation}
\label{eq:C_s_brownian_underdamped}
\begin{array}{lll}
    C^{\text{B}}_\text{s}(t)
 &=&\displaystyle \frac{\lambda^2}{4\Omega}\coth{(\beta(\Omega+i\Gamma)/2)}e^{i\Omega |t|} e^{-\Gamma|t|}\\
   &&\displaystyle -\frac{\lambda^2}{4\Omega}\coth{(\beta(-\Omega+i\Gamma)/2)}e^{-i\Omega |t|} e^{-\Gamma|t|}\\
   &&+\displaystyle\frac{2i}{\beta}\sum_{k>0}J(\omega^\text{M}_{k})e^{-|\omega^\text{M}_{k}||t|},
    \end{array}
\end{equation}
where $\omega^\text{M}_k=2\pi i k /\beta$.
Following Eq.~(\ref{eq:fas}), we can define $f^{\text{B}}_\text{s}(t)=-({\lambda^2}/{2\Omega})\cos(\Omega t)e^{-\Gamma |t|}$ so that the quantum  and classical part of the correlation read
\begin{equation}
\label{eq:C_Q_brownian_underdamped}
\begin{array}{lll}
    C^{\text{B}}_\text{Q}(t)&=&C^{\text{B}}_\text{as}(t)-f^{\text{B}}_\text{s}(t)=\displaystyle\frac{\lambda^2}{2\Omega}e^{-i\Omega t}e^{-\Gamma|t|}\;,
    \end{array}
\end{equation}
and 
\begin{equation}
\label{eq:C_class_brownian_underdamped}
\begin{array}{lll}
       C^{\text{B}}_\text{class}(t)
 &=&\displaystyle \frac{\lambda^2}{4\Omega}\coth{(\beta(\Omega+i\Gamma)/2)}e^{i\Omega |t|} e^{-\Gamma|t|}\\
   &-&\displaystyle \frac{\lambda^2}{4\Omega}\coth{(\beta(-\Omega+i\Gamma)/2)}e^{-i\Omega |t|} e^{-\Gamma|t|}\\
   &-&\displaystyle\frac{\lambda^2}{2\Omega}\cos(\Omega t)e^{-\Gamma |t|}+\frac{2i}{\beta}\sum_{k>0}J^{\text{B}}(\omega^\text{M}_{k})e^{-|\omega^\text{M}_{k}||t|},
    \end{array}
\end{equation}
respectively.
It is interesting to note that, in the $\beta\rightarrow\infty$ limit, simplifications in the classical correlation occur to obtain the usual expression 
\begin{equation}
\label{eq:C_mats_brownian_underdamped}
\begin{array}{lll}
       C^{\text{B}}_\text{class}(t)
 &\overset{\beta\rightarrow\infty}{=}&\displaystyle\frac{i}{\pi}\int_0^\infty dx J^{\text{B}}(ix)e^{-x|t|}\;.
    \end{array}
\end{equation}
which only involve the so-called Matsubara contribution. In Fig.~\ref{fig:brownian_corr}, we show an example of these correlations at zero temperature.

The dynamics of a system coupled to a bath with the spectral density in Eq.~(\ref{eq:spectral_density_main}) can be computed through a traditional pseudomode model in which quantum modes are used to both exactly describe the  quantum contributions \emph{and} to approximate the Matsubara series in Eq.~(\ref{eq:C_class_brownian_underdamped}).  A minimal number of $1+N_\text{mats}$ pseudomodes are needed in the zero temperature limit (where $N_\text{mats}$ is the number of Matsubara modes). On the contrary, the stochastic pseudomode model only requires a \emph{single quantum degree of freedom initially at zero temperature} to model an underdamped  Brownian environment \emph{at any temperature}. More specifically, the model reads
\begin{equation}
\label{eq:dynamics_B}
\dot{\rho}=-i[H_B,\rho]+D_B[\rho]\;,
\end{equation}
\begin{figure}[t!]
\includegraphics[width = \columnwidth]{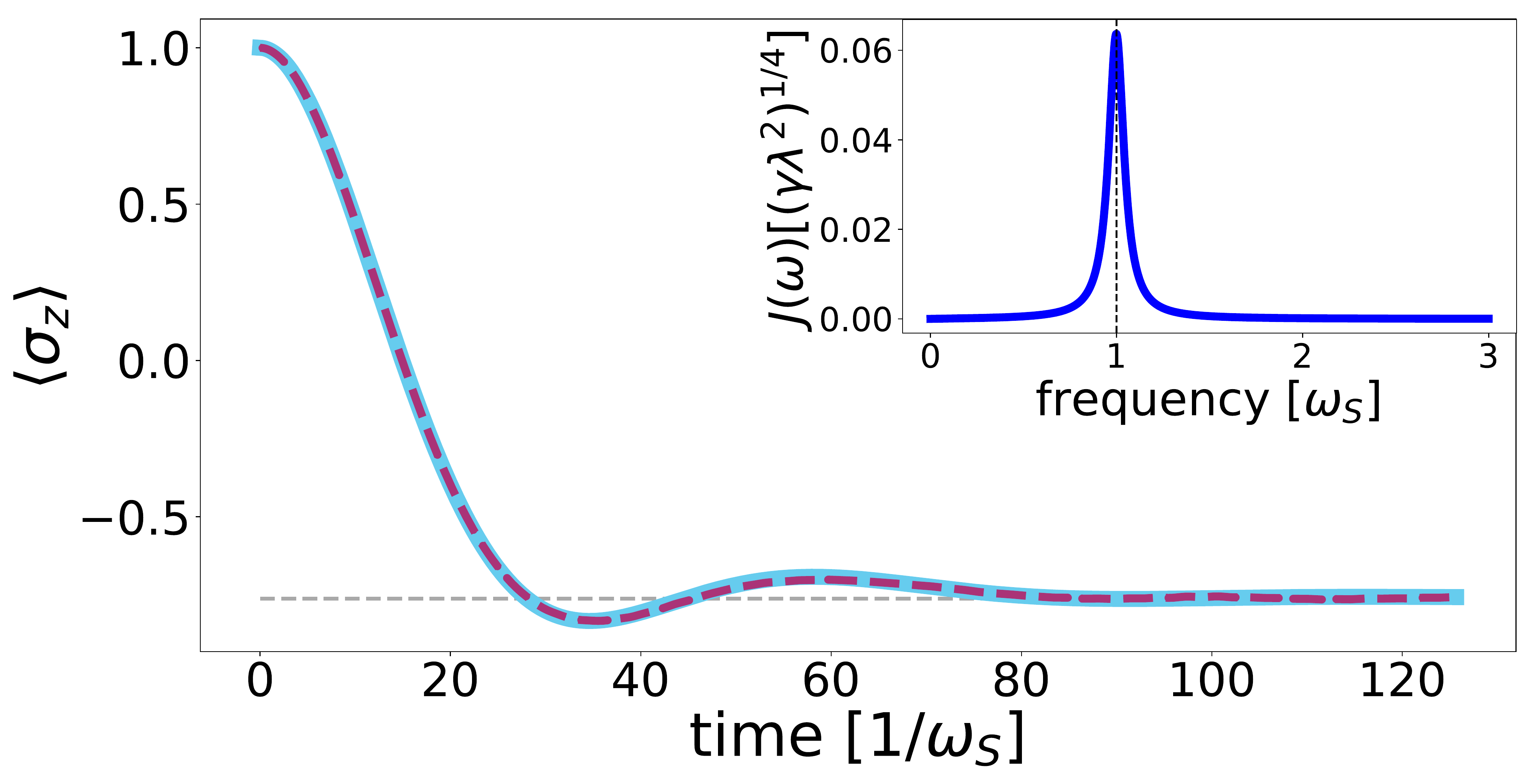} 
\caption{Dynamics of a two level system with Hamiltonian $H_S=(\omega_s/2)\sigma_z$ and coupling operator ${s}=\sigma_x$ strongly coupled to an underdamped Brownian bath at finite temperature. In blue the evolution follows the standard pseudomode model described in Eq.~(\ref{eq:dynamics_B}) in which all degrees of freedom are quantized [for a total of 5 modes, i.e., the ones described in Eq.~(\ref{eq:underdamped_full_PM_param}) and two fitting the Matsubara contribution to the correlation]. 
In dashed-red the evolution of the empirical average in Eq.~(\ref{eq:empirical_average_main}) for $N_\text{stoch}=500$ is evaluated with the stochastic pseudomode model involving a single harmonic oscillator to model the quantum contribution. The standard deviation in light-red from Eq.~(\ref{eq:empirical_sigma}) is, for these parameters, not visible. The horizontal grey dashed line represents the value of the observable at thermal equilibrium. The parameters are $\omega_0=\omega_s$, $\gamma=0.1\omega_s$, $\lambda=(0.2/\sqrt{2\pi})\omega^{3/2}_s$, $\beta=2/\omega_s$, $N_\xi=500$. \label{fig:brownian_dynamics_T}}
\end{figure}
where
\begin{equation}
H_B= H_S + \xi^{\text{B}}(t){s}+\lambda_\text{res}\sigma_x(a+a^\dagger)+\Omega_\text{res}a^\dagger a\;,
\end{equation}
in terms of the  parameters $\lambda_\text{res}=\sqrt{\lambda^2/2\Omega}$, $\Omega_\text{res}=\Omega$ and where
\begin{equation}
D_B[\rho]=\Gamma_\text{res}[2a\rho a^\dagger-a^\dagger a\rho-\rho a^\dagger a]\;,
\end{equation}
as a function of the decay rate $\Gamma_\text{res}=\Gamma$ and at zero temperature. The fields $\xi^{\text{B}}(t)$ can be explicitly defined by using Eq.~(\ref{eq:xi_spectral_representation_2_main}) together with the analytical expression in Eq.~(\ref{eq:c_n_brownian}) for the coefficients $c_n$. To test the method, in Fig. (\ref{fig:brownian_dynamics_T0}) and Fig. (\ref{fig:brownian_dynamics_T}) we show the dynamics of a single two level system with Hamiltonian $H_S=(\omega_s/2)\sigma_z$ and coupling operator ${s}=\sigma_x$. The stochastic model results are compared to those obtained solving the fully quantized pseudomode model both at zero and finite temperature and for different regimes involving strongly and weakly coupled resonances with different broadening.

\begin{figure}[t!]
\includegraphics[width = \columnwidth]{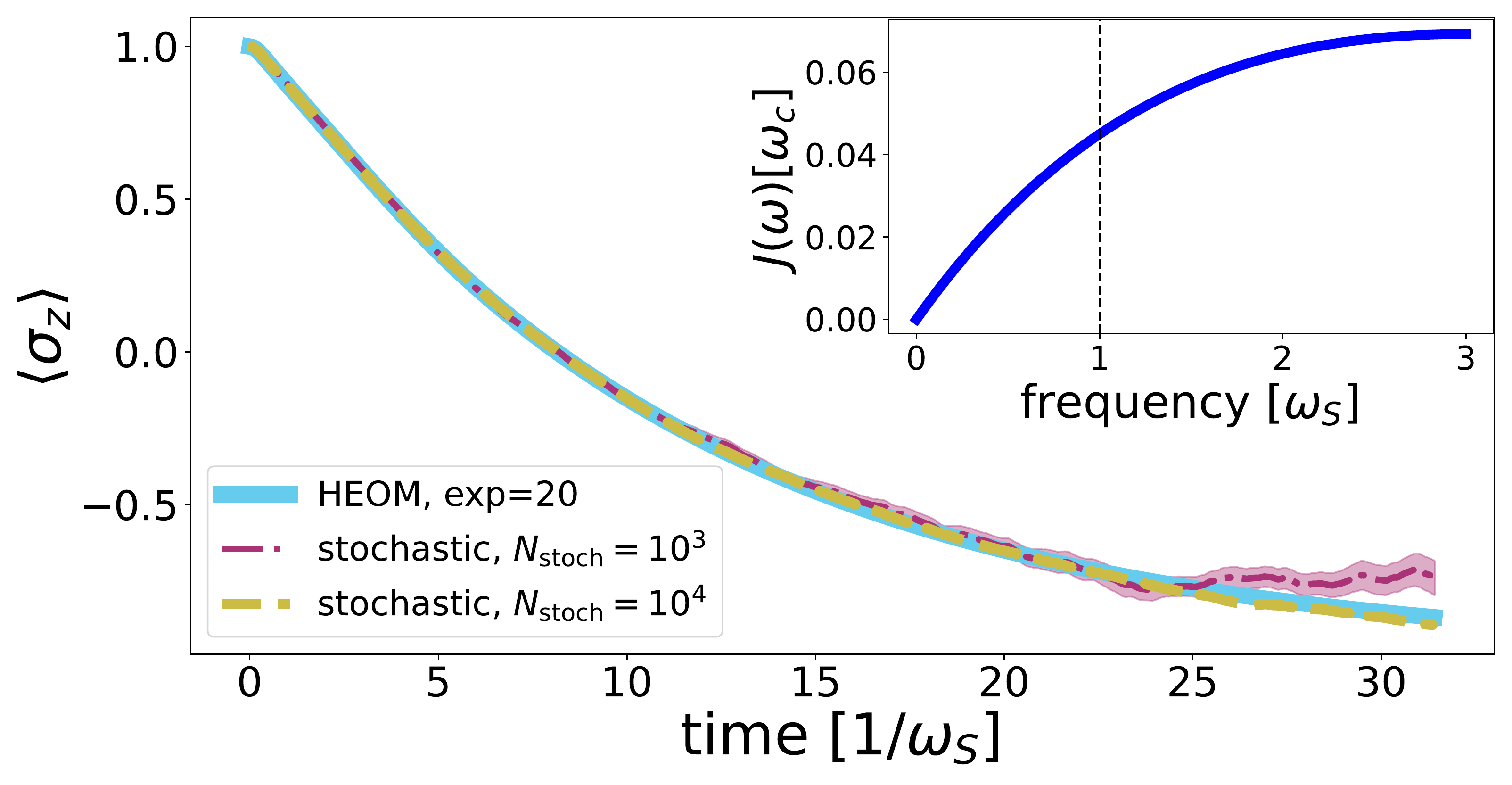} 
\caption{Dynamics of a two level system with Hamiltonian $H_S=(\omega_s/2) \sigma_z$ and coupling operator ${s}=\sigma_x$ coupled to a Ohmic spectral density (inset) at zero temperature.  In dashed-green and dashed-red we used the stochastic model (averaged $N_\text{stoch}=10^3$ and $N_\text{stoch}=10^4$ times, respectively), where  two quantum pseudomodes accounts for the antisymmetric part of the correlation. This result is compared to the solution of the Hierarchical Equations of Motion \cite{Lambert_Bofin} where the full correlation function is approximated as a linear combination of $20$ decaying exponentials.  This value is chosen to check the results of the presented method with respect to a nearly-exact case.
In light red the 
 bound on the estimated standard deviation obtained from Eq.~(\ref{eq:empirical_sigma}).
In summary, parameters are $N^\Omega_\text{PM}=2$, $\alpha = 2/100$, $\omega_c=3\omega_s$, $N_\xi=1000$. 
\label{fig:ohmic_dynamics}}
\end{figure}
\subsection{Ohmic Spectral density with exponential cut-off}
\label{sec:Ohmic}
\begin{figure}[t!]
\includegraphics[width = \columnwidth]{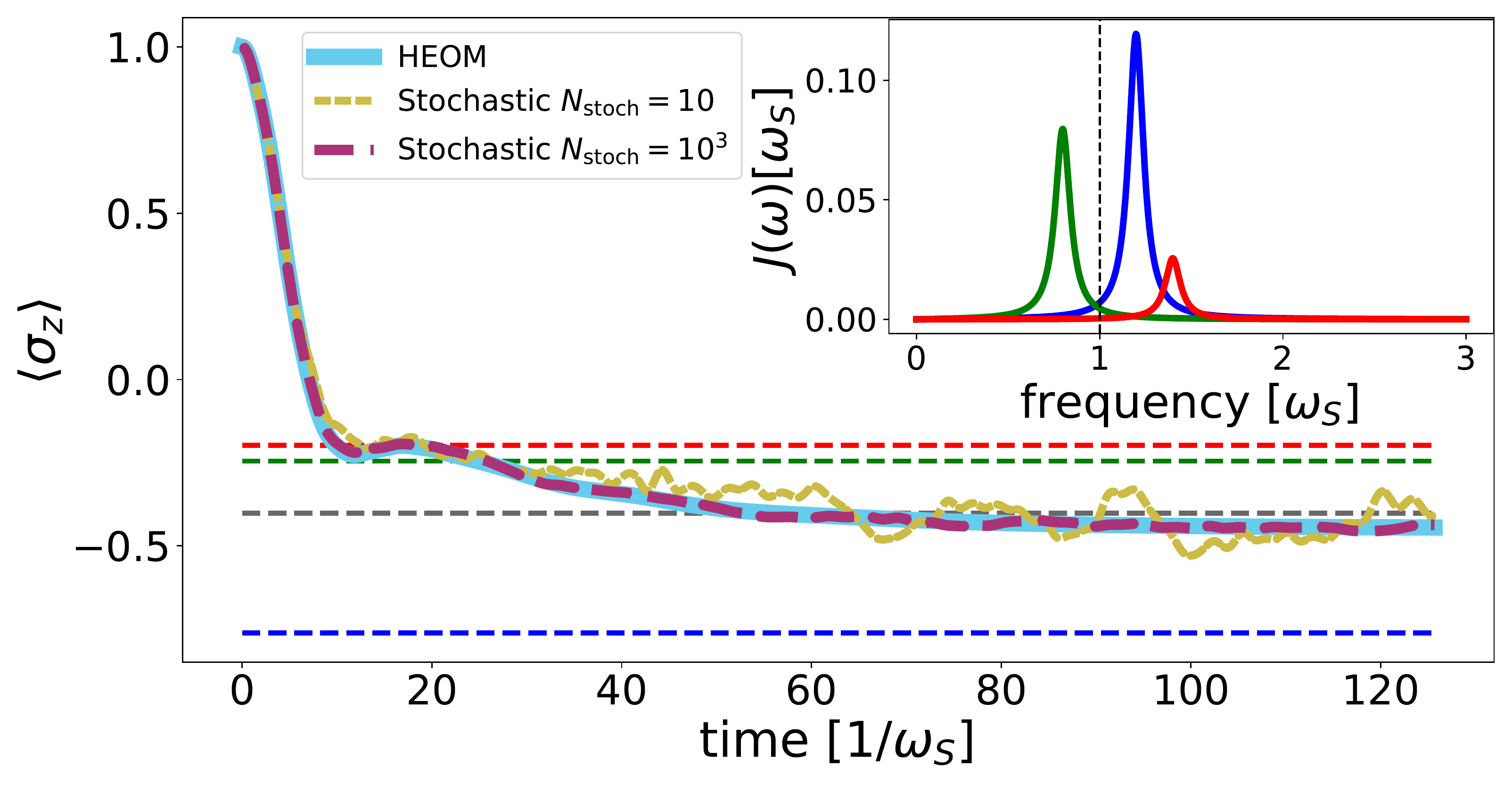} 
\caption{{  Dynamics of a two-level system coupled to three baths characterized by different spectral densities at different temperatures (see inset). In full (cyan) the dynamics is computed using the HEOM. Dashed curves are computed using the stochastic pseudomode model averaging over $N_\text{stoch}=10$ (yellow) and $N_\text{stoch}=10^4$ (red) realizations for the classical noise. The dashed colored horizontal lines correspond to the thermal occupation which the qubit would have at the temperature of each individual bath. In grey, the occupation it would have in the steady state when neglecting hybridization effects with the environment, using the elegant expression given in \cite{e23010077}.}
\label{fig:3baths}}
\end{figure}

In this section we consider the following spectral density
\begin{equation}
    J^\Omega(\omega)=\pi\alpha \omega e^{-\omega/\omega_c}\;,
\end{equation}
which shows a Ohmic behavior at small frequencies characterized by the strength parameter $\alpha\in\mathbb{R}$ and by an exponential decay characterized by the frequency cut-off $\omega_c$ \cite{BrandesDiss}. While the transcendental form of this spectral density does not allow to use the results presented in section \ref{sec:rational} for rational functions, it is still possible to follow the procedure in section \ref{sec:ZeroT_main} to define the stochastic pseudomode model using numerical optimization. Specifically, it is possible to find the parameters $\{a^{\Omega}_j,b^{\Omega}_j,c^{\Omega}_j\}$ which better approximate the antisymmetric part of the correlation 
\begin{equation}
\label{eq:ansatz_Ohmic_main}
\begin{array}{lll}
C^{\Omega}_{\text{as}}(t)  &=&\displaystyle -2i\alpha\omega_{c}^{2}\frac{\omega_{c}t}{(1+\omega_{c}^{2}t^{2})^{2}}\simeq\sum_{j=1}^{N^{\Omega}_\text{PM}}a^{\Omega}_j \sin(b^{\Omega}_j t)e^{-c^{\Omega}_j |t|}.
 \end{array}
\end{equation}
We note that, with respect to the fully quantized pseudomode model, the procedure presented here has the advantage of not having to fit the symmetric part of the correlation which also contains all information about the temperature of the original environment.
In Fig.~\ref{fig:ohmic_dynamics} we show an example in which the dynamics of a two level system is computed by using two pseudomode together with classical stochastic fields. We compared this result against the Hierarchical Equation of Motion solved using the BoFiN package \cite{Lambert_Bofin,Qutip1,Qutip2,PhysRevA.98.063815,Li2022pulselevelnoisy}.
{ \section{Multiple baths at different temperatures}
\label{sec:MultipleBaths}
It is interesting to note that, in the case of an underdamped Brownian spectral density, the Hilbert space of the stochastic pseudomode model augments the system only by a single extra pseudomode initialized in the vacuum for any temperature of the original environment. This should be compared to the deterministic pseudomode method which, as shown in Eq.~(\ref{eq:dynamics_B_app}), requires $3+N_\text{mats}$ pseudomodes \footnote{An alternative representation which uses $2+N_\text{mats}$ is possible by introducing complex temperatures \cite{Paul}} where $N_\text{mats}$ is the number of modes corresponding to a fit with $N_\text{mats}$ exponentials for the Matsubara contribution to the correlation, see Eq.~(\ref{eq:dynamics_B_app}). This might result in a considerable reduction of the Hilbert space when the system is coupled to multiple baths at different temperatures. We analyze this case in Fig.~\ref{fig:3baths}, where a single two-level system is coupled to three independent Brownian baths at different temperature. 

There, we compare the steady state value of $\langle\sigma_z\rangle$ with the ones the spin would have when in equilibrium to each bath. Using the elegant formalism in \cite{e23010077}, we further show that the steady state is different from the one expected for a system weakly coupled to all three baths. In fact, the averaging over the statistics of the field complete the full characterization of the effects due to hybridization with the quantum degrees of the original bath.
To further test the method, in Appendix \ref{app:ohmic}, we present the results for an environment causing pure decoherence, i.e., when the system interaction operator commutes with the system Hamiltonian.}

\begin{figure}[t!]
\includegraphics[width = \columnwidth]{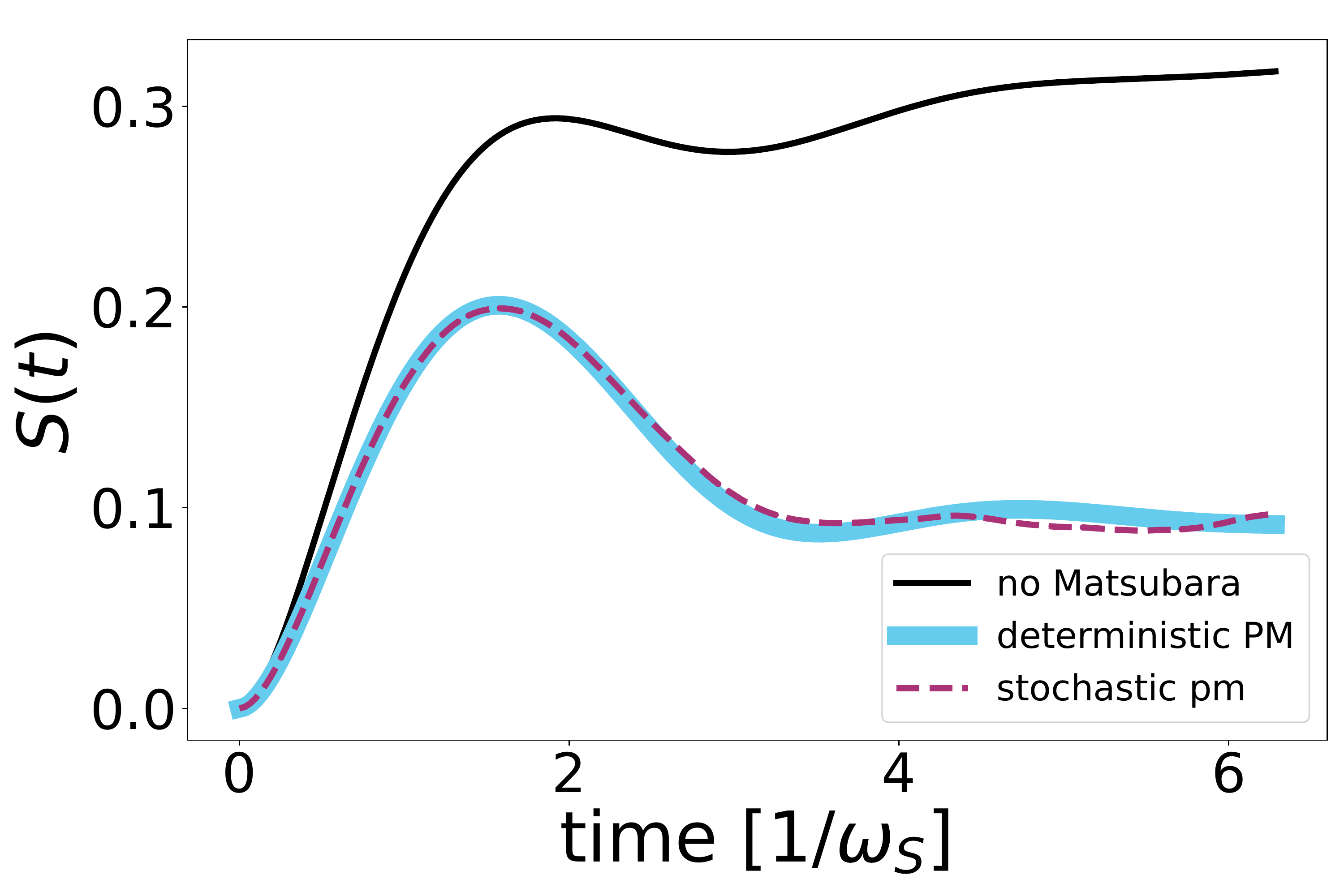} 
{ \caption{Using the example and parameters from \figref{fig:brownian_dynamics_T0}b we calculate the von Neumann entropy of the reduced system state as a function of time. The solid curve shows the result from a deterministic pseudomode model with a single quantum pseudmode and where the Matsubara, or classical, contribution to the bath correlation function is omitted entirely.  The thick solid blue curve is the same but with the Matsubara contribution included as quantum pseudomodes.
In dashed-purple we show the  result for the hybrid stochastic pseudomode model, where the quantum part of the correlation function is the same as that for the solid-black curve, but the Matsubara part is captured by classical stochastic fields, with $N_\text{stoch}=1000$. In both the deterministic and the hybrid-stochastic approaches the addition of the imaginary valued Matsubara contribution reduces the system entropy. \label{fig:S}}}
\end{figure}

{ \section{Von Neumann entropy}
\label{sec:Entropy}
The hybrid pseudomode model described in this article is an effective method, i.e., its only goal is to reproduce the \emph{effects} of the environment on the system, rather than the system-bath full dynamics. It does so by introducing ancillary degrees of freedom which are not directly related to the original environment. This allows to extend the regime of the associated parameters to unphysical values as long as the reduced density matrix of the system remains physical. As a consequence, despite the effective nature of these degrees of freedom, it is still meaningful to estimate relevant quantities which are a function of such a reduced dynamics. One important example is the von Neumann entropy
\begin{equation}
    S(t)=-\text{Tr}_S\rho_S(t)\log{\rho_S(t)}\;,
\end{equation}
which can be used as a measure of the information contained in the system and the correlation between bipartite systems. For example, in our context, it can be considered as a measure of system-bath entanglement when the initial state of the bath is at zero temperature, {  and the initial state of the system is a pure state}. 

In normal circumstances, one might expect that the presence of stochastic noise  might drive a system to a higher entropy state, i.e., it would increase its information disorder. It is then interesting to test whether this intuition remains true for the kind of noise needed to model the classical contribution to the correlation function of a quantum bath considered here. In fact, in the case of a zero-temperature bath characterized by an underdamped Brownian spectral density, the driving field is purely imaginary. In Fig.~\ref{fig:S}, we show that this unphysical field cause a decrease on the  von Neumann  entropy of the reduced state of the system. 

While the effective nature of the fields prevents a direct physical interpretation, the meaning of this result can still be found in the description of the properties associated with different contributions to the correlation function of a physical environment. For example, any approximation which neglects the Matsubara contribution of a Brownian environment at zero temperature will correspond to an overestimation of the disorder in the reduced state of the system, {  due to a breaking of the detailed a balance condition of that environment}. In turn, by using unphysical, imaginary fields, it is possible to correctly reduce such a disorder to the correct physical balance.}
\section{Conclusions}
In this article, we analyzed a decomposition of the effects of a Gaussian Bosonic bath on a system into a quantum and classical contributions. We used this decomposition to define a stochastic version of the Bosonic pseudomode method in which classical noise complements a discrete set of harmonic degrees of freedom to model non-Markovian open quantum systems. This ultimately results in a reduction of the number of ancillary quantum resources needed for the classical simulation of the original environment, which can become critical {  to model highly structured open quantum systems in the presence of multiple baths at different temperature.}.

Furthermore, all temperature effects of the original environment can be encoded into properties of the classical noise, allowing the remaining quantum degrees of freedom to be, initially, at zero temperature. For a class of rational spectral densities, all parameters of the stochastic pseudomode model are explicitly provided without the need of any best fit optimization for all temperatures of the original bath, thereby simplifying the modeling stage. 

We tested the method for Ohmic spectral densities with polynomial and exponential decay against the { deterministic} pseudomode method and the hierarchical equations of motion, respectively. 
 { The possibility to encode all temperature effects in the statistics of the driving field even for multiple baths leads to improved handling of the computational resources with respect to the fully deterministic pseudomode model. We further observed interesting features related to the unphysicality of this field which, despite its noisy nature, decreases the von Neumann entropy of the system-environment partition. This shows that the simplicity in the interpretation of this method can lead to new interesting ways to explore non-Markovian effects such as the entanglement structure  in open quantum systems.
 
 }

\begin{figure}[t!]
\includegraphics[width = \columnwidth]{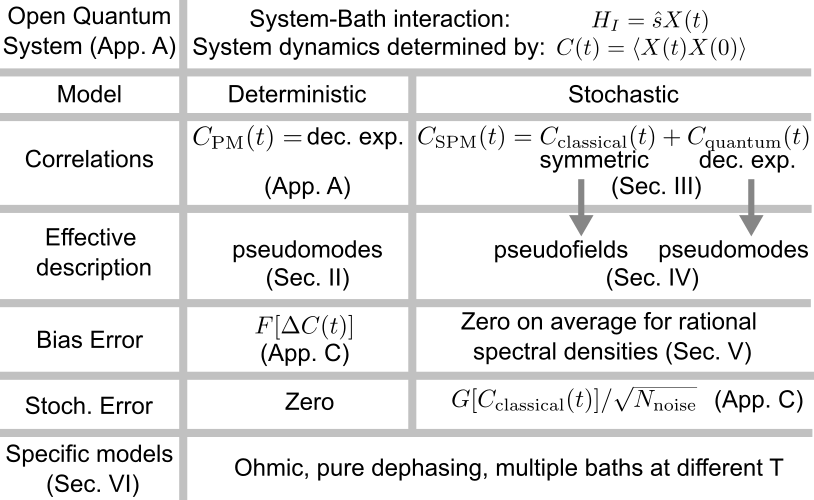} 
\caption{
 \label{fig:outline}{ Summary. Given a linear interaction between a system and a Gaussian environment, the system dynamics is a function of the correlation of the environmental coupling operator. Different models rely on different approximations for the correlation. For example, the deterministic pseudomode model assumes an ansatz given by a sum of decaying exponentials which introduces a bias error. By using a  classical/quantum decomposition, it is possible to consider an additional stochastic driving to model, on average, more general correlations. In this case, the observables are stochastic variable whose variance scales as the inverse of the number of noise realizations. Numerical examples for different specific models are provided.}}
\end{figure}
In summary, {  our hybrid method allows one to model the non-perturbative effects of a Bosonic bath on a system's dynamics with an approach that balances conceptual simplicity (it only requires averaging over the solutions of a Lindblad-like equation) and numerical accuracy, while using a limited amount of quantum resources with respect to} {  deterministic algorithms such as the regular pseudomode model and the regular HEOM}. 

Beyond numerical simulations, { despite the purely effective nature of the degrees of freedom involved,} the quantum-classical decomposition presented here {  could} also be useful in analyzing the interplay between quantum and classical effects induced by environments in the context of quantum thermodynamics \cite{Calabrese, Adesso}, quantum control \cite{Lambert_Bofin,Brif_2010,Mangaud}, and quantum transport \cite{Plenio,Rebentrost_2009,Scholes_harvesting,Lambert_harvesting}. {  In this direction, we provide evidence that the imaginary nature of the stochastic field characterizes effects which tend to decrease the von Neumann entropy of the reduced system state.}

\acknowledgements
M.C. acknowledges support from NSFC (Grants No.~12050410264 and No.~11935012) and NSAF (Grant No.~U1930403). N.L. acknowledges support from the Information Systems Division, RIKEN, for use of their facilities.

\newpage
\appendix

\section{Influence Superoperator}
\label{app:influenceSuperoperator}
In this section, we derive an expression for the influence superoperator (see also \cite{Ma_2012,Aurell,Cirio2021}) appearing in Eq.~(\ref{eq:rhoS}) which encodes all the effects of a bosonic Gaussian environment linearly interacting with a quantum system. The proof does not make use of any path-integral technique.

\subsection{A Canonical Derivation}
Here, we present a canonical (i.e., purely operator-based) derivation of the Bosonic influence superoperator.
  We consider system $S$ coupled to bosonic environment $B$ as
\begin{equation}
\label{eq:HSIB}
H=H_{S}+H_{I}+H_{B}\;,
\end{equation}
where $H_{S}$ is the system Hamiltonian and $H_{B}=\sum_{k}\omega_{k}b_{k}^{\dagger}b_{k}$
characterizes the energy $\omega_{k}$ of each environmental mode
$b_{k}$. We suppose the system-environment coupling to be in the
form $H_{I}={s}X$, where ${s}$ is a system operator and
$X$ is an interaction operator linearly expanded by $\{b_k^\dagger,b_k\}$ . Meanwhile we assume
the initial state can be denoted as $\rho(0)=\rho_{S}(0)\otimes\rho_{B}(0)$,
where $\rho_{S}(0)$ is the initial state of the system, $\rho_{B}(0)$
is a Gaussian state of the bosonic environment (for example, a thermal equilibrium state at inverse temperature $\beta$). We further suppose that the initial state of the environment is invariant under the free dynamics, i.e., $\rho_B(0)=U_0(t)\rho_B(0)U_0^\dagger(t)$ in terms of the free propagator $U_{0}(t)=\text{exp}[-i(H_{S}+H_{B})t]$

In the interaction picture, the Shr\"{o}dinger equation 
\begin{equation}
\dot{\rho}=-i[H_{I}(t),\rho]\;,
\end{equation}
can be formally solved as
\begin{equation}
\label{eq:formal_1}
\rho(t)=\mathcal{T}\left[e^{-i\int_{0}^{t}d\tau H_{I}^{\times}(\tau)}\right]\rho(0)\;,
\end{equation}
where $H_{I}(t)=U_{0}^{\dagger}(t)H_{I}U_{0}(t)$, and where we further defined $O^{\times}[\cdot]=[O,\cdot]$ as the superoperator associated with the commutation with the operator $O$. The time-ordering operator $\mathcal{T}$ acts on superoperators as
\begin{equation}
\begin{array}{lll}
\displaystyle\mathcal{T}\left[O^{\times}(t_{1})O^{\times}(t_{2})\right]&=&\displaystyle O^{\times}(t_{1})O^{\times}(t_{2})\theta(t_{1}-t_{2})\\
&&\displaystyle+O^{\times}(t_{2})O^{\times}(t_{1})\theta(t_{2}-t_{1})\;,
\end{array}
\end{equation}
where $\theta$ is the step function defined as $\theta(x)=1$ for $x\geq 0$ and zero otherwise. 

We are now interested in analyzing the reduced density matrix for the system, i.e., $\rho_S(t)=\text{Tr}[\rho(t)]$ which, using Eq.~(\ref{eq:formal_1}), takes the form
\begin{widetext}
\begin{equation}
\label{eq:formal_2}
\rho_{S} =\sum_{n=0}^{\infty}\frac{(-i)^{n}}{n!}\mathrm{Tr}_{B}\left[\mathcal{T}\int_{0}^{t}dt_{1}\cdots\int_{0}^{t}dt_{n}H_{I}^{\times}(t_{1})\cdots H_{I}^{\times}(t_{n})\rho(0)\right]\;.
\end{equation}
\end{widetext}
We can now consider the explicit form of the interaction Hamiltonian in the interaction frame, i.e., $H_{I}(t)={s}(t)\otimes X(t)$, where ${s}(t)=e^{iH_{S}t}{s}e^{-iH_{S}t}$ and
$X(t)=e^{iH_{B}t}Xe^{-iH_{B}t}$ and use it in Eq.~(\ref{eq:formal_2}). In this way, the first non-trivial term can be written as
\begin{equation}
\label{eq:HISX}
 H_{I}^{\times}(t)\left[\rho_{S}(0)\otimes\rho_{B}(0)\right]=
\sum_{i=1}^{2}\mathcal{S}_{t}^{i}[\rho_{S}(0)]\otimes\mathcal{X}_{t}^{i}[\rho_{B}(0)]\;,
\end{equation}
where $\mathcal{S}_{t}^{1}[\cdot]={s}(t)\cdot$, $\mathcal{S}_{t}^{2}[\cdot]=-\cdot {s}(t)$, $\mathcal{X}_{t}^{1}[\cdot]=X(t)\cdot$, and $\mathcal{X}_{t}^{2}[\cdot]=\cdot X(t)$. Using this notation in Eq.~(\ref{eq:formal_2}), the reduced density matrix takes the form
\begin{equation}
\label{eq:rhoS_intermediate}
\rho_S(t)=\sum_{n=0}^\infty \frac{(-i)^n}{n!}\rho_n(t)\;,
\end{equation}
where $\rho_0(t)=\text{Tr}_B[\rho(0)]$ and
\begin{widetext}
\begin{equation}
\label{eq:rho_n}
{\rho}_{n}(t)=  \int_{0}^{t}dt_{1}\cdots\int_{0}^{t}dt_{n}\sum_{i_1,\cdots,i_n}\mathrm{Tr}_{B}\left[\mathcal{T}\mathcal{X}_{t_1}^{i_{1}}\cdots\mathcal{X}_{t_n}^{i_{n}}\rho_{B}(0)\right]\mathcal{T}\mathcal{S}_{t_1}^{i_{1}}\cdots\mathcal{S}_{t_n}^{i_{n}}\rho_{S}(0)\;,
\end{equation}
\end{widetext}
for $n>0$. To make progress, we now take advantage of the Gaussianity of the state $\rho_{B}(0)$ which translates into the following version of Wick's theorem for superoperators, see section \ref{app:Wick_superoperators},
\begin{equation}
\label{eq:wickforsuper-operator}
\mathrm{Tr}\left[\mathcal{T}\mathcal{X}_{t_1}^{i_{1}}\cdots\mathcal{X}_{t_{2N}}^{i_{2N}}\rho_{B}\right]=\sum_{C\in P_{2N}}\prod_{(l,m)\in C}\mathrm{Tr}\left[\mathcal{T}\mathcal{X}_{t_l}^{i_{l}}\mathcal{X}_{t_m}^{i_{m}}\rho_{B}\right],
\end{equation}
where the traces are intended on the bath $B$, where we omitted the time label in the state $\rho_B(0)$ and where $P_{2N}$ is the collection of all possible sets (full contractions $C$) whose elements are $N$ disjoint pairs (i.e., contractions) taken from $\{1,\cdots,2N\}$. The cardinality of $P_{2N}$ is $(2N-1)!!=(2N)!/(2^N N!)$, i.e., the number of different ways to fully contract the set $\{1,\cdots,2N\}$. We further have that $\mathrm{Tr}_{B}\left[\mathcal{T}\mathcal{X}_{t_1}^{i_{1}}\cdots\mathcal{X}_{t_{2N+1}}^{i_{2N+1}}\rho_{B}\right]=0$. We refer to section \ref{app:Wick_superoperators} for a detailed proof of this version of the Wick's theorem written in terms of time-ordered superoperators.

Using Wick's theorem in Eq.~(\ref{eq:rho_n}), we obtain
\begin{widetext}
\begin{equation}
\label{eq:expF_app}
\begin{array}{lll}
\rho_{2N}(t) & =&\displaystyle\sum_{C\in P_{2N}}\int_{0}^{t}dt_{1}\cdots\int_{0}^{t}dt_{2N}\prod_{(l,m)\in C}\sum_{i_{l},i_{m}}\mathrm{Tr}_{B}\left[\mathcal{T}\mathcal{X}_{t_{l}}^{i_{l}}\mathcal{X}_{t_{m}}^{i_{m}}\rho_{B}(0)\right]\mathcal{T}\left[\mathcal{S}_{t_{1}}^{i_{1}}\cdots\mathcal{S}_{t_{2N}}^{i_{2N}}\right]\rho_{S}(0)\\
 & =&\displaystyle\sum_{C\in P_{2N}}\mathcal{T}\left[\int_{0}^{t}dt_{1}\cdots\int_{0}^{t}dt_{2N}\prod_{(l,m)\in C}\sum_{i_{l},i_{m}}\mathrm{Tr}_{B}\left[\mathcal{T}\mathcal{X}_{t_{l}}^{i_{l}}\mathcal{X}_{t_{m}}^{i_{m}}\rho_{B}(0)\right]\mathcal{S}_{t_{1}}^{i_{1}}\cdots\mathcal{S}_{t_{2N}}^{i_{2N}}\right]\rho_{S}(0)\\
 & =&\displaystyle\sum_{C\in P_{2N}}\mathcal{T}\left[\int_{0}^{t}dt_{1}\cdots\int_{0}^{t}dt_{2N}\prod_{(l,m)\in C}\left(\sum_{i_{l},i_{m}}\mathrm{Tr}_{B}\left[\mathcal{T}\mathcal{X}_{t_{l}}^{i_{l}}\mathcal{X}_{t_{m}}^{i_{m}}\rho_{B}(0)\right]\mathcal{S}_{t_{l}}^{i_{l}}\mathcal{S}_{t_{m}}^{i_{m}}\right)\right]\rho_{S}(0)\\
 & =&\displaystyle\mathcal{T}\left\{ \sum_{C\in P_{2N}}\left(\int_{0}^{t}dt_{2}\int_{0}^{t}dt_{1}\sum_{i_{2},i_{1}}\mathrm{Tr}_{B}\left[\mathcal{T}\mathcal{X}_{t_{2}}^{i_{2}}\mathcal{X}_{t_{1}}^{i_{1}}\rho_{B}(0)\right]\mathcal{S}_{t_{2}}^{i_{2}}\mathcal{S}_{t_{1}}^{i_{1}}\right)^{N}\rho_{S}(0)\right\} \\
 & =&\displaystyle\mathcal{T}\left\{ \frac{(2N)!}{N!}\left(\frac{1}{2}\int_{0}^{t}dt_{2}\int_{0}^{t}dt_{1}\sum_{i_{2},i_{1}}\mathrm{Tr}_{B}\left[\mathcal{T}\mathcal{X}_{t_{2}}^{i_{2}}\mathcal{X}_{t_{1}}^{i_{1}}\rho_{B}(0)\right]\mathcal{S}_{t_{2}}^{i_{2}}\mathcal{S}_{t_{1}}^{i_{1}}\right)^{N}\rho_{S}(0)\right\}\\
 & =&\displaystyle\mathcal{T}\left\{ \frac{(2N)!}{N!}\left(\int_{0}^{t}dt_{2}\int_{0}^{t_{2}}dt_{1}\sum_{i_{2},i_{1}}\mathrm{Tr}_{B}\left[\mathcal{X}_{t_{2}}^{i_{2}}\mathcal{X}_{t_{1}}^{i_{1}}\rho_{B}(0)\right]\mathcal{S}_{t_{2}}^{i_{2}}\mathcal{S}_{t_{1}}^{i_{1}}\right)^{N}\rho_{S}(0)\right\}\;.
\end{array}
\end{equation}
\end{widetext}
Using this result into Eq.~(\ref{eq:rhoS_intermediate}), we obtain
\begin{widetext}
\label{eq:expF_app_2}
\begin{equation}
    \begin{array}{lll}
\rho_{S}(t) & =&\displaystyle\mathcal{T}\left\{ \sum_{N=0}^{\infty}\frac{(-1)^{N}}{N!}\left(\int_{0}^{t}dt_{2}\int_{0}^{t_{2}}dt_{1}\sum_{i_{2},i_{1}}\mathrm{Tr}_{B}\left[\mathcal{X}_{t_{2}}^{i_{2}}\mathcal{X}_{t_{1}}^{i_{1}}\rho_{B}(0)\right]\mathcal{S}_{t_{2}}^{i_{2}}\mathcal{S}_{t_{1}}^{i_{1}}\right)^{N}\right\} \rho_{S}(0)\\
 & =&\displaystyle\mathcal{T}\;\mathrm{Exp}\left\{-\int_{0}^{t}dt_{2}\int_{0}^{t_{2}}dt_{1}\sum_{i_{2},i_{1}}\mathrm{Tr}_{B}\left[\mathcal{X}_{t_{2}}^{i_{2}}\mathcal{X}_{t_{1}}^{i_{1}}\rho_{B}(0)\right]\mathcal{S}_{t_{2}}^{i_{2}}\mathcal{S}_{t_{1}}^{i_{1}}\right\}\rho_{S}(0)\\
 &\equiv&\displaystyle\mathcal{T}e^{\mathcal{F}(t,{s},C(t))}\rho_S(0)\;.
 \end{array}
 \end{equation}
 in terms of the superoperator
 \begin{equation}
 \label{eq:functional}
 \begin{array}{lll}
\mathcal{F}(t,{s},C(t))[\cdot] &=&\displaystyle -\int_{0}^{t}dt_{2}\int_{0}^{t_{2}}dt_{1}\sum_{i_{2},i_{1}}\mathrm{Tr}_{B}\left[\mathcal{X}_{t_{2}}^{i_{2}}\mathcal{X}_{t_{1}}^{i_{1}}\rho_{B}(0)\right]\mathcal{S}_{t_{2}}^{i_{2}}\mathcal{S}_{t_{1}}^{i_{1}}[\cdot]\\
 &=&\displaystyle -\int_{0}^{t}dt_{2}\int_{0}^{t_{2}}dt_{1}C(t_{2},t_{1})\left[{s}(t_{2}){s}(t_{1})[\cdot]-{s}(t_{1})[\cdot]{s}(t_{2})\right] +C(t_{1},t_{2})\left[[\cdot]{s}(t_{1}){s}(t_{2})-{s}(t_{2})[\cdot]{s}(t_{1})\right]\\
 &=&\displaystyle\int_{0}^{t}dt_{2}\int_{0}^{t_{2}}dt_{1}\;C(t_{2}-t_{1})[{s}(t_{1})\cdot,{s}(t_{2})]-C(t_{1}-t_{2})[\cdot{s}(t_{1}),{s}(t_{2})]\;,
 \end{array}
\end{equation}
\end{widetext}
which proves Eq.~(\ref{eq:rhoS}) and which depends on the correlation function  
\begin{equation}
\label{eq:corr_func}
\begin{array}{lll}
    C(t_{2},t_{1})&=&\mathrm{Tr}_{B}\left[X(t_{2})X(t_{1})\rho_{B}(0)\right]\\
    &=&\displaystyle\sum_k g_k^2\mathrm{Tr}_{B}[(a^\dagger e^{i\omega t_2}+a e^{-i\omega t_2})\\
    &&\times(a^\dagger e^{i\omega t_1}+a e^{-i\omega t_1})\rho_{B}(0)]\\
    &=&\displaystyle\int d\omega \frac{J(\omega)}{\pi}[(2 n(\omega)+1)\cos(\omega t)-i\sin(\omega t)],
    \end{array}
\end{equation}
of the environmental interaction operator $X(t)$. Here, $J(\omega)=\pi\sum_k g_k^2\delta(\omega-\omega_k)$ and $t=t_2-t_1$ so that, using $2n(\omega)+1=\coth{\beta\omega/2}$, we get Eq.~(\ref{eq:correlation}) in the main text. The invariance of the initial environmental state under free dynamics implies the stationarity of the correlation, i.e., 
\begin{equation}
\begin{array}{lll}
C(t_{2},t_{1})&=&\mathrm{Tr}_{B}\left[X(t_{2})U^\dagger_0(t_1)X(0)U_0(t_1)\rho_{B}(0)\right]\\
&=&\mathrm{Tr}_{B}\left[X(t_{2})U^\dagger_0(t_1)X(0)\rho_{B}(0)U_0(t_1)\right]\\
&=&\mathrm{Tr}_{B}\left[U_0(t_1)X(t_{2})U^\dagger_0(t_1)X(0)\rho_{B}(0)\right]\\
&=&\mathrm{Tr}_{B}\left[X(t_{2}-t_1)X(0)\rho_{B}(0)\right]\\
&\equiv&C(t)\;.
\end{array}
\end{equation}

In summary, the effects on the reduced dynamics of a Gaussian bosonic environment linearly coupled to a quantum system can be fully described by an influence superoperator which depends on the system coupling operator ${s}$ and the correlation function $C(t)$ of the environmental coupling operator. This directly implies the equivalence between the reduced dynamics of any open quantum system sharing the same ${s}$ and $C(t)$, see \cite{Tamascelli}.\\

In the following, we are going to further improve on this derivation to allow for a general linear system-bath interaction.

\subsection{Generalized Interaction}
\label{app:generalized_interaction}
It is worth noticing that the result proven in Eq.~(\ref{eq:expF_app}) can be generalized to the case where the system and the environment interact through a more general interaction of the form, i.e., operating the substitution
\begin{equation}
\label{eq:HprimeI}
H_I \mapsto H'_I =\sum_{\alpha}{s}^\alpha X^\alpha\;,
\end{equation}
in Eq.~(\ref{eq:HSIB}), where ${s}^\alpha$ and $X^\alpha$ define a collection of system and environmental interaction operators. 
Under the replacement above, we note that Eq.~(\ref{eq:HISX}) takes the more general form
\begin{equation}
 H_{I}^{'\times}(t)\left[\rho_{S}(0)\otimes\rho_{B}(0)\right]=
\sum_{\bar{\alpha}}\mathcal{S}_{t}^{\bar{\alpha}}[\rho_{S}(0)]\otimes\mathcal{X}_{t}^{\bar{\alpha}}[\rho_{B}(0)]\;,
\end{equation}
where $\bar{\alpha}\equiv(i,\alpha)$ is a multi-index and where $\mathcal{S}_{t}^{(1,\alpha)}[\cdot]={s}^\alpha(t)\cdot$, $\mathcal{S}_{t}^{(2,\alpha)}[\cdot]=-\cdot {s}^\alpha(t)$, $\mathcal{X}_{t}^{(1,\alpha)}[\cdot]=X^\alpha(t)\cdot$, and $\mathcal{X}_{t}^{(2,\alpha)}[\cdot]=\cdot X^\alpha(t)$. 
As a consequence, we can follow all the calculations done in the previous section by simply replacing the upper indexes $i$ with multi-indexes, i.e., $i\mapsto{\alpha'}=(i,\alpha)$ to get
\begin{equation}
\label{eq:appDiffEq}
    \rho'_S(t)=\mathcal{T}e^{\mathcal{F}'(t,\{{s}\},\{C(t)\})}\rho_S(0)\;.
\end{equation}
where $\rho'_S(t)$ is the reduced density matrix for the system when the interaction Hamiltonian with the environment takes the generalized form $H_I'(t)$ above, and where
 \begin{equation}
 \label{eq:functional_generalized}
\mathcal{F}' =-\int_{0}^{t}dt_{2}\int_{0}^{t_{2}}dt_{1}\sum_{\alpha'_1,\alpha'_2}\mathrm{Tr}_{B}\left[\mathcal{X}_{t_{2}}^{\alpha'_{2}}\mathcal{X}_{t_{1}}^{\alpha'_{1}}\rho_{B}(0)\right]\mathcal{S}_{t_{2}}^{\alpha'_{2}}\mathcal{S}_{t_{1}}^{\alpha'_{1}}
\end{equation}
in which we omitted the functional dependencies $(t,\{{s}\},\{C(t)\})$ from $\mathcal{F}'$. We note that these dependencies highlight that in this case the influence superoperator depends on the whole set $\{{s}\}$ of ${s}^\alpha$ operators and the whole set \{C(t)\} of correlation functions $C_{\alpha\beta}$ which can be defined by further noting that
\begin{widetext}
\begin{equation}
\label{eq:gen}
\begin{array}{lll}
\mathcal{F}'&=&-\displaystyle\int_0^t d t_2 \int_0^{t_2} dt_1\mathcal{G}'(t_2,t_1)\\
\mathcal{G}'&=&\displaystyle\sum_{i_1,i_2,\alpha_1,\alpha_2}\mathrm{Tr}_{B}\left[\mathcal{X}_{t_{2}}^{i_2,\alpha_{2}}\mathcal{X}_{t_{1}}^{i_1,\alpha_{1}}\rho_{B}\right]\mathcal{S}_{t_{2}}^{i_2,\alpha_{2}}\mathcal{S}_{t_{1}}^{i_1,\alpha_{1}}\\
&=&\displaystyle\sum_{\alpha,\beta}\text{Tr}_B\left[X^\alpha_{t_2} X^\beta_{t_1}\rho_\beta\right] {s}^\alpha_{t_2}{s}^\beta_{t_1}[\cdot]+\displaystyle\text{Tr}_B\left[\rho_\beta X^\beta_{t_1} X^\alpha_{t_2}\right] [\cdot] {s}^\beta_{t_1}{s}^\alpha_{t_2}-\text{Tr}_B\left[X^\beta_{t_1} \rho_\beta X^\alpha_{t_2}\right] {s}^\beta_{t_1}[\cdot]{s}^\alpha_{t_2}-\displaystyle\text{Tr}_B\left[X^\alpha_{t_2}\rho_\beta X^\beta_{t_1}\right] {s}^\alpha_{t_2}[\cdot] {s}^\beta_{t_1}\\
&=&\displaystyle\sum_{\alpha,\beta}\text{Tr}_B\left[X^\alpha_{t_2} X^\beta_{t_1}\rho_\beta\right] [{s}^\alpha_{t_2},{s}^\beta_{t_1}[\cdot]]-\displaystyle\text{Tr}_B\left[\rho_\beta X^\beta_{t_1} X^\alpha_{t_2}\right] [{s}^\alpha_{t_2},[\cdot] {s}^\beta_{t_1}]\\
&=&\displaystyle\sum_{\alpha,\beta} C^{\alpha\beta}(t_2,t_1) [{s}^\alpha_{t_2},{s}^\beta_{t_1}[\cdot]]-C^{\beta\alpha}(t_1,t_2) [{s}^\alpha_{t_2},[\cdot] {s}^\beta_{t_1}]\;.
\end{array}
\end{equation}
\end{widetext}
Explicitly, we defined the correlation functions as
\begin{equation}
\label{eq:Cab}
    C^{\alpha\beta}(t_2,t_1)=\text{Tr}_B\left[X^\alpha_{t_2} X^\beta_{t_1}\rho_\beta\right]\;.
\end{equation}
In conclusion, the influence superoperator for the generalized interaction Hamiltonian $H_I'$ can be written as
\begin{equation}
\label{eq:Fprime}
    \begin{array}{lll}
    \mathcal{F}'[\cdot]&=&\displaystyle\int_0^t d t_2 \int_0^{t_2} dt_1\sum_{\alpha,\beta}\left\{C^{\beta\alpha}(t_1,t_2) [{s}^\alpha_{t_2},\cdot {s}^\beta_{t_1}]\right. \\
    &&\left.-C^{\alpha\beta}(t_2,t_1) [{s}^\alpha_{t_2},{s}^\beta_{t_1}\cdot]\right\}\;.
    \end{array}
\end{equation}

\subsubsection{Quantum white noise limit}
Here, we exemplify the results above in the case where the environment can be modeled as quantum white noise (i.e., frequency-independent spectral density non-zero for both positive and negative energies, and frequency-independent Bose-Einstein occupation number) and interact with the system in a ``rotating-wave'' fashion, i.e., $H^{\text{r.w.}}_I=s^+ A(0) + s^{-} A^\dagger(0)$, where $s^+=(s^-)^\dagger$ and $A(t)=\sum_k g_k b_k(t)$. In this case, only two out of the four correlations in Eq.~(\ref{eq:Cab}) are non-zero, i.e., $C^{++}(t_2,t_1)=C^{--}(t_2,t_1)=0$ and
\begin{equation}
    \begin{array}{lll}
         C^{+-}(t_2,t_1)&=&\displaystyle\text{Tr}_B\left[A^\dagger_{t_2}A_{t_1}\rho_\beta\right]  =\displaystyle\sum_k g_k^2 n_k e^{i\omega_k t}\\
         &=&\displaystyle 2\Gamma n\delta(t)\\
         
         C^{-+}(t_2,t_1)&=&\displaystyle\text{Tr}_B\left[A_{t_2}A^\dagger_{t_1}\rho_\beta\right]=\sum_k g_k^2 (n_k+1) e^{-i\omega_k t}\\
          &=&\displaystyle 2\Gamma (n+1)\delta(t) \;,
    \end{array}
\end{equation}
where $t=t_2-t_1$ and where we used $\int_{-\infty}^\infty d\omega e^{i\omega t}=2\pi\delta(t)$ together with the quantum white noise conditions
\begin{equation}
    \begin{array}{lll}
    J(\omega)&=&\displaystyle\pi\sum_{k}g_k^2\delta(\omega-\omega_k)=\Gamma\\
    n_k &=&n \;.
    \end{array}
\end{equation}
By using these expressions into Eq.~(\ref{eq:Fprime}), we obtain
\begin{equation}
    \begin{array}{lll}
    \mathcal{G}^\prime[\cdot]&=&\displaystyle C^{+-}(t)[s^-_{t_2},\cdot s^+_{t_1}]-C^{-+}(t)[s^-_{t_2}, s^+_{t_1}\cdot]\\
    &&\displaystyle+C^{-+}(t)[s^+_{t_2},\cdot s^-_{t_1}]-C^{+-}(t)[s^+_{t_2}, s^-_{t_1}\cdot]\\
    
    &=&\displaystyle\Gamma\delta(t)\left\{n[s^-_{t_1},\cdot s^+_{t_1}]-(n+1)[s^-_{t_1}, s^+_{t_1}\cdot]\right.\\
    &&\displaystyle+\left. (n+1)[s^+_{t_1},\cdot s^-_{t_1}]-n[s^+_{t_1}, s^-_{t_1}\cdot]\right\}\\
    
    &=&\displaystyle
    \Gamma\delta(t)\left\{n\left[2 s_{t_1}^-\cdot s_{t_1}^+-\cdot s_{t_1}^+ s_{t_1}^--s_{t_1}^+s_{t_1}^-\cdot\right]\right.\\
    &&+\left.(n+1)\left[2 s_{t_1}^+\cdot s_{t_1}^--\cdot s_{t_1}^- s_{t_1}^+-s_{t_1}^-s_{t_1}^+\cdot\right]\right\}\;,
    \end{array}
\end{equation}
which, once inserted in Eq.~(\ref{eq:appDiffEq}) gives a Lindblad equation in the Shr\"{o}dinger picture, i.e.,
\begin{equation}
    \dot{\rho}^\prime_S(t)=-i[H_s,\rho^\prime_S(t)]+ \Gamma D[\rho^\prime_S(t);s^-]\;,
\end{equation}
in terms of the dissipator
\begin{equation}
    \begin{array}{lll}
    D[\cdot;O]&=&\displaystyle
n\left[2 O\cdot O^\dagger-\cdot O^\dagger O-O^\dagger O\cdot\right]\\
    &&+(n+1)\left[2 O^\dagger\cdot O-\cdot OO^\dagger-OO^\dagger\cdot\right]\;,
    \end{array}
\end{equation}
where $O$ is a generic system operator.

\subsection{Classical Environment}
\label{app:classical_environment}
The results presented in the previous section include the possibility that part of the environment is made out of a classical stochastic process. To explicitly see this, we can consider, in Eq.~(\ref{eq:HSIB}), the replacement 
\begin{equation}
\label{eq:Bplusxi_class}
    H_I\mapsto H_I+H_I^\text{class}\;,
\end{equation}
 where $H_I^\text{class}={s}\xi(t)$ in terms of a classical stochastic field $\xi(t)$ which complements the original quantum environment $B$ with additional classical noise. In the following, we will denote the average with respect to the underlying probability distribution of the field $\xi(t)$ by $\mathbb{E}[\cdot]$. We will further suppose the stochastic process $\xi(t)$ to be Gaussian (i.e., all statistical moments depend on the second order one by Wick's theorem), stationary (i.e., the second order moment is invariant under translation in time of both its augments) and to have zero mean (i.e., $\mathbb{E}[\xi(t)]=0$).
With the replacement in Eq.~(\ref{eq:Bplusxi_class}), all the reasoning done in section \ref{app:influenceSuperoperator} continues to hold as long as we operate the substitution $\text{Tr}_B\mapsto\mathbb{E}~\text{Tr}_B$. As a result, the correlation appearing in Eq.~(\ref{eq:corr_func}) takes a further contribution 
\begin{equation}
\label{eq:class_corr_env}
C(t_2,t_1)\mapsto C(t_2,t_1)+C_\text{class}(t_2,t_1)\;,
\end{equation}
where
\begin{equation}
    C_\text{class}(t_2,t_1) = \mathbb{E}[\xi(t_2)\xi(t_1)]\;.
\end{equation}
The hypothesis of stationarity for the process $\xi(t)$ implies that $C_\text{class}(t_2,t_1)= C(t)$ for $t=t_2-t_1$. Furthermore, since the field $\xi(t)$ is classical it has trivial commutation relations ($[\xi(t_2),\xi(t_1)]=0$), and the correlation satisfies the additional constraint
\begin{equation}
\begin{array}{lll}
C_\text{class}(t)&=&\displaystyle\mathbb{E}[\xi(t_2)\xi(t_1)]=\displaystyle \mathbb{E}[\xi(t_1)\xi(t_2)]\\
&=&C_\text{class}(-t)\;,
\end{array}
\end{equation}
i.e., it is symmetric under time-reversal. 

Even more explicitly, using the same reasoning as above, it is possible to generalize the interaction term considered in section \ref{app:generalized_interaction} with the addition of a classical stochastic field, i.e., further generalizing Eq.~(\ref{eq:HprimeI}) using
\begin{equation}
    H'_I\mapsto H'_I+H_I^\text{class}\;.
\end{equation}
This ultimately results in an additional term in the influence superoperator of the form
\begin{equation}
\mathcal{F}'\mapsto \mathcal{F}'+\mathcal{F}(t,{s},C_\text{class}(t))\;.
\end{equation}

\subsection{Wick's theorem for time-ordered superoperators}
\label{app:Wick_superoperators}
To optimize space, whenever not explicitly stated, all traces in this section are assumed to be over the bath $B$.
Common derivations of Wick's theorem (see \cite{Rammer}, pag. 243) show how to reduce $n-$point correlation functions involving fields linear in bosonic creation and annhilation operators in terms of 2-point correlation functions. Using the notation presented in this article, this allows, for example, to write
\begin{equation}
\label{eq:usualWick}
\langle{X}(t_1)\cdots{X}(t_{2N})\rangle=\sum_{C\in P_{2N}}\prod_{(l,m)\in C}\mathrm{Tr}_{B}\left[{X}(t_l){X}(t_m)\rho_{B}\right]
\end{equation}
where $\langle\cdot\rangle\equiv \text{Tr}_B[\cdot\rho_B(0)]$ and where $\rho_B(0)$ is a Gaussian state such that $\mathrm{Tr}_{B}[X(t)\rho_B]=0$, and where $P_{2N}$ is the collection of all possible sets (full contractions $C$) whose elements are $N$ disjoint pairs (i.e., contractions) taken from $\{1,\cdots,2N\}$. We further have that $\mathrm{Tr}_{B}\left[\mathcal{X}(t_1)\cdots\mathcal{X}(t_{2N+1})\rho_{B}\right]=0$.
In this section, we show that Wick's theorem can also be written in terms of time-ordered superoperators linear in the fields $X(t)$. More specifically, we are interested in correlations taking the form
\begin{equation}
C_{2N}=\mathrm{Tr}_{B}\left[\mathcal{T}\mathcal{X}_{t_1}^{i_{1}}\cdots\mathcal{X}_{t_{2N}}^{i_{2N}}\rho_{B}\right]\;,
\end{equation}
where $\mathcal{X}_{t}^{1}[\cdot]=X(t)\cdot$ and $\mathcal{X}_{t}^{2}[\cdot]=\cdot X(t)$. To begin, we can write
\begin{equation}
C_{2N}=\mathrm{Tr}_{B}\left[\mathcal{X}_{t_{P_T(1)}}^{i_{P_T(1)}}\cdots\mathcal{X}_{t_{P_T(2N)}}^{i_{P_T(2N)}}\rho_{B}\right]\;,
\end{equation}
where $P_T$ is a permutation of $\{1,\cdots,2N\}$ such that $t_{P_T(1)}\geq\cdots\geq t_{P_T(2N)}$. Now, depending on the value of their upper indexes, the superoperators act on the left or right on the density matrix so that
\begin{equation}
\begin{array}{lll}
C_{2N}&=&\displaystyle\mathrm{Tr}\left[\prod_{j:i_{P_T(j)}=1}\mathcal{X}_{t_{P_T(j)}}^{i_{P_T(j)}}\prod_{k:i_{P_T(k)}=2}\mathcal{X}_{t_{P_T(k)}}^{i_{P_T(k)}}\rho_{B}\right]\\
&=&\displaystyle\mathrm{Tr}\left[\prod_{k:i_{\bar{P}_T(k)}=2}X(t_{P_T(k)})\prod_{j:i_{P_T(j)}=1}X(t_{P_T(j)})\rho_{B}\right]
\end{array}
\end{equation}
where we used the cyclic property of the trace to bring the operators on the right of the density matrix back on the left. We further noted that this procedure has caused the fields labeled by $k$ indexes to be anti-ordered in time and indicated this by the overbar in $\bar{P}_T(k)$.
It is now possible to invoke the usual version of Wick's theorem given in Eq.~(\ref{eq:usualWick}). In doing so, we obtain three different type of contractions: (i) contractions between indexes $j$  (which we will denote as $j_1$ and $j_2$) whose contribution to Wick's theorem can be written as 
\begin{equation}
\label{eq:wick_1}
\begin{array}{lll}
    \text{Tr}[X(t_{P_T(j_1)})X(t_{P_T(j_2)})\rho_B]&=&\text{Tr}[\mathcal{X}^1_{t_{P_T(j_1)}}\mathcal{X}^1_{t_{P_T(j_2)}}\rho_B]\\
    &=&\text{Tr}[\mathcal{T}\mathcal{X}^1_{t_{j_1}}\mathcal{X}^1_{t_{j_2}}\rho_B]\;,
    \end{array}
\end{equation}
(ii) contractions between indexes $k$  (which we will denote as $k_1$ and $k_2$) whose contribution to Wick's theorem can be written as 
\begin{equation}
\label{eq:wick_2}
\begin{array}{lll}
    \text{Tr}[X(t_{\bar{P}_T(k_1)})X(t_{\bar{P}_T(k_2)})\rho_B]&=&\text{Tr}[\mathcal{X}^2_{t_{{P}_T(k_1)}}\mathcal{X}^2_{t_{{P}_T(k_2)}}\rho_B]\\
    &=&\text{Tr}[\mathcal{T}\mathcal{X}^2_{t_{k_1}}\mathcal{X}^2_{t_{k_2}}\rho_B]\;,
    \end{array}
\end{equation}
and $(iii)$ contractions between $j$ and $k$ indexes whose contribution to Wick's theorem can be written as 
\begin{equation}
\label{eq:wick_3}
\begin{array}{lll}
    \text{Tr}_B[X(t_{\bar{P}_T(k)})X(t_{P_T(j)})\rho_B]&=&\text{Tr}_B[\mathcal{X}^2_{t_{\bar{P}_T(k)}}\mathcal{X}^1_{t_{\bar{P}_T(j)}}\rho_B]\\
    &=&\text{Tr}_B[\mathcal{T}\mathcal{X}^2_{t_{k}}\mathcal{X}^1_{t_{j}}\rho_B]\;,
    \end{array}
\end{equation}
where we noted that, in this last case, the time-ordering is irrelevant as the superoperators commute.
As a consequence, using Wick's theorem for operators in Eq.~(\ref{eq:usualWick}) together with Eqs.~(\ref{eq:wick_1}), (\ref{eq:wick_2}), and \ref{eq:wick_3}) we can write
\begin{equation}
\label{eq:wickforsuper-operator_proof}
\mathrm{Tr}\left[\mathcal{T}\mathcal{X}_{t_1}^{i_{1}}\cdots\mathcal{X}_{t_{2N}}^{i_{2N}}\rho_{B}\right]=\sum_{C\in P_{2N}}\prod_{(l,m)\in C}\mathrm{Tr}\left[\mathcal{T}\mathcal{X}_{t_l}^{i_{l}}\mathcal{X}_{t_m}^{i_{m}}\rho_{B}\right],
\end{equation}
which proves Eq.~(\ref{eq:wickforsuper-operator}). Since the number of superoperators present in the correlation is equal to the number of corresponding operators, the hypothesis $\text{Tr}_B[X(t)\rho_B]=0$ together with Wick's theorem for operators allow us to also conclude that $\mathrm{Tr}_{B}\left[\mathcal{T}\mathcal{X}_{t_1}^{i_{1}}\cdots\mathcal{X}_{t_{2N}}^{i_{2N+1}}\rho_{B}\right]=0$.

\section{Pseudomode model}
\label{app:pseudomode_model}
In this section we provide details about the Pseudomode model. In particular, we derive the pseudoShr\"{o}dinger equation in Eq.~(\ref{eq:pseudoShr\"{o}dinger}) which supports the main correspondence presented in Eq.~(\ref{eq:PM_dynamics}).\\

Following \cite{Tamascelli,Lambert}, in order to characterize a pseudomode model, we proceed in two logical steps. First, we define an open quantum system for which there exists a specific interaction operator whose  correlation function can be expressed as a sum of decaying exponentials. As a consequence, this open quantum system can be used to define a structured environment which simulates physical baths whose correlation function can be expressed in such a functional form. Secondly, we show that the reduced density matrix for a system interacting with this environment can be equivalently computed by solving the pseudomode Lindblad master equation in Eq.~(\ref{eq:pseudoShr\"{o}dinger}).
\subsubsection{Open Quantum System}
\label{app:open_quantum_system}
We consider a model in which the system directly couples with $N_\text{PM}$ pseudomodes each of which interacts with its own residual environment made out of bosonic modes as
\begin{equation}
\label{eq:HT}
 H_T =H_\mathrm{PM}+H^I_\mathrm{PM-RE}+H_\mathrm{RE}\;,
\end{equation}
where 
\begin{equation}
\label{eq:Xsimple}
\begin{array}{lll}
    H_\mathrm{PM}&=&H_\mathrm{S}+ X {s}+\sum_{j=1}^{N_\text{PM}} \Omega_j a_j^\dagger a_j\;,\\
     X &=&\sum_j\lambda_j X_j\;.
    \end{array}
\end{equation}
Here, $H_S$ is the system Hamiltonian and we considered $N_\text{PM}$ pseudomodes $a_j$ having frequency $\Omega_j\in\mathbb{C}$. The system-pseudomode interaction is characterized by  a generic system operator ${s}$ and an interaction operator $ X $, where $\lambda_j\in\mathbb{C}$ and  $X_j=a_j+a_j^\dagger$. We further suppose that each each pseudomode $a_j$ is associated with an independent residual environment ($\text{RE}$) made out by a collection of bosonic modes $b_{jk}$ having frequency $\omega_{jk}$ so that $H_{\text{RE}}=\sum_j\sum_k \omega_{jk}b_{jk}^\dagger b_{jk}$. To highlight their importance, we now describe the interaction of each pseudomode with its residual environment and the initial condition in the "PM-RE" space in more detail.
 \begin{enumerate}
\item We assume the interaction with the $j$th pseudomode to be written in following rotating-wave  form: 
\begin{equation}
 H_\mathrm{PM-RE}^{I,j}=\sum_k g_{jk}(a_j b^\dagger_{jk}+a_j^\dagger b_{jk})\;, 
\end{equation}
so that $H_\mathrm{PM-RE}^\mathrm{I}=\sum_{j=1}^{N_\mathrm{PM}} H_\mathrm{PM-RE}^{I,j}$.
\item  We assume the spectral density characterising the $j$th residue environment to be a constant and defined for both negative and positive frequencies, i.e.,
\begin{equation}
\label{eq:Jj}
J_j(\omega)=\pi\sum_{k} g^2_{jk}\delta(\omega-\omega_{jk}) \equiv\Gamma_j\;.  
\end{equation}
\item We assume the initial state in the PM-RE space to be 
\begin{equation}
\label{eq:rhoofpmre}
    \rho_{\text{PM-RE}}(0)\propto\prod_{j=1}^{N_\text{PM}}\text{exp}\left[{-\beta_j\Omega_j a^\dagger_j a_j-\beta_{jk}\sum_k\omega_{jk}b^\dagger_{jk}b_{jk}}\right],
\end{equation}
where we omitted the factor needed to ensure $\text{Tr}_{\text{PM-RE}}\rho_{\text{PM-RE}}(0)=1$.
In order to allow the invariance of the initial state under the free dynamics in PM-RE space, i.e., under the Hamiltonian
\begin{equation}
\begin{array}{lll}
 H_\mathrm{PM}^\mathrm{free}&=&\displaystyle H_\mathrm{PM-RE}^I+H_\mathrm{RE}+\sum_{j=1}^{N_\mathrm{PM}}\Omega_j a_j^\dagger a_j\\
&=&\displaystyle\sum_{j=1}^{N_\mathrm{PM}}\left[\Omega_j a_j^\dagger a_j+\sum_k \omega_{jk} b^\dagger_{jk}b_{jk}\right.\\
&&\displaystyle\left.+g_{jk}(a_j b^\dagger_{jk}+a_j^\dagger b_{jk})\right]\;,
 \end{array}
\end{equation}
we consider a quantum white noise setting in which the expectation value of the occupation of each pseudomode is frequency independent, i.e.,
\begin{equation}
\label{eq:excitationnumber}
\langle b_{jk}^\dagger b_{jk}\rangle=\langle a_j^\dagger a_j\rangle=n_j\;.   
\end{equation}
Here the expectation values are taken with respect to Eq.~(\ref{eq:rhoofpmre}), which further implies the constraints
\begin{equation}
 \beta_j\Omega_j=\beta_{jk}\omega_{jk}=\mathrm{ln}\frac{1+n_j}{n_j}\;.
\end{equation}
Using these relations, Eq.~(\ref{eq:rhoofpmre}) can be written as
$\rho_{\text{PM-RE}}(0)\propto\prod_{j=1}^{N_\text{PM}}\text{exp}\left[-\mathrm{ln}(1+1/n_j){N}_\mathrm{PM-RE}^j\right]$, where ${N}_\mathrm{PM-RE}^j=a_j^\dagger a_j+\sum_k b_{jk}^\dagger b_{jk}$ is the total number of excitations of the jth pseudomode and its residue environment. Since the  Hamiltonian $H_\mathrm{PM-RE}^I$ describes a rotating-wave interaction, and there is no interaction between the pseudomodes $a_j$, we have $[H_\mathrm{PM-RE}^\mathrm{free},\sum_j{N}_\mathrm{PM-RE}^j]=0$, which, in turn, implies
\begin{equation}\label{eq:steadyrho}
[H_\mathrm{PM-RE}^\mathrm{free},\rho_\mathrm{PM-RE}(0)]=0\;,    
\end{equation}
i.e., the initial state is invariant under the free PM-RE dynamics.
\end{enumerate}

We now consider the correlation function
\begin{equation}
    C(t+s,s)=\text{Tr}_\mathrm{PM-RE}[ X (t+s) X (s)\rho_\mathrm{PM-RE}(0)]\;,
\end{equation}
where $ X (t)=U^\dagger(t) X U(t)$, and $U(t)=e^{-i H_\mathrm{PM}^\mathrm{free}t}$. Using Eq.~(\ref{eq:steadyrho}), we can prove that this correlation is stationary, i.e,
\begin{equation}\begin{split}
\label{eq:correlationforpmre}
C(t+s,s)=&\mathrm{Tr}[U^\dagger(t+s) X U(t) X U(s)\rho_\mathrm{PM-RE}(0)]\\
=&\mathrm{Tr}[U^\dagger(t) X U(t) X U(s)\rho_\mathrm{PM-RE}(0)U^\dagger(s)]\\
=&\mathrm{Tr}[ X (t) X (0)\rho_\mathrm{PM-RE}(0)]:=C(t)\;,
\end{split}
\end{equation}
where, for notational convenience, we omitted the labels $\text{PM}\otimes\text{ RE}$ under trace operator. We now compute $C(t)$ explicitly by considering the Heisenberg equations of motion
\begin{equation}
\label{eq.heisenbergPM-RE}
\begin{array}{lllll}
\dot{a}_j(t)&=&i[H_\mathrm{PM}^\mathrm{free},a_j]&=&\displaystyle-i\left(\Omega_j a_j+\sum_k g_{jk}b_{jk}\right)\\
\dot{b}_{jk}(t)&=&i{[H_\mathrm{PM}^\mathrm{free},b_{jk}]}&=&\displaystyle-i\left(g_{jk}a_j+\omega_{jk}b_{jk}\right)\;,
\end{array}
\end{equation}
with initial conditions $a_j(0)$, $b_{jk}(0)$  at $t=0$. By introducing the Laplace transformation $a_j[s]$ and $b_{jk}[s]$ of the operators, we can write 
\begin{equation}
    a_j[s]=F_j[s]a_j(0)-iF_j[s]\sum_k G_{jk}[s]b_{jk}(0)\;,
\end{equation}
in terms of
\begin{equation}
    \begin{array}{lll}
    F_j[s]&=&\displaystyle\frac{1}{s+i\Omega_j+\displaystyle\sum_{k}\frac{g^2_{jk}}{s+i\omega_{jk}}}\\
    &=&\displaystyle\frac{1}{s+i\Omega_j+\displaystyle\frac{1}{i\pi}\int d\omega \frac{J_j(\omega)}{\omega-i s}}\\
     &=&\displaystyle\frac{1}{s+i\Omega_j+\Gamma_j\text{sg(t)}}\\
     
     G_{jk}[s]&=&\displaystyle\frac{g_{jk}}{s+i\omega_{jk}}\;,

    \end{array}
\end{equation}
where we used Eq.~(\ref{eq:Jj}) and where $\text{sg}(t)=t/|t|$, for $t\neq 0$. We can now use the inverse Laplace transform to find
\begin{equation}
\label{eq:fua}
\begin{array}{lll}
    a_j(t)&=&a_j(0) e^{-i\Omega_j t-\Gamma_j|t|}\\
    &&-i\displaystyle\sum_k g_{jk}b_{jk}(0)\int_0^t d\tau e^{-i\Omega_j \tau-\Gamma_j|\tau|}e^{-i\omega_{jk}(t-\tau)}\;,
    \end{array}
\end{equation}
where we used the convolution theorem.
Thanks to the stationarity condition in Eq.~(\ref{eq:correlationforpmre}), we can use the formula above to directly compute the correlation as
\begin{equation}
\label{eq:correlationbyheisenberge}
\begin{array}{lll}
    C(t)&=&\displaystyle\text{Tr}[ X (t)  X (0)]\\
    &=&\displaystyle\sum_j\lambda_j^2\left(\text{Tr}[a^\dagger_j(t)a_j(0)]+\text{Tr}[a_j(t)a^\dagger_j(0)]\right)\\
      &=&\displaystyle\sum_j\lambda_j^2 \left[n_j e^{i\Omega_j t}+(n_j+1)e^{-i\Omega_j t}\right]e^{-\Gamma_j|t|}\;. 
    \end{array}
\end{equation}
Despite having already reached the main goal of this section with the expression above, it is interesting to also compute the correlation directly, i.e., without assuming (or, in our case, previously proving) translational invariance in time. This will highlight the role of the quantum fluctuations in the residual environment in order to effectively obtain stationarity. To do this, we write, with the help of Eq.~(\ref{eq:fua}),
\begin{widetext}
\begin{equation}
\begin{array}{lll}
    C(t_2,t_1)  &=&\text{Tr}[ X (t_2)  X (t_1)]\\
    &=&\displaystyle\sum_j\lambda_j^2\left\{e^{i\Omega_j(t_2-t_1)-\Gamma_j(|t_2|+|t_1|)}\text{Tr}[a^\dagger_j(0)a_j(0)]+e^{-i\Omega_j(t_2-t_1)-\Gamma_j(|t_2|+|t_1|)}\text{Tr}[a(t_2) a^\dagger(t_1)]\right\}\\
    &&+\displaystyle\sum_j\lambda_j^2\sum_k g^2_{jk}\text{Tr}[b^\dagger_{jk}(0) b_{jk}(0)]\int_0^{t_2}d\bar{\tau}\int_0^{t_1}d{\tau} e^{i\omega_{jk}(t_2-\bar{\tau}-t_1+\tau)}e^{i\Omega_j \bar{\tau}-\Gamma_j|\bar{\tau}|}e^{-i\Omega_j {\tau}-\Gamma_j|{\tau}|}\\
        &&+\displaystyle\sum_j\lambda_j^2\sum_k g^2_{jk}\text{Tr}[b^{jk}(0)b^\dagger_{jk}(0)]\int_0^{t_2}d\bar{\tau}\int_0^{t_1}d{\tau} e^{i\omega_{jk}(t_2-\bar{\tau}-t_1+\tau)}e^{i\Omega_j \bar{\tau}-\Gamma_j|\bar{\tau}|}e^{-i\Omega_j {\tau}-\Gamma_j|{\tau}|}\;.
    
    \end{array}
\end{equation}
\end{widetext}
Using Eq.~(\ref{eq:excitationnumber}) and Eq.~(\ref{eq:rhoofpmre}), the last two expressions can be written as $\pi$ times an integral over frequencies which can be evaluated giving rise to a $2\pi\delta(\bar{\tau}-[t_2-t_1+\tau])$. This allows to further compute one of the integrals in time to obtain, assuming $t_2>t_1$,
\begin{widetext}
\begin{equation}
\begin{array}{lll}
    C(t_2,t_1)  &=&\text{Tr}[ X (t_2)  X (t_1)]\\
    &=&\displaystyle\sum_j\lambda_j^2\left\{n_j e^{i\Omega_j(t_2-t_1)}+(n_j+1)e^{-i\Omega_j(t_2-t_1)}\right\}e^{-\Gamma_j(|t_2|+|t_1|)}\\
    &&+\displaystyle2\sum_j\lambda_j^2 \Gamma_j \left\{n_j\int_0^{t_1}d{\tau}e^{i\Omega_j (t_2-t_1)-\Gamma_j|t_2-t_1+\tau|}e^{-\Gamma_j|{\tau}|}+(1+n_j)\int_0^{t_1}d{\tau}e^{-i\Omega_j (t_2-t_1)-\Gamma_j|t_2-t_1+\tau|}e^{-\Gamma_j|{\tau}|}\right\}\\
    &=&\displaystyle\sum_j\lambda_j^2\left\{n_j e^{i\Omega_j(t_2-t_1)}+(n_j+1)e^{-i\Omega_j(t_2-t_1)}\right\}e^{-\Gamma_j(t_2-t_1)}\;,
    \end{array}
\end{equation}
\end{widetext}
which, indeed, shows how the fluctuations in the residual bath variables are essential to generate the $t_2-t_1$ dependence for stationarity.

\subsubsection{Lindblad Master Equation}
\label{app:lindblad_master_equation}
Here, we show that the reduced system dynamics can be computed by solving a Lindblad master equation in the  S+PM space. We do this by considering the Hamiltonian in Eq.~(\ref{eq:HT}) and by tracing the residue environment. 
This can be done using the results in section \ref{app:generalized_interaction}, which rely on the following correlations for the residue environment operator $B=\sum_{j,k}g_{j,k}b_{j,k}$, i.e.,
\begin{equation}
    \begin{array}{lll}
C_1^\mathrm{RE}(t_2,t_1)&=&\displaystyle\mathrm{Tr}_B\left[B^\dagger(t_2) B(t_1)\rho_B\right]\\
&=&\displaystyle\sum_{j,k} g_{j,k}^2e^{i\omega_{k,j}(t_2-t_1)}\mathrm{Tr}_B\left[b_{j,k}^\dagger b_{j,k}\rho_B\right]\\
&=&\displaystyle\frac{1}{\pi}\sum_j\int d\omega J_j(\omega)n_je^{i\omega t}\\
&=&\displaystyle2\sum_j\Gamma_j n_j\delta(t)\;,\\
C_2^\mathrm{RE}(t_2,t_1)&=&\displaystyle\mathrm{Tr}_B\left[B(t_2) B^\dagger(t_1)\rho_B\right]\\
&=&\displaystyle2\sum_j\Gamma_j(1+n_j)\delta(t)\;.
\end{array}\end{equation}
where $B(t)=e^{iH_\mathrm{RE}t}Be^{-iH_\mathrm{RE}t}$ and $t=t_2-t_1$.  The reduced dynamics in the S-PM space can be obtained using Eq.~(\ref{eq:gen}), i.e. $\rho_S(t)=\mathcal{T}e^{\mathcal{F}^\mathrm{RE}(t)}\rho_S(0)$. The corresponding $\mathcal{G}^\mathrm{RE}$ is
\begin{equation}\begin{array}{ll}
\mathcal{G}^\mathrm{RE}(t_2,t_1)=&\displaystyle\sum_{j=1}^{N_\mathrm{PM}}2\Gamma_j n_j\delta(t)D(a_j(t_2),a_j(t_1))\\
&\displaystyle+2\Gamma_j(1+n_j)\delta(t)D(a_j^\dagger(t_2),a_j^\dagger(t_1))\;,
\end{array}\end{equation}
and the corresponding $\mathcal{F}^\mathrm{RE}$ is
\begin{equation}
\label{eq:fsuperoperator}
\begin{array}{lll}
\mathcal{F}^\mathrm{RE}(t')&=&\displaystyle-\int_0^{t'}d t_2\int_0^{t_2}d t_1\mathcal{G}(t_2,t_1)\\
&=&\displaystyle\sum_{j=1}^{N_\mathrm{PM}}\int_0^{t'}d t_2-\Gamma_j(1+n_j)D(a_j^\dagger(t_2),a_j^\dagger(t_2))\\
&&\displaystyle-\Gamma_j n_j D(a_j(t_2),a_j(t_2))\;,
\end{array}\end{equation}
where we defined
\begin{equation}
    D(a,b)[\cdot]=[a,b^\dagger\cdot]-[a^\dagger,\cdot b]\;,
\end{equation}
and the Heisenberg operator $a_j(t)=e^{i H_\mathrm{PM}t}a_j e^{-iH_\mathrm{PM}t}$. Going back to the Schrondinger picture, the dynamics in the S-PM space takes the following Lindblad form 
\begin{equation}
\label{eq:lindforspm}
 \dot{\rho}_\mathrm{SPM}=-i[H_\mathrm{PM},\rho_\mathrm{SPM}]+L[\rho_\mathrm{SPM}]\;, 
\end{equation}
where 
\begin{equation}\begin{split}
  L[\cdot]=&\sum_j\Gamma_j(1+n_j)\left(2a_j[\cdot]a_j^\dagger-\{a_j^\dagger a_j,\cdot\}\right)\\
  &\sum_j\Gamma_j n_j \left(2a_j^\dagger[\cdot]a_j- \{a_ja_j^\dagger,\cdot\}\right)\;.  
\end{split}
\end{equation}
To finish this section, we show that the correlation in Eq.~(\ref{eq:correlationbyheisenberge}) can be computed using the influence superoperator formalism. To do this, we consider the free dynamics in the PM-RE space as described by the Hamiltonian $H_\mathrm{PM}^\mathrm{free}$. The translational invariance of $C(t)$ in Eq.~(\ref{eq:correlationforpmre}) allows to write it as 
\begin{equation}
    \begin{split}
     \label{eq:lindbladcorrelation}
       C(t) =&\mathrm{Tr_{PM\otimes RE}}[ X U(t) X \rho_\mathrm{PM-RE}(0)U^\dagger(t)]\\
       =&\mathrm{Tr_{PM}}[ X \mathrm{Tr_{RE}}[U(t) X \rho_\mathrm{PM-RE}(0)U^\dagger(t)]]\;,
    \end{split}
\end{equation}
 which in turn, allows to use the results in section \ref{app:generalized_interaction} with the unphysical initial condition $ X \rho_\mathrm{PM-RE}(0)$  We can then use Eq.~(\ref{eq:gen}) where the influence superoperator can be calculated using the similar steps as those used to get Eq.~(\ref{eq:fsuperoperator}). We can then write 
 \begin{equation}\begin{split}\label{eq:slindcorrelation}
  C(t)&=\mathrm{Tr_{PM}}[ X e^{\mathcal{L}_\mathrm{PM}t}[ X \rho_\mathrm{PM}(0)]]\\
  &=\mathrm{Tr_{PM}}[e^{\mathcal{L}^\dagger_\mathrm{PM}t}[ X ] X \rho_\mathrm{PM}(0)]\;,
\end{split}
\end{equation}
where $\rho_\mathrm{PM}(t)=\mathrm{Tr_{RE}}[U(t)\rho_\mathrm{PM}(0)\otimes \rho_\mathrm{RE}(0)U^\dagger(t)]$, $
\mathcal{L}_\mathrm{PM}[\cdot]=-i\sum_{j=1}^\mathrm{N_{PM}}[\Omega_ja_j^\dagger a_j,\cdot]+L[\cdot]$ and where
 $\mathcal{L}^\dagger_\mathrm{PM}$ is the adjoint operator of $\mathcal{L}_\mathrm{PM}$
\begin{equation}\begin{split}
\mathcal{L}^\dagger_\mathrm{PM}[\cdot]=&\sum_{j=1}^\mathrm{N_{PM}}i[\Omega_ja_j^\dagger a_j,\cdot]+\Gamma_j n_j \left(2a_j[\cdot]a_j^\dagger- \{a_ja_j^\dagger,\cdot\}\right)\\
&+\Gamma_j(1+n_j)\left(2a_j^\dagger[\cdot]a_j-\{a_j^\dagger a_j,\cdot\}\right)\;.
\end{split}
\end{equation}
By using the identities
\begin{equation}\begin{array}{lll}
\mathcal{L}_\mathrm{PM}^\dagger[a_j]&=&-(i\Omega_j+\Gamma_j)a_j\\  
\mathcal{L}_\mathrm{PM}^\dagger[a_j^\dagger]&=&(i\Omega_j-\Gamma_j)a_j^\dagger\;,
\end{array}\end{equation}
the correlation can be written as
\begin{equation}\label{eq:dec_app}\begin{array}{lll}
 C(t)
 &=&\displaystyle\sum_{j=1}^\mathrm{N_{PM}} \mathrm{Tr_{PM}}[e^{\mathcal{L}_\mathrm{PM}^\dagger t}[X_j]\sum_{j=1}^\mathrm{N_{PM}}X_j\rho_\mathrm{PM}(0)]\\
 &=&\displaystyle\sum_{j=1}^\mathrm{N_{PM}} \lambda_j^2\{e^{(-i\Omega_j-\Gamma_j)t}\langle a_j a_j^\dagger\rangle+e^{(i\Omega_j-\Gamma_j)t}\langle a_j^\dagger a_j\rangle\} \\
 &=&\displaystyle\sum_{j=1}^\mathrm{N_{PM}} \lambda_j^2e^{-\Gamma_jt}\left(e^{-i\Omega_jt}(n_j+1)+e^{i\Omega_jt}n_j\right)\;.
\end{array}
\end{equation}
We note that we could have also obtained the same result by using the more general expression given by  Eq.~(6) in \cite{Tamascelli}, i.e., 
\begin{equation}
    C(t+s,s)=\mathrm{Tr_{PM}}\{X e^{\mathcal{L}t}\left[X e^{\mathcal{L}s}[\rho_\mathrm{PM}(0)]\right]\}\;,
\end{equation}
where $X=\sum_{j=1}^\mathrm{N_{PM}}\lambda_j x_j$. Since the initial state is invariant under $\mathcal{L}$, i.e., $e^{\mathcal{L}t}\rho_\mathrm{PM}(0)=\rho_\mathrm{PM}(0)$, this is equivalent to Eq.~(\ref{eq:slindcorrelation}).

\section{Complementing Pseudomodes with Classical Stochastic Fields}
\label{app:stoch_PM_zeroT}
In this section, we analyze how to replace some of the quantum degrees of freedom in the pseudomode model with classical stochastic fields.

We consider a modification of the pseudomode model presented in Eq.~(\ref{eq:HT}) defined by adding classical stochastic noise to the system, i.e., 
\begin{equation}
\label{eq:Bplusxi}
    H^\xi_T= H^\xi_\text{PM} +H_{\text{PM-RE}}^\text{I}+H_{\text{RE}}\;,
\end{equation}
where $H^\xi_\text{PM}=H_\text{PM}+{s}\xi(t)$. Here, the field $\xi(t)$ is supposed to have all the properties outlined in section \ref{app:classical_environment}, i.e., Gaussianity, stationarity, and zero mean. 
We then proceed by characterizing this model following the two logical steps already used in section \ref{app:pseudomode_model}, i.e., we first find the correlation of the interacting operator of the model and then describe the master equation which can be used to simulate the reduced density matrix.

The correlation of the model in Eq.~(\ref{eq:Bplusxi}) can be found by simply considering the results provided in  section \ref{app:open_quantum_system} and section \ref{app:classical_environment}. In fact, Eq.~(\ref{eq:class_corr_env}) tells us that the stochastic open quantum system in Eq.~(\ref{eq:Bplusxi}) has correlations given by
\begin{equation}
\label{eq:CPM_Cclass}
    {C}_\text{PM+class}=C_\text{PM}+C_\text{class}\;,
\end{equation}
where 
\begin{equation}
\label{eq:CPM_Cclass_explicit}
    \begin{array}{lll}
      C_\text{PM}(t_2,t_1)&=&\displaystyle\text{Tr}_{\text{PM-RE}}[X_\text{PM}(t_2)X_\text{PM}(t_1)\rho_{\text{PM-RE}}(0)]\\
      &=&\displaystyle\sum_{j=1}^{N_\text{PM}} \lambda_j^2 [(1+n_j)e^{-i\Omega_j t}+n_j e^{i\Omega_j t}]e^{-\Gamma_j |t|}\\
        C_\text{class}(t_2,t_1) &=&\displaystyle \mathbb{E}[\xi(t_2)\xi(t_1)]\;,
    \end{array}
\end{equation}
with $X_\text{PM}(t)=\sum_{j=1}^{N_\text{PM}}\lambda_j X_j(t)$ and where $C_\text{PM}$ is the correlation of the pure pseudomode model, see Eq.~(\ref{eq:correlationbyheisenberge}).

In our second step, we present a master equation to compute the reduced system dynamics. To do this, we can simply adapt the results presented in section \ref{app:lindblad_master_equation} by incorporating the action of the stochastic field into the system, i.e., by considering $H_\mathrm{PM}\mapsto H_\mathrm{PM}+{s}\xi(t)$. This leads to the Lindblad equation for S-PM system
\begin{equation}
\dot{\rho}^{\xi}(t)=-i[H_\text{PM}+{s}\xi(t),\rho^{\xi}(t)]+\sum_{j=0}^{N_\text{PM}} D_j[\rho^{\xi}(t)]\;,
\end{equation}
where
\begin{equation}\begin{split}
D_j[\cdot]=&\Gamma_j[(1+n_j)(2a_j\cdot a_j^\dagger-a_j^\dagger a_j\cdot -\cdot a^\dagger_j a_j)\\
&+n_j(2a^\dagger_j\cdot a_j-a_j a^\dagger_j\cdot -\cdot a_j a^\dagger_j)]\;.
\end{split}\end{equation}
Each Lindblad operator $D_j$ only acts on $j$th pseudomode's space. The reduced system dynamics can be computed by tracing the pseudomodes and averaging over the stochastic noise, i.e., 
\begin{equation}
\label{eq:exact_average}
    \rho_S(t)=\mathbb{E}\left[\text{Tr}_\text{PM}[\rho^{\xi}(t)]\right]\;.
\end{equation}
In practice, it might not  be possible to compute this exact average over the stochastic field. In these cases, the equation above must be replaced with an empirical average as
\begin{equation}
\label{eq:empirical_average}
    {\rho}^\xi_S(t;N_\text{stoch})=\frac{1}{N_\text{stoch}}\sum_{j=1}^{N_\text{stoch}}\text{Tr}_\text{PM}[\rho^{\xi_j}(t)]\;,
\end{equation}
where $\xi_j(t)$ indicates one of the $N_\text{stoch}$ realizations of the stochastic process. 
This highlights the fact that, in order to take advantage of this formalism, the resulting density matrix is now a random variable. In turn, this mean that, on top of the deterministic sources of error due to potential imprecision in modeling the correlation of the original environment, here we also need to take into account for further uncertainties due to the statistical properties of the stochastic variable $\bar{\rho}^\xi_S(t)$. The analysis of these deterministic and stochastic errors will be focus of the next section.
\subsection{Error Analysis}
It is important to note that, by replacing the exact average in Eq.~(\ref{eq:exact_average}) with the ``empirical'' one in Eq.~(\ref{eq:empirical_average}) we are effectively introducing a potential error in the model. Intuitively, we can divide the sources of imprecision in the pseudomode model into two classes. One is a ``bias'' due to errors in approximating the correlatoin of the original environment $C(t)$ with ${C}_\text{PM}$ in Eq.~(\ref{eq:CPM_Cclass}).The second source of error is due to the stochastic nature of Eq.~(\ref{eq:empirical_average}). In the following we estimate more precisely the combined effects of these two error contributions. 

To be more specific, we are going to consider physical quantities in the form of the empirical average $O^\xi(t;N_\text{stoch})=\text{Tr}_{S} \left[{O}_S\rho^\xi_S(t)\right]$ for a generic system operator ${O}$ and we can further define $O^\xi(t)\equiv O^\xi(t;1)$. Using Eq.~(\ref{eq:empirical_average}) we can further write
\begin{equation}
\label{eq:noiseRandom}
O^\xi(t;N_\text{stoch})=\frac{1}{N_\text{stoch}}\sum_{j=1}^{N_\text{stoch}}\text{Tr}_{S} \left[{O}\rho^\xi_S(t)\right]\;,
\end{equation}
which better highlights its nature of stochastic variable. Its expectation value is
\begin{equation}
\label{eq:Obar}
    {O}(t)=\mathbb{E}[O^\xi(t;N_\text{stoch})]\;,
\end{equation}
which is the stochastic-errors free estimate of the true expectation value $O_\text{true}(t)$ and only affected by the bias error in reproducing the original correlation function. This can be quantified more explicitly by defining the following measure for the total error as
\begin{equation}
    \Delta O(t)=\mathbb{E}[|O^\xi(t;N_\text{stoch})-O_\text{true}(t)|]\;,
\end{equation}
which, using the triangle inequality, takes the form
\begin{equation}
    \Delta O(t)\leq \Delta_\text{stoch} O(t)+\Delta_\text{bias} O(t)\;,
\end{equation}
where 
\begin{equation}
\begin{array}{lll}
      \Delta_\text{stoch} O(t)&=&\displaystyle\mathbb{E}[|O^\xi(t;N_\text{stoch})-{O}(t)|]\\
     \Delta_\text{bias} O(t)&=&\displaystyle|{O}(t)-O_\text{true}(t)|\;.
\end{array}
\end{equation}
In the following, we analyze these two sources of errors separately.
\subsubsection{Bias}
\label{app:bias}
The bias error is due to the inaccuracy in reconstructing the original correlation function, i.e., it depends on the difference $\Delta C(t)=C(t)-{C}_\text{PM+class}$ where $C(t)$ is the correlation of the original environment and ${C}_\text{PM+class}$ is the pseudomode model one in Eq.~(\ref{eq:CPM_Cclass}). While this bias can in principle originate from both classical and quantum-pseudomodes contributions, the classical part can usually be defined to achieve any accuracy by increasing the number of modes in the spectral expansion of the stochastic fields. As a consequence, the bias error is usually dominated by the functional form of the pseudomode correlation as a finite sum of decaying exponentials, see Eq.~(\ref{eq:CPM_Cclass_explicit}).

To estimate the bias, we closely follow \cite{Mascherpa} and re-present the derivation of one of its results. We do this for self-consistency and to explicitly avoid using any reference to path-integral techniques. We start by interpreting ${O}(t)$ in Eq.~(\ref{eq:Obar}) as obtained by tracing over an environment made out of the tensor product of the original bath $B$ and an auxiliary ``error bath'' $B_\text{err}$. This auxiliary bath is supposed to be Gaussian and to have correlations given by $C_\text{err}(t)={C}_\text{PM+class}(t)-C(t)$. This allows us to consider $S'=S+B$ (system + original bath) as a new system and use the results in section \ref{app:influenceSuperoperator}. We have
\begin{equation}
    \Delta_\text{bias} O(t)=|\text{Tr}_{S'}{O}(t)\mathcal{T}e^{\Delta \mathcal{F}}\rho(0)-\text{Tr}_{S'}\mathcal{T}{O}(t)\rho(0)|\;,
\end{equation}
where ${O}(t)=e^{iHt}{O}e^{-iHt}$ with the Hamiltonian defined in Eq.~(\ref{eq:full_Hamiltonian}) in the $S+B$ space, and where $\Delta\mathcal{F}=\mathcal{F}[t,{s},C_\text{err}(t)]$ is the result of tracing out $B_\text{err}$ using the definition in Eq.~(\ref{eq:functional}). Here, $\rho(0)=\rho_S(0)\rho_B$, see definitions in section (\ref{sec:pseudomode_model}). By expanding the exponential and by the triangle inequality we get
\begin{align}
\begin{array}{lll}
    \Delta_\text{bias} O(t)&=&\displaystyle|\sum_{n=1}^\infty\frac{1}{n!}\text{Tr}_{S'}{O}(t)\mathcal{T}(\Delta \mathcal{F})^n\rho(0)|\\
    &\leq&\displaystyle\sum_{n=1}^\infty\frac{1}{n!}|\text{Tr}_{S'}{O}(t)\mathcal{T}(\Delta \mathcal{F})^n\rho(0)|\;.
    \end{array}
\end{align}
Using Eq.~(\ref{eq:functional}), we can write $\Delta\mathcal{F}$ as
\begin{equation}
    \Delta\mathcal{F}=\int_0^t dt_2\int_0^{t_2} dt_1\sum_{j=1}^4 (\Delta C)_{j} \hhat{s}_j\;,
\end{equation}
where, explicitly, $(\Delta C)_{j}=C_\text{err}(t_2-t_1)(\delta_{j1}+\delta_{j2})-C_\text{err}(t_1-t_2)(\delta_{j3}+\delta_{j4})$, and where $\hhat{s}_1[\cdot]={s}(t_1)[\cdot]{s}(t_2)$, $\hhat{s}_2[\cdot]=-[\cdot]{s}(t_2){s}(t_1)$, $\hhat{s}_3[\cdot]={s}(t_1){s}(t_2)[\cdot]$, and $\hhat{s}_4[\cdot]=-{s}(t_2)[\cdot]{s}(t_1)$. This leads to
\begin{widetext}
\begin{equation}
    \begin{array}{lll}
    \Delta_\text{bias} O(t)&\leq&\displaystyle\sum_{n=1}^\infty\frac{1}{n!}\sum_{j_1,\cdots j_n}\left(\int_0^t dt_2\int_0^{t_2} dt_1\right)^n|\text{Tr}_{S'}{O}(t)\mathcal{T}(\Delta C)_{j_1}\cdots(\Delta C)_{j_n}\hhat{s}_{j_1}\cdots \hhat{s}_{j_n}\rho(0)|\\
    &\leq&\displaystyle\sum_{n=1}^\infty\frac{1}{n!}\sum_{j_1,\cdots j_n}\left(\int_0^t dt_2\int_0^{t_2} dt_1\right)^n|(\Delta C)_{j_1}|\cdots|(\Delta C)_{j_n}|~|\text{Tr}_{S'}{O}(t)\mathcal{T}\hhat{s}_{j_1}\cdots \hhat{s}_{j_n}\rho(0)|\\
    &\leq&\displaystyle\sum_{n=1}^\infty\frac{1}{n!}\sum_{j_1,\cdots j_n}\left(\int_0^t dt_2\int_0^{t_2} dt_1\right)^n|(\Delta C)_{j_1}|\cdots|(\Delta C)_{j_n}|~||{O}(t)\mathcal{T}\hhat{s}_{j_1}\cdots \hhat{s}_{j_n}\rho(0)||_1\;,
    \end{array}
\end{equation}
\end{widetext}
in which we used an abuse in notation in labeling time-dependencies to achieve a leaner notation and where we used the fact that the trace norm $||\cdot||_1$ does not increase under partial trace, see Eq.~(17) in \cite{Rastegin} and section \ref{app:trace_norm}. We now suppose to act with the time ordering so that all the superoperators $\hhat{s}$ are correctly ordered. We then note that each term $\hhat{s}_{j_1}\cdots \hhat{s}_{j_n}\rho(0)$ involves $2n$ operators ${s}(t)$ some of which are going to be placed before and some after $\rho(0)$. We can then make use of Eq.~(\ref{eq:Holder}) multiple times to write
\begin{widetext}
\begin{equation}
    \begin{array}{lll}
    \Delta_\text{bias} O(t)
    &\leq&\displaystyle\sum_{n=1}^\infty\frac{1}{n!}\sum_{j_1,\cdots j_n}\left(\int_0^t dt_2\int_0^{t_2} dt_1\right)^n|(\Delta C)_{j_1}|\cdots|(\Delta C)_{j_n}|~||{O}||_\infty||{s}||^{2n}_{\infty}||\rho(0)||_1\\
    &=&\displaystyle||{O}||_\infty\sum_{n=1}^\infty\frac{||{s}||^{2n}_{\infty}}{n!}
   \left(\int_0^t dt_2\int_0^{t_2} dt_1| \sum_{j}(\Delta C)_{j}|\right)^n\\
    
    &=&\displaystyle ||{O}||_\infty\sum_{n=1}^\infty\frac{4^n||{s}||^{2n}_{\infty}}{n!}
    \left(\int_0^t dt_2\int_0^{t_2} dt_1| C_\text{err}(t_2,t_1)|\right)^n\\

    &=&\displaystyle ||{O}||_\infty\left\{ \text{exp}\left[4||{s}||_{\infty}\int_0^t dt_2\int_0^{t_2} dt_1| C_\text{err}(t_2,t_1)|\right] - 1\right\}\;,
    \end{array}
\end{equation}
\end{widetext}
see Eq.~(6) in \cite{Mascherpa}. Here, we used the fact that Shatten norms are unitarily invariant, $||\rho(0)||_1=1$, and the fact that each $\Delta C_{j}$ takes the form given in Eq.~(\ref{eq:correlation}) so that $|\Delta C_{j}|=|C_\text{err}(t_2,t_1)|$.

\subsubsection{Stochastic Error}
\label{app:stochastic_error}
In this subsection, we consider the stochastic error
\begin{equation}
\label{eq:delta_Stoch}
    \Delta_\text{stoch} O(t)=\displaystyle\mathbb{E}[|O^\xi(t;N_\text{stoch})-\mathbb{E}[O^{\xi}(t)]|]\;.
\end{equation}
Unfortunately, it is not possible to analyze this uncertainty using the same techniques as used in the previous section. In fact, we cannot directly consider the influence superoperator in the context of the ``empirical'' average Eq.~(\ref{eq:empirical_average}) of the stochastic noise since, in this case, Wick's theorem only hold on average. For this reason, we need to take one step further back to Eq.~(\ref{eq:rhoS_intermediate}) to write
\begin{widetext}
\begin{equation}
\label{eq:all_moments}
\begin{array}{lll}
\Delta_\text{stoch} O&=&\displaystyle\mathbb{E}\left[\left|\sum_{n=0}^\infty \frac{(-i)^n}{n!} \int_{0}^{t}dt_{1}\cdots\int_{0}^{t}dt_{n}\left(\frac{1}{N_\text{stoch}}\sum_{j=1}^{N_\text{stoch}}\xi^j_{t_1}\cdots\xi^j_{t_n}-\mathbb{E}[\xi_{t_1}\cdots\xi_{t_n}]\right)\text{Tr}_{S''}\left[{O}(t)\mathcal{T}{s}_{t_1}^{\times}\cdots{s}_{t_n}^{\times}\rho_{S}(0)\right]\right|\right]\\

&\leq&\displaystyle\mathbb{E}\left[\sum_{n=0}^\infty \frac{1}{n!} \int_{0}^{t}dt_{1}\cdots\int_{0}^{t}dt_{n}\left|\frac{1}{N_\text{stoch}}\sum_{j=1}^{N_\text{stoch}}\xi^j_{t_1}\cdots\xi^j_{t_n}-\mathbb{E}[\xi_{t_1}\cdots\xi_{t_n}]\right|\left|\text{Tr}_{S''}\left[{O}(t)\mathcal{T}{s}_{t_1}^{\times}\cdots{s}_{t_n}^{\times}\rho_{S}(0)\right]\right|\right]\\

&\leq&\displaystyle\sum_{n=0}^\infty \frac{1}{n!} \int_{0}^{t}dt_{1}\cdots\int_{0}^{t}dt_{n}\sqrt{\mathbb{E}\left|\frac{1}{N_\text{stoch}}\sum_{j=1}^{N_\text{stoch}}\xi^j_{t_1}\cdots\xi^j_{t_n}-\mathbb{E}[\xi_{t_1}\cdots\xi_{t_n}]\right|^2}\left|\text{Tr}_{S''}\left[{O}(t)\mathcal{T}{s}_{t_1}^{\times}\cdots{s}_{t_n}^{\times}\rho_{S}(0)\right]\right|\;,
\end{array}
\end{equation}
\end{widetext}
where we omitted the time-dependence on the left hand-side and where $S''$ is the composition of the system $S$ and the { deterministic} pseudomode open quantum system. We also made use of the inequality $\mathbb{E}[|X|]\leq \sqrt{E[|X|^2]}$. This result shows us that the stochastic error depends on the average of the difference between the empirical moments and the true ones. In order to obtain a more direct bound we now proceed using rather aggressive inequalities which will limit the result to short times. To start, we write
\begin{widetext}
\begin{equation}
    \mathbb{E}\left|\frac{1}{N_\text{stoch}}\sum_{j=1}^{N_\text{stoch}}\xi^j_{t_1}\cdots\xi^j_{t_n}-\mathbb{E}[\xi_{t_1}\cdots\xi_{t_n}]\right|^2=\frac{1}{N^2_\text{noise}}\mathbb{E}\left[\sum_{j,k}\overline{\xi^j_{t_1}\cdots\xi^j_{t_n}}\xi^k_{t_1}\cdots\xi^k_{t_n}\right]-\left|\mathbb{E}[\xi_{t_1}\cdots\xi_{t_n}]\right|^2\;,
\end{equation}
\end{widetext}
and note that contributions of Wick's theorem on the first term in which $k$ and $j$ fields are never contracted will end up simplifying with the term $|\mathbb{E}[\xi_{t_1}\cdots\xi_{t_n}]|^2$ (also considering that for odd $n$ the expectation value $\mathbb{E}[\xi_{t_1}\cdots\xi_{t_n}]$ is just zero). We are then left with contributions in which at least one contraction happens between  $k$ and $j$ fields which generates a $\delta_{jk}$ and are consequently proportional to $1/N_\text{stoch}$ which already allows us to conclude that
\begin{equation}
\label{eq:Delta_stoch_O(1/N)}
    \Delta_\text{stoch} O(t)=O\left(\frac{1}{\sqrt{N_\text{stoch}}}\right)\;.
\end{equation}
Furthermore, each of the resulting correlation terms arising from the contractions considered above can be upper bounded by $C^\text{max}_\text{class}=\max_{t'\in[0,t]}\tilde{C}_\text{class}(t')$, where we used the definition in Eq.~(\ref{eq:C_tilde_complex_average}). We note that, for real correlations ${C}_\text{class}(t')\in\mathbb{R}$, we have $C^\text{max}_\text{class}=\max_{t'\in[0,t]}{C}_\text{class}(t')$. With these definitions, we can write
\begin{equation}
\begin{array}{lll}
    \Delta_\text{stoch} O&\leq&\displaystyle\frac{||{O}||_\infty}{\sqrt{N_\text{stoch}}}\sum_{n=1}^\infty \frac{\sqrt{(2n-1)!!}(2||{s}||_\infty t\sqrt{C^\text{max}_\text{class}})^n}{n!}\\
    
    &\leq&\displaystyle\frac{||{O}||_\infty}{\sqrt{N_\text{stoch}}}F(||{s}||_\infty t\sqrt{2C^\text{max}_\text{class}})\;,
    \end{array}
\end{equation}
where we omitted the time dependence on the left hand-side and in terms of the function 
\begin{equation}
    F(x)=\sum_{n=1}^\infty\sqrt{\frac{(2n){  !}}{(n!)^3}} x^n\;,
\end{equation} 
which is sub-exponential in the asymptotic limit,
and where we used the same techniques as in section \ref{app:bias} to analyze the term $\left|\text{Tr}_{S''}\left[{O}(t)\mathcal{T}{s}_{t_1}^{\times}\cdots{s}_{t_n}^{\times}\rho_{S}(0)\right]\right|$ and also considered that $(2n-1)!!=(2n)!/(2^n n!)$ overestimates the number of contractions when $n$ is even. The practical applicability of this expression is limited within a time scale given by $1/(||{s}||_\infty\sqrt{2C^\text{max}_\text{class}})$. Whenever this expression did not give a useful bound, it is always possible to compute a empirical variance following the bound
\begin{equation}
\label{eq:before_stochVar}
\begin{array}{lll}
    \Delta^2_\text{stoch} O(t)&\leq&\displaystyle\mathbb{E}[\left|O^\xi(t;N_\text{stoch})-\mathbb{E}[O^\xi(t)]\right|^2]\\
    &=&\displaystyle\mathbb{E}\{[O^\xi(t;N_\text{stoch})]^2\}-\mathbb{E}
    ^2[O^\xi(t;N_\text{stoch})]\;,
    \end{array}
\end{equation}
where we used  the inequality $\mathbb{E}^2[|X|]\leq E[|X|^2]$ and the fact that $\mathbb{E}[O^\xi(t)]|^2]$ is the expectation value of the random variable $O^\xi(t;N_\text{stoch})$, which can be directly checked using the definition in Eq.~(\ref{eq:noiseRandom}).

In abstract terms, given a random variable $X$ having finite expectation value $\mu_X=\mathcal{E}[X]$ and finite variance $\sigma^2_X=\mathbb{E}[X^2]-\mathbb{E}^2[X]$, it is possible to estimate $\mu_X$ by making a single extraction of the empirical avarage random variable 
\begin{equation}
Y=\frac{1}{N}\sum_{k=1}^N X_k\;. 
\end{equation}
In fact, $\mathbb{E}[Y]=\mu_X$ and
\begin{equation}
\begin{array}{lll}
\sigma_Y^2&=&\displaystyle\mathbb{E}[Y^2]-\mathbb{E}^2[Y]\\
&=&\displaystyle\frac{1}{N^2}\sum_{k,\bar{k}}\mathbb{E}[X_kX_{\bar{k}}]-\mu^2_X\\
&=&\displaystyle\frac{N \mathbb{E}[X^2]+(N^2-N)\mathbb{E}^2[X]}{N^2}-\mu^2_X\\
&=&\displaystyle\frac{E[X^2]-E^2[X]}{N}=\frac{\sigma_X^2}{N}\;.
\end{array}
\end{equation}
For large $N$ a single extraction of $Y$ gives a good approximation to $\mu_X$. The variance $\sigma^2_X$ can instead be estimated by a single extraction of the random variable 
\begin{equation}
Z=\frac{1}{N}\sum_k X_k^2-\left(\frac{1}{N}\sum_k X_k\right)^2\;.
\end{equation}
In fact
\begin{equation}
\label{eq:EZ}
    \begin{array}{lll}
    \mathbb{E}[Z]&=&\displaystyle\frac{1}{N}\sum_k \mathbb{E}[X_k^2]-\frac{1}{N^2}\mathbb{E}[X_k X_{\bar{k}}]\\
    &=&\displaystyle\frac{1}{N}\sum_k \mathbb{E}[X_k^2]-\frac{1}{N}\mathbb{E}[X]+\frac{N^2-N}{N}\mathbb{E}^2[X]\\
    &=&\displaystyle\left(1-\frac{1}{N}\right)(\mathbb{E}[X^2]-\mathbb{E}^2[X])\\
    &=&\displaystyle\left(1-\frac{1}{N}\right)\sigma_X^2\;.
    \end{array}
\end{equation}
The variance $\sigma^2_Z$ of $Z$ depends on the higher momentum of the variable $X$ and goes to zero as $N\rightarrow\infty$. This can be proven without explicitly computing all terms explicitly. In fact, we have
\begin{equation}
    \begin{array}{lll}
        \mathbb{E}[Z^2]&=&\displaystyle\frac{1}{N^2}\sum_{k\bar{k}}\mathbb{E}[X_k^2 X^2_{\bar{k}}]+\frac{1}{N^4}\sum_{k\bar{k}l\bar{l}}\mathbb{E}[X_k X_{\bar{k}}X_l X_{\bar{l}}]\\
        &=&-\displaystyle\frac{2}{N^3}\sum_{k\bar{l}}\mathbb{E}[X_k^2 K_l X_{\bar{l}}]\;.
    \end{array}
\end{equation}
The leading order in this expression (constant in $N$) is obtained when all the labels of the stochastic variables are different from each other. Such a term can be directly computed and it is equal to $\mathbb{E}^2[X]+\mathbb{E}^4[X]-2\mathbb{E}[X^2]\mathbb{E}^2[X]=\sigma_X^4=\mathbb{E}^2[Z]$, where the last equality does not take into account terms $O(1/N)$. The remaining terms are higher order in $1/N$ and depend on higher momentum of the $X$ random variable. Supposing these momentum to be finite, we then obtain the result that
\begin{equation}
\label{eq:sigmaZ}
    \sigma^2_{Z}=\mathbb{E}[Z^2]-\mathbb{E}^2[Z]=O\left(\frac{1}{N}\right)\;.
\end{equation}
To resume, these results show that a single extraction of the variables $Y$ and $Z$ can be used to estimate the expectation value and variance of the variable $X$.

In our case, we can identify $X\mapsto O^\xi(t)\equiv O^\xi(t;1)$ and $Y\mapsto O^\xi(t;N_\text{stoch})$, which leads to
\begin{equation}
    \Delta^2_\text{stoch}O(t)\leq \frac{\sigma^2_{O^\xi(t)}}{N_\text{stoch}}=\frac{\mathbb{E}[Z]}{N_\text{stoch}}+O\left(\frac{1}{N^2_\text{noise}}\right)\;.
\end{equation}
where we used Eq.~(\ref{eq:EZ}) to relate the right hand-side to the expectation value of $Z\mapsto \frac{1}{N}\sum_k [O^\xi(t)_k]^2-(\frac{1}{N}\sum_k O^\xi(t)_k)^2$, which can be estimated numerically. By Chebyshev's inequality, the probability that $Z$ is $k\sigma_Z$ distant from the expectation value $\mathbb{E}[Z]$ is less than $1/k^2$. Because of Eq.~(\ref{eq:sigmaZ}), this translates the possibility of writing that
\begin{equation}
    \Delta^2_\text{stoch}O(t)\leq \frac{\tilde{Z}}{N_\text{stoch}}+O\left(\frac{1}{N^{3/2}_\text{noise}}\right)\;,
\end{equation}
with probability almost 1. Here $\tilde{Z}$ is a single realization of the random variable $Z$.

To finish, it is instructive to explicitly compute the first terms in Eq.~(\ref{eq:all_moments}) under the same considerations done below such expression. At second order, we have
\begin{equation}
\begin{array}{lll}
    \Delta^2_\text{stoch} O(t)&\leq&||{O}||_\infty[\mu_1(t)+\mu_2(t)]\;,
    \end{array}
    \end{equation}
where 
\begin{widetext}
\begin{equation}
\begin{array}{lll}
    \mu_1(t)&=&\displaystyle 2||{s}||_\infty\int_0^t dt_1\sqrt{\frac{1}{N^2_\text{noise}}\sum_{j,k}\mathbb{E}[\bar{\xi}_{t_1}^j\xi_{t_1}^k]-|E[\xi_{t_1}]|^2}\\
    &=&\displaystyle\frac{2||{s}||_\infty \sqrt{\tilde{C}_\text{class}(0)} t}{\sqrt{N_\text{stoch}}}\\
    
    \mu_2(t)&=&\displaystyle \frac{4||{s}||^2_\infty}{2!}\int_0^t dt_1\int_0^t dt_2\sqrt{\frac{1}{N^2_\text{noise}}\sum_{j,k}\mathbb{E}[\overline{\xi_{t_2}^j\xi_{t_1}^j}\xi_{t_2}^k\xi_{t_1}^k]-|E[\xi_{t_1}\xi_{t_2}]|^2}\\
    
    &=&\displaystyle \frac{2||{s}||^2_\infty}{\sqrt{N_\text{stoch}}}\int_0^t dt_1\int_0^t dt_2\sqrt{\tilde{C}^2_\text{class}(0)+\tilde{C}^2_\text{class}(t_2-t_1)}\;.
    
    \end{array}
    \end{equation}
    \end{widetext}

\subsubsection{Some properties of the trace norm}
\label{app:trace_norm}
In this section, we report two important properties of the trace norm. To begin, we start by defining the Schatten p-norm as
\begin{equation}
    ||A||_p=\left(\sum_{j=1}^{N_A}|\sigma^p_j(A)|\right)^{1/p}\;,
\end{equation}
where $A:\mathbb{R}^{N_A}\rightarrow\mathbb{R}^{N_A}$, $N_A\in\mathbb{N}$, $N_A>0$. Here, $\sigma_j(A)$ is the $j$th singular value of $A$, i.e., the $j$th eigenvalue of the operator $|A|=(A^\dagger A)^{1/2}$. We further denote
\begin{equation}
    ||A||_\infty=\max_{j}\sigma_j(A)\;.
\end{equation}
In our case, in order to apply this formalism, we consider a regularization of the total Hilbert space (for example, by truncating the number and dimension of all harmonic Hilbert spaces). \\

Given these definitions, we first state the H\"{o}lder inequality for Schatten norms, see \cite{Bathia}, Eq.~(IV.42), pag. 95, which reads
\begin{equation}
\label{eq:Holder}
    ||AB||_1\leq||A||_p ||B||_q\;,
\end{equation}
where $A,B$ are generic operators, where $p,q,\in\mathbb{N}$ ($p,q\geq 1$) with $1/p+1/q=1$, and where $||\cdot||_p$ is the Schatten $p$-norm.\\

For self consistency and to better relate to the present content, we now also report the proof given in \cite{Rastegin} showing that the trace norm does not increase under partial trace.
Following \cite{Rastegin}, we define an operator $Q$ acting on a Hilbert space $\mathcal{H}_A\otimes\mathcal{H}_B$ of dimension $n_A n_B$. We are going to consider the partial trace with respect to $B$, i.e., $Q_A=\text{Tr}_{B}[Q]$. We further consider an orthonormal basis $\ket{e_i}_A$ and $\ket{f_\alpha}_B$ with $i=0,\cdots, n_A-1$ and $\alpha=0,\cdots,n_B-1$. Using these definitions, the operator $Q$ can be written as
\begin{equation}
    Q=\sum_{\alpha,\beta=0}^{n_B-1}\sum_{i,j=0}^{n_A-1}\ketbra{e_i}{e_j} Q_{ij}^{\alpha\beta}\ketbra{f_\alpha}{f_\beta}\;,
\end{equation}
in terms of $Q_{ij}^{\alpha\beta}\in\mathbb{C}$.
We now define the following generalization $X$ and $Z$ of the Pauli operators acting on the space $B$ as
\begin{equation}
    \begin{array}{lll}
    Z\ket{f_\alpha}&=&e^{2\pi i \alpha/n_{B}}\ket{f_\alpha}\\
    X\ket{f_\alpha}&=&\ket{f_{\alpha+1}}\;,
    \end{array}
\end{equation}
where the ``closed boundary conditions'' $\ket{f_{n_A}}=\ket{f_0}$ are intended. Taking the conjugate of the definitions above, we get $\bra{f_\alpha}(Z^\dagger Z)\ket{f_\beta}=\bra{f_\alpha}(X^\dagger X)\ket{f_\beta}=\delta_{\alpha\beta}$, impliying that both $X$ and $Z$ are unitary. 
We also have
\begin{equation}
    \begin{array}{lll}
    Z(Q) &:=& \displaystyle\frac{1}{n_B}\sum_{\eta=0}^{n_B-1}(\mathbb{I}_A\otimes Z^\eta)Q(\mathbb{I}_A\otimes Z^\eta)^\dagger\\
    &=&\displaystyle\sum_{\eta=0}^{n_B-1}\sum_{\alpha,\beta=0}^{n_B-1}\sum_{i,j=0}^{n_A-1}\frac{e^{2\pi i \eta(\alpha-\beta)/n_{B}}}{n_B}\\
    &&\displaystyle\times\ketbra{e_i}{e_j} Q_{ij}^{\alpha\beta}\ketbra{f_\alpha}{f_\beta}\\
    &=&\displaystyle\sum_{\alpha,\beta=0}^{n_B-1}\sum_{i,j=0}^{n_A-1}\ketbra{e_i}{e_j} Q_{ij}^{\alpha\beta}\delta_{\alpha\beta}\ketbra{f_\alpha}{f_\beta}\;,
    \end{array}
\end{equation}
and
\begin{equation}
\label{eq:temp_a}
    \begin{array}{lll}
    X(Z(Q)) &:=& \displaystyle\sum_{\eta=0}^{n_B-1}(\mathbb{I}_A\otimes X^\eta)Z(Q)(\mathbb{I}_A\otimes X^\eta)^\dagger\\
    &=&\displaystyle\sum_{\eta=0}^{n_B-1}\sum_{\alpha,\beta=0}^{n_B-1}\sum_{i,j=0}^{n_A-1}\bra{e_j} Q_{ij}^{\alpha\beta}\delta_{\alpha\beta}\ket{f_{\alpha+\eta}}\\
    &&\times \ketbra{e_i}{f_{\beta+\eta}}\\
    &=&\displaystyle\sum_{\alpha,\beta=0}^{n_B-1}\sum_{i,j=0}^{n_A-1}\ketbra{e_i}{e_j} \sum_\alpha Q_{ij}^{\alpha\alpha}\mathbb{I}_B\\
    &=&\text{Tr}_B(Q)\otimes\mathbb{I}_B\;.
    \end{array}
\end{equation}
At the same time, we also have
\begin{equation}
\label{eq:temp_b}
    \begin{array}{lll}
    X(Z(Q)) &=& \displaystyle\frac{1}{n_B}\sum_{\eta,\eta'=0}^{n_B-1}(\mathbb{I}_A\otimes X^{\eta'} Z^\eta)Q(\mathbb{I}_A\otimes Z^{\eta\dagger} X^{\eta'\dagger})\\
     &=& \displaystyle\frac{1}{n_B}\sum_{\eta,\eta'=0}^{n_B-1}(\mathbb{I}_A\otimes X^{\eta'} Z^\eta)Q(\mathbb{I}_A\otimes X^{\eta'} Z^{\eta})^\dagger\;.
    \end{array}
\end{equation}
Taking the norm of both Eq.~(\ref{eq:temp_a}) and Eq.~(\ref{eq:temp_b}), and considering that $X$ and $Z$ are unitary, we get
\begin{equation}
\label{eq:temp_1}
    ||\text{Tr}_B(Q)\otimes\mathbb{I}_B||_p\leq \frac{n_B^2}{n_B}||Q||_p=n_B ||Q||_p\;.
\end{equation}
At the same time, the singular values of $\text{Tr}_B(Q)\otimes\mathbb{I}_B$ are $n_B-$degenerate which gives
\begin{equation}
\label{eq:temp_2}
    \begin{array}{lll}
        ||\text{Tr}_B(Q)\otimes\mathbb{I}_B||_p&=&\displaystyle  \left[n_B\sum_j\sigma^p_j(\text{Tr}_B(Q))\right]^p\\
        &=&\displaystyle n_B^{1/p}||\text{Tr}_B(Q)||_p\;.
    \end{array}
\end{equation}
Together, Eq.~(\ref{eq:temp_1}) and Eq.~(\ref{eq:temp_2}) imply
\begin{equation}
  n_B^{1/p}  ||\text{Tr}_B(Q)||_p\leq n_B||Q||_p\;,
\end{equation}
which leads to
\begin{equation}
 ||\text{Tr}_B(Q)||_p\leq n^{(p-1)/p}_B||Q||_p\;,
\end{equation}
so that, for $p=1$, 
\begin{equation}
 ||\text{Tr}_B(Q)||_1\leq ||Q||_1\;.
\end{equation}

\subsection{Spectral Representation of Classical Correlations}
\label{app:spectral_representation}
In this section, we introduce a spectral representation to define a stationary Gaussian stochastic fields $\xi(t)$ characterized by a vanishing mean and by a symmetric correlation function $C(t')$. In this section, to add extra clarity, we use $t$ to denote times within the domain $t\in[0,T]$, while we use and extra apostrophe to denote variables which can be written as the difference between two times which implies $t'\in[-T,T]$. 

Our main goal is to define  the field $\xi(t)$ in such a way to satisfy $\mathbb{E}[\xi(t)]=0$ and  $\mathbb{E}[\xi(t_2)\xi(t_1)]=C(t_2-t_1)$ where $C(-t')=C(t')$.

Since we are interested in simulating the dynamics of a system for times within the domain $[0,T]$, we will take advantage of the structure of $L^2(-T,T)$ functional spaces (characterizing the domain of the difference between two times, i.e., $t'=t_2-t_1\in[-T,T]$) by defining the inner product $\langle f,g\rangle_T$ between to functions $f,g:[-T,T]\rightarrow \mathbb{C}$ as
\begin{equation}
    \langle f,g\rangle_T=\frac{1}{2T}\int_{-T}^T d\tau \bar{f}(\tau)g(\tau)\;,
\end{equation}
where the overbar denotes complex conjugation. In this space, we can define the orthonormal basis $\phi_n(t')=e^{i n\pi t'/T}$ which fulfills
\begin{equation}
    \langle\phi_n,\phi_m\rangle_T=\delta_{n,m}\;,
\end{equation}
and it has the property that
\begin{equation}
\label{eq:temp_phi}
    \phi_n(t_2-t_1)=\phi_n(t_2)\phi_{-n}(t_1)\;.
\end{equation}
With this formalism at hand, a stationary correlation can be decomposed as
\begin{equation}
\label{eq:dec_C}
    C(t') = \sum_{n=-\infty}^\infty c_n \phi_n(t')\;,
\end{equation}
where $c_n=\langle\phi_n,C\rangle_T$. The symmetry $C(t')=C(-t')$, together with $\phi_n(-t')=\phi_{-n}(t')$ implies the constraint $c_{-n}=c_n$ which simply reflects the fact that the function $C(t')$ can be expanded using cosine functions and that
\begin{equation}
\label{eq:cn_app}
    c_n=\frac{1}{2T}\int_{-T}^Td\tau\cos(n\pi \tau/T)C(t)\;.
\end{equation}
 Using Eq.~(\ref{eq:temp_phi}), we can further write
\begin{equation}
\label{eq:temp_ct2t1}
\begin{array}{lll}
    C(t_2-t_1) &=& \displaystyle\sum_{n=-\infty}^\infty c_n \phi_n(t_2)\phi_{-n}(t_1)\\
    &=&\displaystyle \sum_{n,m=-\infty}^\infty \phi_m(t_2) C_{mn}\phi_{n}(t_1)\;,
    \end{array}
\end{equation}
where $C_{mn}=c_m\delta_{n,-m}$. This matrix is symmetric because of the constraint $c_{-n}=c_n$ and hence we can diagonalize it as
\begin{equation}
\label{eq:temp_cmn}
    C_{mn}=\sum_{j=-\infty}^\infty O_{m j}d_j O_{jn}\;,
\end{equation}
where $O_{mj}$ is the (possibly complex) orthogonal matrix made out of the $m$th element of the $j$th eigenvector with eigenvalue $d_j$ of the matrix $C_{mn}$. In this case, the eigenvectors are just trigonometric functions which can be expressed as
\begin{equation}
\label{eq:eigenvectors_O}
\begin{array}{lll}
    O_{mj}&=&\delta_{j0}\delta_{m0}+\theta(j)[\delta_{j,m}+\delta_{j,-m}]/\sqrt{2}\\
    &&-\theta(-j)[\delta_{j,m}-\delta_{j,-m}]/\sqrt{2}\;,
    \end{array}
\end{equation}
 corresponding to the eigenvalues 
 \begin{equation}
 \label{eq:eigenvalues_O}
     d_j=\delta_{j0}c_0 + c_j[\theta(j)-\theta(-j)]\;,
 \end{equation} 
 where $\theta(x)=1$ for $x>0$ and zero otherwise. We note that the sign in front of the eigenvectors in Eq.~(\ref{eq:eigenvectors_O}) can be chosen arbitrarily. As we will see later, the minus sign in the second line of Eq.~(\ref{eq:eigenvectors_O}) is designed to allow a cleaner expression for the fields. Plugging Eq.~(\ref{eq:temp_cmn}) into Eq.~(\ref{eq:temp_ct2t1}), we can write
\begin{equation}
\label{eq:dec_C_CdC}
\begin{array}{lll}
    C(t_2-t_1) &=&\displaystyle \sum_{n,m,j=-\infty}^\infty \phi_m(t_2)  O_{m j}d_j O_{jn}\phi_{n}(t_1)\\
    &=&\displaystyle \sum_{j=-\infty}^\infty \phi^C_j(t_2)  d_j \phi^C_{j}(t_1)\;.
    \end{array}
\end{equation}
where 
\begin{equation}
\label{eq:phiCC}
    \begin{array}{lll}
        \phi_j^C(t)&=&\displaystyle\sum_{n=-\infty}^\infty O_{jn}\phi_{n}(t)\\
        &=&\displaystyle\delta_{j0}+\sqrt{2}[\theta(j)\cos(n\pi t/T)+\theta(-j)i\sin(n\pi t/T)].
    \end{array}
\end{equation}
We explicitly note that the last step in the above equation only holds when the eigenvectors can be organized in a, possibly complex, orthogonal matrix which is a consequence of the symmetry under time reversal of the original correlation. \emph{This is a manifestation of the fact that the construction showed in here is prevented for ``non-classical'' correlations, i.e., correlations which are not symmetric under time-reversal}.
Using Eq.~(\ref{eq:eigenvalues_O}) and Eq.~(\ref{eq:phiCC}) into Eq.~(\ref{eq:dec_C_CdC}), we can further be more explicit and write
\begin{equation}
\label{eq:coscossinsin}
\begin{array}{lll}
    C(t_2-t_1) &=&\displaystyle c_0+2\sum_{n=1}^\infty c_n[\cos(n\pi t_2/T)  \cos(n\pi t_1/T)\\
    &&-i^2\sin(n\pi t_2/T)  \sin(n\pi t_1/T)]\\
    &=&\displaystyle c_0+2\sum_{n=1}^\infty c_n[\cos(n\pi t_2/T)  \cos(n\pi t_1/T)\\
    &&+\sin(n\pi t_2/T)  \sin(n\pi t_1/T)]\;.
    \end{array}
\end{equation}
whose correctness can be directly checked by noting that Eq.~(\ref{eq:dec_C}) can be written as
\begin{equation}
\label{eq:actual_corr}
    C(t')=c_0+2\sum_{n=1}^\infty c_n\cos[n\pi(t_2-t_1)/T]\;,
\end{equation}
which, indeed, corresponds to Eq.~(\ref{eq:coscossinsin}) by using the trigonometric identity $\cos(\alpha-\beta)=\cos(\alpha)\cos(\beta)+\sin(\alpha)\sin(\beta)$. As already noted above, the whole construction of this section would be prevented if the correlation has a term proportional to a sine function as the analog of the trigonometric identity above contains non-factorizable products of cosines and sines.

We are now in a position to finally define the stochastic variable $\xi(t)$ as the following spectral representation
\begin{equation}
\label{eq:xi_spectral_representation}
    \xi(t)=\sum_{j=-\infty}^\infty\xi_j \sqrt{d_j}\phi_j^C(t)\;,
\end{equation}
where $\xi_j\in\mathcal{N}(0,1)$ are independent Gaussian variables with unit variance and zero mean. This immediately implies that $\mathbb{E}[\xi(t)]=0$. Furthermore, we can use Eq.~(\ref{eq:dec_C_CdC}) to show that
\begin{equation}
\label{eq:true_E}
    \begin{array}{lll}
    \mathbb{E}[\xi(t_2)\xi(t_1)]&=&\displaystyle\sum_{j,j'}\mathbb{E}[\xi_j\xi_{j'}]\sqrt{d_j d_{j'}}\Phi_j^C(t_2)\Phi_{j'}^C(t_1)\\
    &=&\displaystyle\sum_{j}d_j \Phi_j^C(t_2)\Phi_{j}^C(t_1)\\
    &=&C(t_2-t_1)\;.
    \end{array}
\end{equation}
Using Eq.~(\ref{eq:eigenvalues_O}) and Eq.~(\ref{eq:phiCC}) in Eq.~\ref{eq:xi_spectral_representation}, we can write the fields in the following more epxlicit form
\begin{equation}
\label{eq:xi_spectral_representation_2}
    \xi(t)=\sqrt{c_0} \xi_0+\sum_{n=1}^\infty\sqrt{2c_n}[\xi_n \cos(n\pi t/T)+\xi_{-n}\sin(n\pi t/T)].
\end{equation}
Here, we note that the plus sign in front of the sine is due to our previous sign choice in Eq.~(\ref{eq:eigenvectors_O}). We can further check explicitly that, indeed these fields have the correct statistics by writing
\begin{equation}
    \begin{array}{lll}
    \mathbb{E}[\xi(t_2)\xi(t_1)] &=&\displaystyle c_0+2\sum_{n=1}^\infty c_n[\cos(n\pi t_2/T)  \cos(n\pi t_1/T)\\
    &&+\sin(n\pi t_2/T)  \sin(n\pi t_1/T)]\\
   &=& \displaystyle c_0+2\sum_{n=1}^\infty c_n\cos[n\pi(t_2-t_1)/T]\;,
    \end{array}
\end{equation}
where we again used the identity $\cos(\alpha-\beta)=\cos(\alpha)\cos(\beta)+\sin(\alpha)\sin(\beta)$. It is interesting to note that, despite the fact the presence of \emph{antisymmetric} functions (the sines) in the definition of the fields $\xi(t)$ is essential in reproducing the stationarity of the \emph{symmetric} correlation $C(t')$.

As an example, we compute the spectral representation above for terms in the correlations taking the form of a symmetric exponential, i.e.,
\begin{equation}
C^{\text{decay}}(t_2-t_1)=\alpha^2e^{\omega|t_2-t_1|}\;,
\end{equation}
in terms of the frequency-like parameter $\omega\in\mathbb{C}$. Here the suffix $\text{decay}$ indicates that, usually, correlation of this kind are used to model purely decaying contributions such as  the Matsubara ones (for $\omega\in\mathbb{R}$ and $\omega<0$). However, they can also be uesd to model resonant effects through $\cos(\omega t)=(e^{i\omega|t|}+e^{-i\omega|t|})/2$). By using Eq.~(\ref{eq:cn_app}), we find
\begin{equation}
c_n^\text{decay}=\frac{1}{2T}\int_{-T}^Td\tau~ e^{\omega\tau}\cos\left(\frac{n\pi\tau}{T}\right)=L_n(\omega)\;,
\end{equation}
in terms of the function
\begin{equation}
L_n(\omega)=(e^{\omega T}e^{in\pi}-1)\frac{\omega T}{(\omega T)^2+(n\pi)^2}\;.
\end{equation}

It is interesting to finish this section with an analysis of the first two moments of the empirical expectation value for the correlation, i.e., the random variable
\begin{equation}
    C_\text{class}^{\text{emp}}(t_2,t_1)=1/N_\text{stoch}\sum_j \xi_j(t_2)\xi_j(t_1)\;,
\end{equation}
 which constitutes an estimate of the true average in Eq.~(\ref{eq:true_E}) by averaging over $N_\text{stoch}$ independent realization of the noise $\xi_j(t)$, $j=1,\dots,N_\text{stoch}$. The expectation value is simply the expected value of the correlation
\begin{equation}
\begin{array}{lll}
    \mathbb{E}[C_\text{class}^{\text{emp}}(t_2,t_1)]
    &=&\displaystyle\frac{1}{N_\text{stoch}}\sum_j\mathbb{E}[\xi_j(t_2)\xi_j(t_1)]\\
    &=&C_\text{class}(t_2-t_1)\;.
    \end{array}
\end{equation}
while the variance is
\begin{equation}
\begin{array}{lll}
    \sigma^2_\text{emp}&=&\mathbb{E}\left[\left|C_\text{class}^{\text{emp}}(t_2,t_1)-\mathbb{E}[C_\text{class}^{\text{emp}}(t_2,t_1)]\right|^2\right]\\
    &=&\mathbb{E}[|C_\text{class}^{\text{emp}}(t_2,t_1)|^2]-|C_\text{class}(t_2,t_1)|^2\\
        &=&\displaystyle\frac{1}{N
        ^2_\xi}\sum_{j,k}\mathbb{E}[\bar{\xi}_j(t_2)\bar{\xi}_j(t_1)\xi_k(t_2)\xi_k(t_1)]\\
        &&-|C_\text{class}(t_2,t_1)|^2\\

    &=&\displaystyle\frac{1}{N
        ^2_\xi}\sum_{j,k} \{\mathbb{E}[\bar{\xi}_j(t_2)\bar{\xi}_j(t_1)]\mathbb{E}[{\xi}_k(t_2)\xi_k(t_1)]\\
    &&+\displaystyle\mathbb{E}[\bar{\xi}_j(t_2)\xi_k(t_2)]\mathbb{E}[\bar{\xi}_j(t_1)\xi_k(t_2)]\\
        &&+\displaystyle\mathbb{E}[\bar{\xi}_j(t_2)\xi_k(t_1)]\mathbb{E}[\bar{\xi}_j(t_1)\xi_k(t_2)]\}\\
        &&-|C_\text{class}(t_2,t_1)|^2\\
        
        &=&\displaystyle[|\tilde{C}(0)|^2+|\tilde{C}(t_2-t_1)|^2]/{N_\text{stoch}}\;,
    \end{array}
\end{equation}
where we used Wick's theorem and defined the real version of the correlation $C(t)$ can be defined as
\begin{equation}
\label{eq:C_tilde_complex_average}
    \begin{array}{lll}
    \tilde{C}(t)&=&\mathbb{E}[\bar{\xi}(t_2)\xi(t_1)]\\
    &=&\displaystyle|c_0|+2\sum_{n=1}^\infty|c_n|\cos(n\pi t/T)\;,
    \end{array}
\end{equation}
where we used Eq.~(\ref{eq:xi_spectral_representation_2}). Using Eq.~(\ref{eq:actual_corr}), we finally find
\begin{equation}
\label{E_emp_sigma}
    \sigma^2_\text{emp}\leq[|C_\text{class}(0)|^2+|C_\text{class}(t_2-t_1)|^2]/{N_\text{stoch}}\;,
\end{equation}

\section{A class of rational spectral densities}
To be more specific, we now study this decomposition for a sub-class $\mathcal{M}$ of spectral densities which are real and asymmetric on the real axis and meromorphic in the overall complex plane. We further suppose all $2N_\text{p}>1$ poles (we later show this number needs indeed to be even and suppose there is at least one such couple) to be simple (and not located on the real axis) apart from a pole of order $N_\infty<2 N_\text{p}$ at infinity. We now consider three basic facts about functions in $\mathcal{M}$.
\begin{itemize}
    \item[(i)]The poles of $J(\omega)\in\mathcal{M}$ come in complex-conjugate pairs having complex-conjugate residues. To show this, we consider a region $D$ which contains all the poles. The Schwarz reflection principle tells us that $J(\bar{\omega})=\bar{J}(\omega)$ for $\omega\in D$. Now, for any pole $\omega_k$ in the upper (or lower) complex plane $\mathbb{C}$, we can write the Laurent series $J(\omega)=\sum_{j=-\infty}^\infty a_j (\omega-\omega_k)^j$ where $a_{-1}$ represents the residue at $\omega_k$. Using the Schwarz reflection principle, this implies that $J(\omega)=\sum_{j=-\infty}^\infty \bar{a}_j (\omega-\bar{\omega}_k)^j$, i.e, the point $\bar{\omega}_k$ is also a pole whose residue is the complex conjugate of the one evaluated at $\omega_k$. This immediately implies that the number of poles is even, i.e., it can be written as $2N_\text{p}$. In the following we use the notation $(\omega_k,\omega_{-k})$ to denote the complex-conjugate pairs of poles, i.e, $\omega_{-k}=\bar{\omega}_k$. We will further use $k>0$ ($k<0$) to denote poles in the upper (lower) complex plane so that the full set of poles can be written as $\{\omega_k\}$ for $k\in\mathbb{Z}^0_{N_p}$, where
    $\mathbb{Z}^0_{N_p}=\{-N_p,\dots,-1,1,\dots,N_p\}$ and where all $\omega_k$ are distinct.\\

    \item[(ii)]
    Since $J(\omega)$ is meromorphic with single poles located in the finite complex plane, $\prod_{k\in\mathbb{Z}^0_{N_p}} (\omega-\omega_k) J(\omega)$ is analytical in $\mathbb{C}$ and it has a pole of order $N_\infty$ at infinity. This implies that such a function is a $N_\infty$-order polynomial $p(\omega)$ so that   \begin{equation}
    \label{eq:J_expression_1}
        J(\omega)=  \frac{p(\omega)}{\prod_{k\in\mathbb{Z}^0_{N_p}}(\omega-\omega_k)}\;.
    \end{equation} 
     Imposing that $J(\omega)\in\mathbb{R}$ for $\omega\in\mathbb{R}$ implies that all coefficients in the polynomial $p$ are real, i.e., $\bar{p}(\omega)=p(\bar{\omega})$ (since the roots $\omega_k$ appearing in the denominator come in complex-conjugate pairs).
    It is now possible to write this expression in terms of a sum of rational functions with polynomials of degree one at denominator. To see this, we just note that $1/\prod_k (\omega-\omega_k)=\sum_j{\kappa v_j}/{\prod_{k\neq j}(\omega-\omega_k)}$ where $\sum_k v_k = 0$ and $\kappa=-1/\sum_{k}(v_k \omega_k)$. By proceeding iteratively this way, we find a decomposition of the kind $1/\prod_k (\omega-\omega_k)=\sum_k{ c_k}/{(\omega-\omega_k)}$ in terms of some coefficients $c_k\in\mathbb{C}$. Ultimately this leads to 
    \begin{equation}
    \label{eq:J_expression_2}
    J(\omega)=p(\omega)\sum_{k\in\mathbb{Z}^0_{N_p}} \frac{c_k}{(\omega-\omega_k)}\;.
    \end{equation}
    This expression also allows us to find a more explicit expression for the coefficients $c_k$. In fact, the residue at any of the points $\omega_k$ can be computed both from Eq.~(\ref{eq:J_expression_2}) as $p(\omega_k)c_k$, and from Eq.~(\ref{eq:J_expression_1}) as $p(\omega_k)/\prod_{j\neq k}(\omega_k-\omega_j)$. Equating these two results together, we find that $c_k=1/\prod_{j\neq k}(\omega_k-\omega_j)$. Furthermore, the residue at $\omega_k$ explicitly reads 
    \begin{equation}
    \label{eq:residue_explicit}
    \begin{array}{lll}
        R_k&\equiv&\text{Res}[J(\omega);\omega_k]\\
        &=&p(\omega_k)/\prod_{j\neq k}(\omega_k-\omega_j)\;,
        \end{array}
    \end{equation} 
    so that
    \begin{equation}
    \label{eq:k_minus_k}
        \begin{array}{lll}
        R_{-k}&=&\text{Res}[J(\omega);\omega_{-k}]\\
        &=&\displaystyle\frac{p(\bar{\omega}_k)}{\displaystyle\prod_{j\neq {-k}}(\bar{\omega}_k-\omega_j)}\\
        &=&\displaystyle\frac{p(\bar{\omega}_k)}{(\bar{\omega}_k-\omega_k)\displaystyle\prod_{j\neq {-k},k}(\bar{\omega}_k-\bar{\omega}_j)}\\
        &=&\bar{R}_k\;,
        \end{array}
    \end{equation}
    checking explicitly what proved in (i) and where we used that all poles come in complex-conjugate pairs.
    \item[(iii)] The antisymmetry hypothesis on $J(\omega)$ imposes further constraints on the set poles $\{\omega_k\}$ and on the polynomial $p(\omega)$. Mainly, because of antisymmetry, we need to require(
    \begin{equation}
        \begin{array}{lll}
        ~~~~~~~J(\omega)&=&\displaystyle\frac{J(\omega)-J(-\omega)}{2}\\
        &=&\displaystyle\frac{p(\omega)}{2\displaystyle\prod_{k}(\omega-\omega_k)}-\frac{p(-\omega)}{2\displaystyle\prod_{k}(-\omega-\omega_k)}\;.
        \end{array}
    \end{equation}
    To continue, we can distinguish between two kind of poles. For imaginary $\omega_k$, both denominators contain the same $\omega^2+|\omega_k|^2$ factor. For general complex poles $\omega_k$, we can operate a greates common divisor to show that in Eq.~(\ref{eq:J_expression_1}) for an antisymemtric spectral density the poles come either in complex-conjugate pairs (if $\omega_k$ is purely imaginary) or come in quadruples ($\omega_k$, $-\omega_k$ and their conjugate pairs). Furthermore, the polynomial $p(x)$ must be antisymmetric.  
    
    This has immediate implications in the expression for the residues when the poles come in quadruples. To show this, let us denote by $
    \tilde{k}$ the label such that $\omega_{\tilde{k}}=-\omega_{-k}=-\bar{\omega}_k$ so that $\tilde{k}$ has the same sign as $k$ and so that the quadruples of poles are $(\omega_k,\omega_{-k},\omega_{\tilde{k}},\omega_{\widetilde{(-k)}})$. We have
        \begin{equation}
    \label{eq:k_bark}
        \begin{array}{lll}
        R_{\tilde{k}}&=&\text{Res}[J(\omega);\omega_{\tilde{k}}]\\
        &=&\displaystyle\frac{p(-\bar{\omega}_k)}{\displaystyle\prod_{j\neq {\bar{k}}}(-\bar{\omega}_k-\omega_j)}\\
        &=&\displaystyle\frac{p(-\bar{\omega}_k)\left[\displaystyle\prod_{j\neq {\bar{k}},k,-k,\widetilde{(-k)}}(-\bar{\omega}_k-\bar{\omega}_j)\right]^{-1}}{(-\bar{\omega}_k-\omega_k)(-\bar{\omega}_k+\omega_k)(-\bar{\omega}_k-\bar{\omega}_k)}\\
        
        &=&\displaystyle\frac{p(\bar{\omega}_k)\left[\displaystyle\prod_{j\neq {\bar{k}},k,-k,\widetilde{(-k)}}(\bar{\omega}_k+\bar{\omega}_j)\right]^{-1}}{(\bar{\omega}_k+\omega_k)(\bar{\omega}_k-\omega_k)(\bar{\omega}_k+\bar{\omega}_k)}\\
        &=&\displaystyle\frac{p(\bar{\omega}_k)\left[\displaystyle\prod_{j\neq {\bar{k}},k,-k,\widetilde{(-k)}}(\bar{\omega}_k-\bar{\omega}_j)\right]^{-1}}{(\bar{\omega}_k+\omega_k)(\bar{\omega}_k-\omega_k)(\bar{\omega}_k+\bar{\omega}_k)}\\
        &=&R_{-k}\\
        &=&\bar{R}_k\;,
        \end{array}
    \end{equation}
\end{itemize}
where we used the fact that an even number of labels $j$ is present and that, given the symmetries of the poles, a change in sign in all $\omega_j$ is just formal.

A similar proof holds to show that the residues of poles coming in conjugate pairs $(\omega_{k_\text{i}},\omega_{-k_\text{i}})$ are real. In fact, we have
  \begin{equation}
    \label{eq:k_bark_2}
        \begin{array}{lll}
        R_{k_\text{i}}&=&\text{Res}[J(\omega);\omega_{k_\text{i}}]\\
        &=&\displaystyle\frac{p({\omega}_{k_\text{i}})}{({\omega}_{k_\text{i}}-{\bar{\omega}}_{k_\text{i}})\displaystyle\prod_{k\neq {k_\text{i}}}({\omega}_{k_\text{i}}-\omega_k)}\;,
        \end{array}
    \end{equation}
    where $p({\omega}_{k_\text{i}})/({\omega}_{k_\text{i}}-{\bar{\omega}}_{k_\text{i}})$ is real by the antisymmetry of $p$ and the fact that ${\omega}_{k_\text{i}}$ is imaginary. The remaining roots $k\neq {k_\text{i}}$ come either in couples of in quadrupoles. In both cases $({\omega}_{k_\text{i}}-\omega_k)({\omega}_{k_\text{i}}-\bar{\omega}_k)$ and $({\omega}_{k_\text{i}}-\omega_k)({\omega}_{k_\text{i}}-\bar{\omega}_k)({\omega}_{k_\text{i}}+\omega_k)({\omega}_{k_\text{i}}+\bar{\omega}_k)$ are real by construction.

\begin{figure}[t!]
\includegraphics[width = .9\columnwidth]{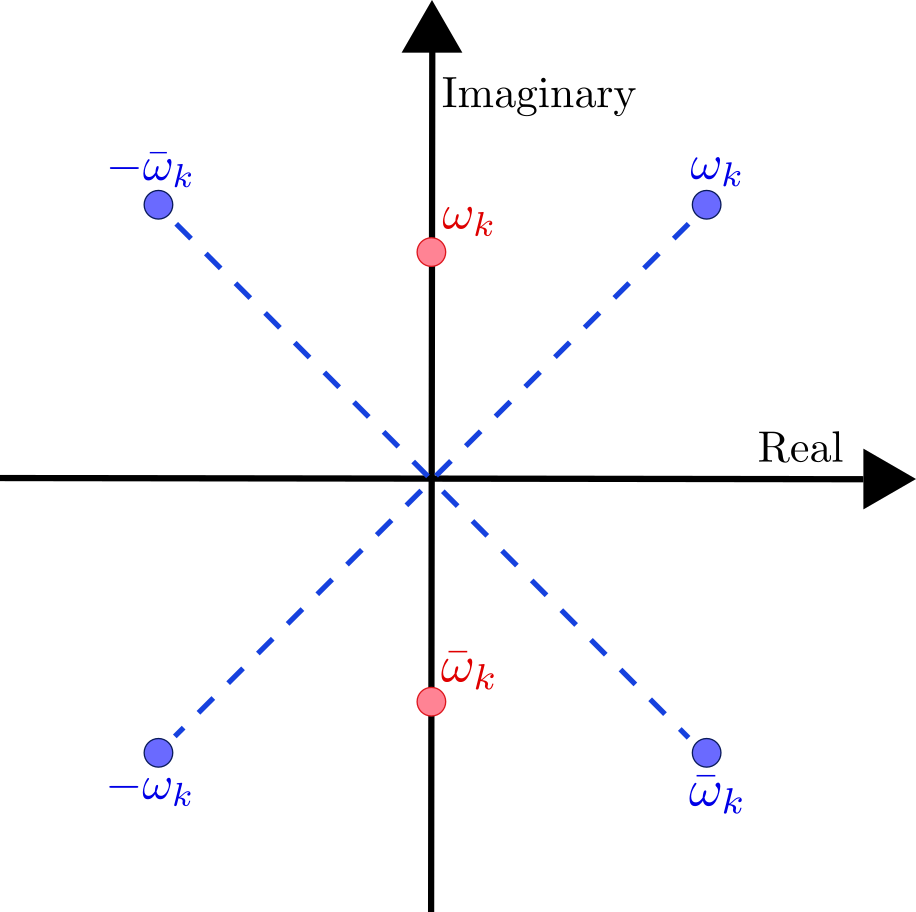} 
\caption{The poles of the spectral densities considered in this section come either in quadruples $(\omega_k,\bar{\omega}_k,-\omega_k,-\bar{\omega}_k)$, represented in blue, or in couples $(\omega_k,\bar{\omega}_k)$, represented in red. \label{fig:poles}}
\end{figure}
We can now compute 
\begin{equation}
    \begin{array}{lll}
    C_\text{as}(t) &=&-\displaystyle\frac{i}{2\pi}\int_{-\infty}^\infty d\omega J(\omega)\sin(\omega t)\\
    &=&\displaystyle\frac{1}{4\pi}\int_{-\infty}^\infty d\omega J(\omega)(e^{-i\omega t}-e^{i\omega t})\;,
    \end{array}
\end{equation}
where we used Eq.~(\ref{eq:C_symm_antisymm_def}) together with the antisymmetry of $J(\omega)$. 
Since the degree $N_\infty$ of the polynomial $p(\omega)$ is smaller than the degree $N_\text{poles}$ of the polynomial in the denominator (i.e., $N_\infty<N_\text{poles})$, \emph{may be replaced by} $N_\infty<N_\text{poles}-1$, Eq.~(\ref{eq:J_expression_1}) implies that $\lim_{\omega\rightarrow\infty}J(\omega)\rightarrow 0$ which allows us to use Jordan lemma to write
\begin{equation}
\begin{array}{lll}
    C_\text{as}(t)&=&-\displaystyle\frac{i\theta(t)}{2}\sum_{k<0}\text{Res}\left[J(\omega)e^{-i\omega t};\omega_k\right]\\
    &&+\displaystyle\frac{i\theta(-t)}{2}\sum_{k>0}\text{Res}\left[J(\omega)e^{-i\omega t};\omega_k\right]\\
    &&-\displaystyle\frac{i\theta(t)}{2}\sum_{k>0}\text{Res}\left[J(\omega)e^{i\omega t};\omega_k\right]\\
    &&+\displaystyle\frac{i\theta(-t)}{2}\sum_{k<0}\text{Res}\left[J(\omega)e^{i\omega t};\omega_k\right]\;,
    \end{array}
\end{equation}
where $\text{Res}[g(\omega);\omega^*]$ denotes the residue of the function $g(\omega)$ at the point $\omega^*$. Now that the Jordan lemma has been used, we can take advantage of Eq.~(\ref{eq:residue_explicit}) and Eq.~(\ref{eq:k_minus_k}) to write
\begin{equation}
\label{eq:C_antisymm_residues}
\begin{array}{lll}
    C_\text{as}(t)
   &=&-\displaystyle\frac{i}{2}\displaystyle\sum_{k>0}\theta(t)[R_k e^{i\omega_k t}+\bar{R}_k e^{-i\bar{\omega}_k t}]\\
   &&+\displaystyle\frac{i}{2}\displaystyle\sum_{k>0}\theta(-t)[R_k e^{-i\omega_k t}+\bar{R}_k e^{i\bar{\omega}_k t}]\;.
    \end{array}
\end{equation}
 We now describe an environment made out of pseudomodes able to exactly reproduce the correlation $C_\text{as}(t)$. To achieve this, we first write Eq.~(\ref{eq:C_antisymm_residues}) in the following slightly more explicit form
 \begin{equation}
 \label{eq:C_anti_PM}
     \begin{array}{lll}
    C_\text{as}(t)
   &=&-\displaystyle\frac{i}{2}\displaystyle\sum_{k>0}\theta(t)[R_k e^{i\omega^\mathcal{R}_k t}+\bar{R}_k e^{-i\omega^\mathcal{R}_k t}]e^{-\omega^\mathcal{I}_k t}\\
   &+&\displaystyle\frac{i}{2}\displaystyle\sum_{k>0}\theta(-t)[R_k e^{-i\omega^\mathcal{R}_k t}+\bar{R}_k e^{i\omega^\mathcal{R}_k t}]e^{\omega^\mathcal{I}_k t}\\
   &=&-\displaystyle i\text{sg}(t)\sum_{k>0}R^\mathcal{R}_k\cos(\omega_k^\mathcal{R}t)e^{-\omega^\mathcal{I}_k|t|}\\
   &&+\displaystyle i\sum_{k>0}R^\mathcal{I}_k\sin(\omega_k^\mathcal{R}t)e^{-\omega^\mathcal{I}_k|t|}\;,
    \end{array}
 \end{equation}
 where the indexes $\mathcal{R}$ and $\mathcal{I}$ denote the real and imaginary part of a variable. 
It is important to note that the sum over poles which belong to the same quadruple $(\omega_k,\omega_{-k},-\omega_k,-\omega_{-k})$ or pure-imaginary pairs $(\omega_k,\omega_{-k}$), if present, can be further reduced/simplified. To do this, we partition the $2N_p$ poles into $N_\text{q}$ quadruples and $N_\text{i}$ imaginary couples so that $2N_p=4N_\text{q}+2N_\text{i}$. We can now write 
 \begin{equation}
 \label{eq:C_anti_PM_2}
     \begin{array}{lll}
    C_\text{as}(t)
   &=&-2\displaystyle i\text{sg}(t)\sum_{{k_\text{q}}=1}^{N_\text{q}}R^\mathcal{R}_{k_\text{q}}\cos(\omega_{k_\text{q}}^\mathcal{R}t)e^{-\omega^\mathcal{I}_{k_\text{q}}|t|}\\
   &&+2\displaystyle i\sum_{{k_\text{q}}=1}^{N_\text{q}}R^\mathcal{I}_{k_\text{q}}\sin(\omega_{k_\text{q}}^\mathcal{R}t)e^{-\omega^\mathcal{I}_{k_\text{q}}|t|}\\

   &&-\displaystyle i\text{sg}(t)\sum_{{k_\text{i}=1}}^{N_\text{i}}R^\mathcal{R}_{k_\text{i}}\cos(\omega_{k_\text{i}}^\mathcal{R}t)e^{-\omega^\mathcal{I}_{k_\text{i}}|t|}\;,
    \end{array}
 \end{equation}
 where we  used Eq.~(\ref{eq:k_bark}), and
 where $k_\text{q}>0$ labels poles coming in quadruples $(\omega_{k_\text{q}},\omega_{-k_\text{q}},-\omega_{k_\text{q}},-\omega_{-k_\text{q}})$ and such that $\text{Re}[\omega_{k_\text{q}}]\geq 0$ and $\text{Im}[\omega_{k_\text{q}}]> 0$. Similarly, $k_\text{i}>0$ labels poles coming in pairs $(\omega_{k_\text{i}},\omega_{-k_\text{i}})$ and such that  $\text{Im}[\omega_{k_\text{i}}]> 0$ and $\text{Re}[\omega_{k_\text{i}}]=0$.

 To characterize a pseudomode model able to reproduce the correlation above, we just need to find the explicit parameters which allows to write $C_\text{PM}(t)=C_\text{as}(t)$, where $C_\text{PM}(t)$ is given in Eq.~(\ref{eq:correlationbyheisenberge}). However, the presence of the sign function in Eq.~(\ref{eq:C_anti_PM}) requires a few further preliminary steps. First, we write Eq.~(\ref{eq:C_anti_PM}) as
  \begin{equation}
  \label{eq:C_antisymm_temp}
     \begin{array}{lll}
    C_\text{as}(t)
   &=&-\displaystyle i\text{sg}(t)\sum_{k>0}R^\mathcal{R}_k\cos(\omega_k^\mathcal{R}t)(e^{-\omega^\mathcal{I}_k|t|}-e^{-W|t|})\\
   &&-\displaystyle i\text{sg}(t)\sum_{k>0}R^\mathcal{R}_k\cos(\omega_k^\mathcal{R}t)e^{-W|t|}\\
   &&+\displaystyle i\sum_{k>0}R^\mathcal{I}_k\sin(\omega_k^\mathcal{R}t)e^{-\omega^\mathcal{I}_k|t|}\;,
    \end{array}
 \end{equation}
 where we introduced an arbitrary frequency scale $W$. We can now take advantage of the identity
 \begin{equation}
     \text{sg}(t)=\frac{\exp[-a(|t|-t)]-\exp[-a(t+|t|)]}{1-\exp[-2a|t|]}\;,
 \end{equation}
 where $a$ is an arbitrary non-zero frequency. By imposing $a\rightarrow(W-\omega_k^\mathcal{I})/2$ for each different $k$, we get
   \begin{equation}
     \begin{array}{lll}
    C_\text{as}
   &=&-\displaystyle i\sum_{k>0}R^\mathcal{R}_k\cos(\omega_k^\mathcal{R}t)[e^{(W-\omega^\mathcal{I}_k)t/2}-e^{-(W-\omega^\mathcal{I}_k)t/2}]\\
   &&\displaystyle\times e^{-(W+\omega^\mathcal{I}_k)|t|/2}-\displaystyle i\text{sg}(t)\sum_{k>0}R^\mathcal{R}_k\cos(\omega_k^\mathcal{R}t)e^{-W|t|}\\
   &&+i\displaystyle \sum_{k>0}R^\mathcal{I}_k\sin(\omega_k^\mathcal{R}t)e^{-\omega^\mathcal{I}_k|t|}\\
   
   &=&\displaystyle 2\sum_{k>0}R^\mathcal{R}_k\cos(\omega_k^\mathcal{R}t)\sin\left[\frac{-i(W-\omega^\mathcal{I}_k)t}{2}\right]\\
   &&\times e^{-(W+\omega^\mathcal{I}_k)|t|/2}\\
   &&-\displaystyle i\text{sg}(t)\sum_{k>0}R^\mathcal{R}_k\cos(\omega_k^\mathcal{R}t)e^{-W|t|}\\
   &&+\displaystyle i\sum_{k>0}R^\mathcal{I}_k\sin(\omega_k^\mathcal{R}t)e^{-\omega^\mathcal{I}_k|t|}\;,
    \end{array}
 \end{equation}
 where we omitted the time-dependence on the left hand-side.
 Using the identity $\sin(x)\cos(y)=[\sin(x+y)+\sin(x-y)]/2$, we finally obtain
  \begin{equation}
  \label{eq:C_as(t)_before_PM}
     \begin{split}
    C_\text{as}
   &=\displaystyle \sum_{k>0}R^\mathcal{R}_k\sin\left[\frac{-i(W-\omega^\mathcal{I}_k)+2\omega_k^\mathcal{R}}{2}t\right]e^{-(W+\omega^\mathcal{I}_k)|t|/2}\\
   +&\displaystyle \sum_{k>0}R^\mathcal{R}_k\sin\left[\frac{-i(W-\omega^\mathcal{I}_k)-2\omega_k^\mathcal{R}}{2}t\right]e^{-(W+\omega^\mathcal{I}_k)|t|/2}\\
   +&\displaystyle i\sum_{k>0}R^\mathcal{I}_k\sin(\omega_k^\mathcal{R}t)e^{-\omega^\mathcal{I}_k|t|}\\
      -&\displaystyle i\text{sg}(t)\sum_{k>0}R^\mathcal{R}_k\cos(\omega_k^\mathcal{R}t)e^{-W|t|}\;.
    \end{split}
 \end{equation}
 As noted before, the sum over poles which belong to the same quadruple $(\omega_k,\omega_{-k},-\omega_k,-\omega_{-k})$ or pure-imaginary pairs $(\omega_k,\omega_{-k}$) can be further reduced/simplified using Eq.~(\ref{eq:k_bark}) to write
 \begin{equation}
  \label{eq:C_as(t)_before_PM_2}
     \begin{array}{lll}
    C_\text{as}
       &=&\displaystyle 2\sum_{k_\text{q}=1}^{N_\text{q}}\left\{iR^\mathcal{I}_{k_\text{q}}\sin(\omega_{k_\text{q}}^\mathcal{R}t)e^{-\omega^\mathcal{I}_{k_\text{q}}|t|}\right.\\
   &+&\displaystyle R^\mathcal{R}_{k_\text{q}}\sin\left[\frac{-i(W-\omega^\mathcal{I}_{k_\text{q}})+2\omega_{k_\text{q}}^\mathcal{R}}{2}t\right]e^{-(W+\omega^\mathcal{I}_{k_\text{q}})|t|/2}\\
   &+&\left.\displaystyle R^\mathcal{R}_{k_\text{q}}\sin\left[\frac{-i(W-\omega^\mathcal{I}_{k_\text{q}})-2\omega_{k_\text{q}}^\mathcal{R}}{2}t\right]e^{-(W+\omega^\mathcal{I}_{k_\text{q}})|t|/2}\right\}\\

   &+&\displaystyle \sum_{k_\text{i}=1}^{N_\text{i}}\left\{iR^\mathcal{I}_{k_\text{i}}\sin(\omega_{k_\text{i}}^\mathcal{R}t)e^{-\omega^\mathcal{I}_{k_\text{i}}|t|}\right.\\
   &+&\left.2\displaystyle R^\mathcal{R}_{k_\text{i}}\sin\left[\frac{-i(W-\omega^\mathcal{I}_{k_\text{i}})}{2}t\right]e^{-(W+\omega^\mathcal{I}_{k_\text{i}})|t|/2}\right\}\\

      &&-\displaystyle i\text{sg}(t)\sum_{k>0}R^\mathcal{R}_k\cos(\omega_k^\mathcal{R}t)e^{-W|t|}\;,
    \end{array}
 \end{equation}
where we omitted the time-dependence on the left hand-side.
 We further note, for $W$ much bigger than any inverse time scale present in the model, the last term is effectively non-zero only at $t=0$. However, in the context presented here, the correlation can always be interpreted as a \emph{distribution}, i.e., as a functional over functions of time (in the domain $[0,t]$). This can be seen explicitly by expanding Eq.~(\ref{eq:appDiffEq}) in series and using it to compute the expectation value of a system observable. In this case, the correlation always appears inside an integral which further involves functions of time related to the system, i.e., it can be interpreted as a linear operator over the space of such functions over the time domain. As a distribution, the correlation can always be redefined over discrete domains, thereby justifying the neglecting of the last term in the previous expression to write
  \begin{equation}
  \label{eq:C_as(t)_before_PM_3}
     \begin{array}{lll}
    C_\text{as}
       &=&\displaystyle 2\sum_{{k_\text{q}}=1}^{N_\text{q}}\left\{iR^\mathcal{I}_{k_\text{q}}\sin(\omega_{k_\text{q}}^\mathcal{R}t)e^{-\omega^\mathcal{I}_{k_\text{q}}|t|}\right.\\
   &+&\displaystyle R^\mathcal{R}_{k_\text{q}}\sin\left[\frac{-i(W-\omega^\mathcal{I}_{k_\text{q}})+2\omega_{k_\text{q}}^\mathcal{R}}{2}t\right]e^{-(W+\omega^\mathcal{I}_{k_\text{q}})|t|/2}\\
   &+&\left.\displaystyle R^\mathcal{R}_{k_\text{q}}\sin\left[\frac{-i(W-\omega^\mathcal{I}_{k_\text{q}})-2\omega_{k_\text{c}}^\mathcal{R}}{2}t\right]e^{-(W+\omega^\mathcal{I}_k)|t|/2}\right\}\\

   &+&\displaystyle 2\sum_{k_\text{i}=1}^{N_\text{i}}\displaystyle R^\mathcal{R}_{k_\text{i}}\sin\left[\frac{-i(W-\omega^\mathcal{I}_{k_\text{i}})}{2}t\right]e^{-(W+\omega^\mathcal{I}_{k_\text{i}})|t|/2}\;,
    \end{array}
 \end{equation}
 where we omitted the time-dependence on the left hand-side.
The resulting expression for the antisymmetric correlation can then be expressed by the following in the form
  \begin{equation}
  \label{eq:C_as(t)_before_PM_4}
    C_\text{as}(t)=\sum_{j=1}^{N_\text{as}} g^\text{as}_j\sin(\Omega^\text{as}_j t)e^{-\Gamma^\text{as}_j|t|}\;,
 \end{equation}
in terms of the parameters $g^\text{as}_j,\Omega^\text{as}_j,\Gamma^\text{as}_j\in\mathbb{C}$ which can be explicitly read by direct comparison with Eq.~(\ref{eq:C_as(t)_before_PM_3}). Here, $N_\text{as}=3N_\text{q}+N_\text{i}$ denotes the maximum number of terms in this decomposition. This immediately implies the possibility to use $N^\text{rational}_\text{PM}=N_\text{as}$ pseudomodes to reproduce the effects of this correlations. However, in order to reproduce sine functions with the expression for the pseudomode correlation given in Eq.~(\ref{eq:correlationbyheisenberge}), with a minimal (i.e., $N_\text{as}$) number of modes, one needs to impose what is, arguably, the highest level of unphysicality to the bosonic occupation numbers. In fact, by labeling generic pseudomodes parameters by an asterisk, reproducing a sine function requires $n^*=-1/2$ which does not allow an obvious truncation scheme for the thermal state which describes the initial state of the pseudomode (since its temperature would be required to satisfy $\beta^*\Omega^*=i\pi$). While alternative solutions might be found (as simple as increasing the number of pseudmodes), in this article we chose a strategy based on the fact that we are always free to reshape the antisymmetric part of the correlation by adding a symmetric function as long as we compensate it with classical stochastic fields, see section \ref{app:stoch_PM_zeroT}. This allows to model Eq.~(\ref{eq:C_as(t)_before_PM_4}) using $N^\text{rational}_\text{PM}$ pseudomodes initially at zero temperature.
 By explicitly defining the following symmetric function  as described in section \ref{sec:ZeroT_main}
 \begin{equation}
 \label{eq:fs_app}
 \begin{array}{lll}
     f_\text{s}&=&-i\displaystyle \sum_{j=1}^{N^\text{rational}_\text{PM}} g^\text{as}_j\cos(\Omega^\text{as}_j t)e^{-\Gamma^\text{as}_j|t|}\\
      &=&\displaystyle 2\sum_{{k_\text{q}}=1}^{N_\text{q}}\left\{R^\mathcal{I}_{k_\text{q}}\cos(\omega_{k_\text{q}}^\mathcal{R}t)e^{-\omega^\mathcal{I}_{k_\text{q}}|t|}\right.\\
   &-&i\displaystyle R^\mathcal{R}_{k_\text{q}}\cos\left[\frac{-i(W-\omega^\mathcal{I}_{k_\text{q}})+2\omega_{k_\text{q}}^\mathcal{R}}{2}t\right]e^{-(W+\omega^\mathcal{I}_{k_\text{q}})|t|/2}\\
   &-&i\left.\displaystyle R^\mathcal{R}_{k_\text{q}}\cos\left[\frac{-i(W-\omega^\mathcal{I}_{k_\text{q}})-2\omega_{k_\text{c}}^\mathcal{R}}{2}t\right]e^{-(W+\omega^\mathcal{I}_k)|t|/2}\right\}\\

   &-&\displaystyle 2i\sum_{k_\text{i}=1}^{N_\text{i}}\displaystyle R^\mathcal{R}_{k_\text{i}}\cos\left[\frac{-i(W-\omega^\mathcal{I}_{k_\text{i}})}{2}t\right]e^{-(W+\omega^\mathcal{I}_{k_\text{i}})|t|/2}\;,
    \end{array}
 \end{equation}
 where we omitted the time-dependence on the left hand-side, we obtain, using Eq.~(\ref{eq:class_Q}), the following quantum contribution to the correlation
 \begin{equation}
     \begin{array}{lll}
    C_\text{Q}(t)
       &=&C_\text{as}(t)-f_\text{s}(t)\\
       &=&\displaystyle -2\sum_{{k_\text{q}}=1}^{N_\text{q}}\left\{R^\mathcal{I}_{k_\text{q}}e^{-i\omega_{k_\text{q}}^\mathcal{R}t}e^{-\omega^\mathcal{I}_{k_\text{q}}|t|}\right.\\
   &+&i\displaystyle R^\mathcal{R}_{k_\text{q}}e^{-i[\omega_{k_\text{q}}^\mathcal{R}-i(W-\omega^\mathcal{I}_{k_\text{q}})/2]t}e^{-(W+\omega^\mathcal{I}_{k_\text{q}})|t|/2}\\
   &+&i\left.\displaystyle R^\mathcal{R}_{k_\text{q}}e^{-i[-\omega_{k_\text{q}}^\mathcal{R}-i(W-\omega^\mathcal{I}_{k_\text{q}})/2]t}e^{-(W+\omega^\mathcal{I}_{k_\text{q}})|t|/2}\right\}\\

   &+&\displaystyle 2i\sum_{k_\text{i}=1}^{N_\text{i}}\displaystyle R^\mathcal{R}_{k_\text{i}}e^{-i[-i(W-\omega^\mathcal{I}_{k_\text{i}})/2]t}e^{-(W+\omega^\mathcal{I}_{k_\text{i}})|t|/2}\;,
    \end{array}
 \end{equation}
 which corresponds to a decomposition like the one presented in Eq.~(\ref{eq:dec_app}) which can be modeled by pseudomodes initially at zero temperature.

 \subsection{Symmetric Correlation}
In this subsection we explicitly compute the symmetric part of the correlation for the subset of spectral densities $\mathcal{M}$. We do this for completeness as the stochastic fields needed to reproduce this part of the correlation can be computed without reference to any analytical form, see section \ref{app:spectral_representation}.

We want to compute, see Eq.~(\ref{eq:C_symm_antisymm_def}),
\begin{equation}
    \begin{array}{lll}
    C_\text{s}(t)&=&\displaystyle\frac{1}{2\pi}\int_{-\infty}^\infty d\omega~J(\omega)\coth(\beta\omega/2)\cos(\omega t)\\
    &=&\displaystyle\frac{1}{4\pi}\int_{-\infty}^\infty d\omega~J(\omega)\coth(\beta\omega/2)(e^{-i\omega t}+e^{i\omega t}),
    \end{array}
\end{equation}
for $J(\omega)\in\mathcal{M}$. We have
\begin{equation}
\begin{array}{lll}
    C_\text{s}(t)&=&-\displaystyle\frac{i\theta(t)}{2}\sum_{k<0}\text{Res}\left[J(\omega)\coth(\beta\omega/2)e^{-i\omega t};\tilde{\omega}_k\right]\\
    &&+\displaystyle\frac{i\theta(-t)}{2}\sum_{k>0}\text{Res}\left[J(\omega)\coth(\beta\omega/2)e^{-i\omega t};\tilde{\omega}_k\right]\\
    &&+\displaystyle\frac{i\theta(t)}{2}\sum_{k>0}\text{Res}\left[J(\omega)\coth(\beta\omega/2)e^{i\omega t};\tilde{\omega}_k\right]\\
    &&-\displaystyle\frac{i\theta(-t)}{2}\sum_{k<0}\text{Res}\left[J(\omega)\coth(\beta\omega/2)e^{i\omega t};\tilde{\omega}_k\right].
    \end{array}
\end{equation}
Here, we denoted by $\tilde{\omega}_k$ the collection of both the poles $\omega_k$ of $J(\omega)$ as described in the previous section and the Matsubara poles $\omega^\text{M}_k=2\pi i k/\beta$, $k\in\mathbb{Z}-\{0\}$ of $\coth(\beta\omega/2)$. $0$ is not a pole for the composite $J(\omega)\coth\beta\omega/2$ because it's cancelled by asymmetric $p(\omega)$ whose degree is larger than 1.

If simple poles of $J(\omega)$ and Matsubara poles have no intersection, using the fact that $
{\rm Res}_{\omega_k^{\rm M}}\coth(\beta\omega/2)=2/\beta$, we can write
\begin{equation}
\label{eq:C_symm_res_mats}
\begin{array}{lll}
    C_\text{s}(t)
   &=&\displaystyle\frac{i}{2}\displaystyle\sum_{k>0}\theta(t)[{R}^\beta_k e^{i\omega_k t}-R^{\prime\beta}_k e^{-i\bar{\omega}_k t}]\\
   &&+\displaystyle\frac{i}{2}\displaystyle\sum_{k>0}\theta(-t)[R^\beta_k e^{-i\omega_k t}-R^{\prime\beta}_k e^{i\bar{\omega}_k t}]\\
   &&+\displaystyle\frac{2i}{\beta}\sum_{k>0}J(\omega^\text{M}_{k})e^{-|\omega^\text{M}_{k}||t|}\;,
    \end{array}
\end{equation}
where $R^\beta_k=R_k \coth(\beta\omega_k/2)$ and $R^{\prime\beta}_k=\bar{R}_k \coth(\beta\bar{\omega}_k/2)$ which is just the conjugate of $R^\beta_k$ for real $\beta$. We also took advantage of the fact that $\omega^\text{M}_{-k}=-\omega^\text{M}_{k}$. For real temperatures, we can further write
\begin{align}
\label{eq:C_symm_res_mats_realT}
\begin{array}{lll}
    C_\text{s}(t)
   &=&-\displaystyle\sum_{k>0}[{R}^{\beta,\mathcal{R}}_k \sin{(\omega^{R}_k |t|)}+{R}^{\beta,\mathcal{I}}_k \cos{(\omega^{R}_k t)}] e^{-\omega_k^\mathcal{I}|t|}\\
   &&+\displaystyle\frac{2i}{\beta}\sum_{k>0}J(\omega^\text{M}_{k})e^{-|\omega^\text{M}_{k}||t|}\;.
    \end{array}
\end{align}
By partitioning the sum into poles belonging to couples and quadrupoles, we have
\begin{equation}
\begin{array}{lll}
    C_\text{s}
      &=&-2\displaystyle\sum_{{k_\text{q}}}[{R}^{\beta,\mathcal{R}}_{k_\text{q}} \sin{(\omega^{R}_{k_\text{q}} |t|)}+{R}^{\beta,\mathcal{I}}_{k_\text{q}} \cos{(\omega^{R}_{k_\text{q}} t)}] e^{-\omega_{k_\text{q}}^\mathcal{I}|t|}\\

            &&-\displaystyle\sum_{{k_\text{i}=1}}^{N_\text{i}}{R}^{\beta,\mathcal{I}}_{k_\text{i}}e^{-\omega_{k_\text{i}}^\mathcal{I}|t|}+\displaystyle\frac{2i}{\beta}\sum_{k>0}J(\omega^\text{M}_{k})e^{-|\omega^\text{M}_{k}||t|}\;,
    \end{array}
\end{equation}
where $k_\text{q}=1,\dots,{N_\text{q}}$, and where we used Eq.~(\ref{eq:k_bark}). Noting that  Eq.~(\ref{eq:k_bark}) implies that $R^{\beta,\mathcal{I}}_{k_\text{i}}=\text{Im}[R_{k_\text{i}}\coth(\beta\omega_{k_\text{i}}/2)]=-i R_{k_\text{i}}\coth(\beta\omega_{k_\text{i}}/2)=-iR^{\beta}_{k_\text{i}}$, and using the fact that $\omega^\mathcal{I}_{k_\text{i}}=-i\omega_{k_\text{i}}$, together with $\cos(\omega t)=\cos(\omega|t|)$ for all $\omega$, we can further write
\begin{equation}
\begin{array}{lll}
    C_\text{s}
      &=&\displaystyle\sum_{{k_\text{q}=1}}^{N_\text{q}}[i R^\beta_{k_\text{q}}e^{i\omega_{k_\text{q}}|t|}-i \bar{R}^\beta_{k_\text{q}}e^{-i\bar{\omega}_{k_\text{q}}|t|}]\\

            &&+i\displaystyle\sum_{{k_\text{i}=1}}^{N_\text{i}}{R}^{\beta}_{k_\text{i}}e^{i\omega_{k_\text{i}}|t|}+\displaystyle\frac{2i}{\beta}\sum_{k>0}J(\omega^\text{M}_{k})e^{-|\omega^\text{M}_{k}||t|}\;,
    \end{array}
\end{equation}
Using Eq.~(\ref{eq:class_Q}) and Eq.~(\ref{eq:fs_app}), we can also find the classical contribution to the correlation function as
\begin{equation}
     \begin{array}{lll}
    C_\text{class}(t)
           &=&C_\text{s}(t)+f_\text{s}(t)\\
          &=&\displaystyle\sum_{{k_\text{q}=1}}^{N_\text{q}}[i R^\beta_{k_\text{q}}e^{i\omega_{k_\text{q}}|t|}-i \bar{R}^\beta_{k_\text{q}}e^{-i\bar{\omega}_{k_\text{q}}|t|}]\\

            &+&i\displaystyle\sum_{{k_\text{i}=1}}^{N_\text{i}}{R}^{\beta}_{k_\text{i}}e^{i\omega_{k_\text{i}}|t|}+\displaystyle\frac{2i}{\beta}\sum_{k>0}J(\omega^\text{M}_{k})e^{-|\omega^\text{M}_{k}||t|}\\
        
       &+&\displaystyle \sum_{{k_\text{q}}=1}^{N_\text{q}}\left\{R^\mathcal{I}_{k_\text{q}}(e^{i\omega_{k_\text{q}}|t|}+e^{-i\bar{\omega}_{k_\text{q}}|t|})\right.\\
   &-&i\displaystyle R^\mathcal{R}_{k_\text{q}}(e^{i\omega_{k_\text{q}}|t|}+e^{-i\bar{\omega}_{k_\text{q}}|t|})\\
   &-&i\left.\displaystyle R^\mathcal{R}_{k_\text{q}}(e^{(-i\omega_{k_\text{q}}^\mathcal{R}-W)|t|}+e^{(i\omega_{k_\text{q}}^\mathcal{R}-W)|t|})\right\}\\

   &-&\displaystyle i\sum_{k_\text{i}=1}^{N_\text{i}}\displaystyle R^\mathcal{R}_{k_\text{i}}(e^{-\omega^\mathcal{I}_{k_\text{i}}|t|}+e^{-W|t|})\;,
    \end{array}
 \end{equation}
which can be further be written as
\begin{equation}
  \label{eq:C_as(t)_before_PM_33_app}
     \begin{array}{lll}
    C_\text{class}(t)
          &=&\displaystyle\sum_{{k_\text{q}=1}}^{N_\text{q}}[i \Delta R_{k_\text{q}}~e^{i\omega_{k_\text{q}}|t|}-i\Delta' R_{k_\text{q}}e^{-i\bar{\omega}_{k_\text{q}}|t|}]\\

            &+&i\displaystyle\sum_{{k_\text{i}=1}}^{N_\text{i}}\Delta R_{k_\text{i}}~e^{i\omega_{k_\text{i}}|t|}+\displaystyle\frac{2i}{\beta}\sum_{k>0}J(\omega^\text{M}_{k})e^{-|\omega^\text{M}_{k}||t|}\;,

    \end{array}
 \end{equation}
  in terms of $\Delta R_{k_\text{x}}=R^\beta_{k_\text{x}}-R_{k_\text{x}}$ and $\Delta' R_{k_\text{x}}=\bar{R}^\beta_{k_\text{x}}-R_{k_\text{x}}$, ($\text{x}=\text{q},\text{i}$) and 
where we used Eq.~(\ref{eq:k_bark_2}) and  neglected terms decaying exponentially in the free parameter $W$.

This explicit expression can also be used to explicitly compute the coefficients for the spectral decomposition of the field $\xi$, see Eq.~(\ref{eq:xi_spectral_representation_2_main}). In fact, we can simply use Eq.~(\ref{eq:C_as(t)_before_PM_33_app}) into Eq.~(\ref{eq:xi_cn}) to obtain
\begin{equation}
\label{eq:c_n_analytical}
\begin{array}{lll}
   c_n
          &=&\displaystyle\sum_{{k_\text{q}=1}}^{N_\text{q}}[i \Delta R_{k_\text{q}}~L({i\omega_{k_\text{q}}})-i\Delta' R_{k_\text{q}}~L_n({-i\bar{\omega}_{k_\text{q}}})]\\

            &+&i\displaystyle\sum_{{k_\text{i}=1}}^{N_\text{i}}\Delta R_{k_\text{i}}~L_n({i\omega_{k_\text{i}}})+\displaystyle\frac{2i}{\beta}\sum_{k>0}J(\omega^\text{M}_{k})L_n({-|\omega^\text{M}_{k}|})\;,

    \end{array}
\end{equation}
where, as in section \ref{app:spectral_representation}, we used the identity
\begin{equation}
\frac{1}{2T}\int_{-T}^Td\tau~ e^{\omega\tau}\cos\left(\frac{n\pi\tau}{T}\right)=L_n(\omega)\;,
\end{equation}
for all $\omega\in\mathbb{C}$, in terms of the function
\begin{equation}
\label{eq:L_N_app}
L_n(\omega)=(e^{\omega T}e^{in\pi}-1)\frac{\omega T}{(\omega T)^2+(n\pi)^2}\;.
\end{equation}

\subsection{Brownian Spectral Density}
We now specialize the results of the previous section for the Brownian spectral density 
\begin{equation}
\label{eq:spectral_density_app}
    J_{B}(\omega)=\frac{\gamma\lambda^{2}\omega}{(\omega^{2}-\omega_{0}^{2})^{2}+\gamma^{2}\omega^{2}}\;,
\end{equation}
characterized by a resonant frequency $\omega_0$, a width $\gamma$ (with dimension of frequency) and an overall strength $\lambda$ (with dimension $\text{frequency}^{3/2}$, so that, for adimensional system coupling ${s}$, the correlation has dimension of energy squared). This spectral density has four poles located at $\pm(\Omega\pm i\Gamma)$, where $\Omega = \sqrt{\omega_0^2-\Gamma^2}$ and $\Gamma = \gamma/2$. This defines an underdamped limit when $\omega_0>\gamma/2$ (poles not located on the imaginary axis) and an overdamped limit when $\omega_0<\gamma/2$ (poles located on the imaginary axis) which we are going to analyze in the following sections. In both cases, the spectral density in Eq.~(\ref{eq:spectral_density_app}) can be written, in the notation of Eq.~(\ref{eq:J_expression_1}) as
\begin{equation}
\label{eq:spectral_density_app2}
    J_{B}(\omega)=\frac{p_B(\omega)}{\prod_{k\in\mathbb{Z}^0_{4}}(\omega-\omega_k)}\;,
\end{equation}
where $p_B(\omega)=\gamma\lambda^2 \omega$ and $\mathbb{Z}^0_{4}=\{-2,-1,1,2\}$ and $\omega_1=\Omega+i\Gamma$, $\omega_2=-\Omega+i\Gamma$, $\omega_{-k}=-\omega_k$.

\subsubsection{Under-damped limit}
The under-damped limit $\Gamma<\omega_{0}$ implies $\Omega>0$ and, using Eq.~(\ref{eq:spectral_density_app2}), we can compute the residues at the two poles $\omega_{1,2}$ on the upper complex plane as
\begin{equation}
\label{eq:residues_brownian}
    \begin{array}{lll}
    \text{Res}_{\omega_1}J_B(\omega)&=&\displaystyle\gamma\lambda^2\frac{\omega_1}{(\omega_1-\omega_2)(\omega_1-\omega_{-1})(\omega_1-\omega_{-2})}\\
    &=&\displaystyle\gamma\lambda^2\frac{\omega_1}{(2i\Gamma)(2\omega_1)(2\Omega)}\\
    &=&\displaystyle-i\frac{\lambda^2}{4\Omega}\\
    
    \text{Res}_{\omega_2}J_B(\omega)&=&\displaystyle\gamma\lambda^2\frac{\omega_2}{(\omega_2-\omega_1)(\omega_2-\omega_{-1})(\omega_2-\omega_{-2})}\\
    &=&\displaystyle\gamma\lambda^2\frac{\omega_2}{(2i\Gamma)(-2\Omega)(2\omega_2)}\\
    &=&\displaystyle i\frac{\lambda^2}{4\Omega}\;.
    \end{array}
\end{equation}
The imaginary nature of the residues above, considerably simplifies the expression in Eq.~(\ref{eq:C_as(t)_before_PM}) for the asymmetric part of the correlation function which reads
\begin{equation}
\begin{array}{lll}
    C_\text{as}(t)&=&\displaystyle-i\frac{\lambda^2}{2\Omega}\sin(\Omega t)e^{-\Gamma |t|}\\
    &=&\displaystyle\frac{\lambda^2}{4\Omega}(e^{-i\Omega t}-e^{i\Omega t})e^{-\Gamma |t|}\;.
    \end{array}
\end{equation}
Similarly, the symmetric part of the correlation in Eq.~(\ref{eq:C_symm_res_mats_realT}) can be written as
\begin{equation}
\begin{array}{lll}
    C_\text{s}(t)
   &=&\displaystyle \frac{\lambda^2}{4\Omega}\coth{(\beta\omega_1/2)}e^{i\omega^{R}_1 |t|} e^{-|\omega_1^I||t|}\\
   &&\displaystyle -\frac{\lambda^2}{4\Omega}\coth{(\beta\omega_2/2)}e^{i\omega^{R}_2 |t|} e^{-|\omega_2^I||t|}\\
   &&+\displaystyle\frac{2i}{\beta}\sum_{k>0}J(\omega^\text{M}_{k})e^{-|\omega^\text{M}_{k}||t|}\\   &=&\displaystyle \frac{\lambda^2}{4\Omega}\coth{(\beta(\Omega+i\Gamma)/2)}e^{i\Omega |t|} e^{-\Gamma|t|}\\
   &&\displaystyle -\frac{\lambda^2}{4\Omega}\coth{(\beta(-\Omega+i\Gamma)/2)}e^{-i\Omega |t|} e^{-\Gamma|t|}\\
   &&+\displaystyle\frac{2i}{\beta}\sum_{k>0}J(\omega^\text{M}_{k})e^{-|\omega^\text{M}_{k}||t|}\;,
    \end{array}
\end{equation}
in terms of the Matsubara frequencies $\omega^\text{M}_k=2\pi ik/\beta$.
In the $\beta\rightarrow\infty$ limit, we have
\begin{equation}
\begin{array}{lll}
    C_s(t)&\overset{\beta\rightarrow\infty}{=}&\displaystyle\frac{\lambda^2}{4\Omega}(e^{i\Omega t}+e^{-i\Omega t})e^{-\Gamma|t|}\\
    &&+\displaystyle\frac{i}{\pi}\int_0^\infty dx J(ix)e^{-x|t|}
      \end{array}
\end{equation}
By defining $R_B=\text{Re}\{\coth[\beta(\Omega+i\Gamma)/2)\}$ and $I_B=\text{Im}\{\coth[\beta(\Omega+i\Gamma)/2)\}$ and writing the full correlation function as
\begin{equation}
\begin{array}{lll}
    C(t)&=&\displaystyle C_\text{s}(t)+C_\text{as}(t)\\
    &=&\displaystyle\frac{\lambda^2}{2\Omega}\left(\frac{R_B+1}{2}e^{-i\Omega t}+\frac{R_B-1}{2}e^{i\Omega t}\right)e^{-\Gamma|t|}\\
    &&\displaystyle+\frac{\lambda^2}{4\Omega}\left(I_B e^{-(-i\Omega+\Gamma) |t|}-I_B e^{-(i\Omega+\Gamma) |t|}\right)\\
    
   &&\displaystyle+\displaystyle\frac{2i}{\beta}\sum_{k>0}J(\omega^\text{M}_{k})e^{-|\omega^\text{M}_{k}||t|}\;.
    \end{array}
\end{equation}
{ From this expression, it is possible to define a completely deterministic model on the lines of the one proposed in \cite{Lambert}. This can be done by introducing three ``resonant'' modes $a_j$, $j=1,2,3$ and $N_\text{mats}$ ``Matsubara'' modes $a_j$, $j=4,\cdots 3+N_\text{mats}$ as}
\begin{equation}
\label{eq:dynamics_B_app}
\dot{\rho}^{ \text{det}}=-i[H_B^{ \text{det}},\rho^{ \text{det}}]+D_{B}^{ \text{det}}[\rho^{ \text{det}}]\;,
\end{equation}
where
\begin{equation}
\begin{array}{lll}
H_B^{ \text{det}}&=&\displaystyle H_S+\sum_{j=1}^{3{ +N_\text{mats}}}\lambda_j(a_j+a_j^\dagger){ {s}}+\Omega_j a_j^\dagger a_j\\

D_B^{ \text{det}}[\rho]&=&\displaystyle{{ \sum_{j=1}^{3+N_\text{mats}}}}\Gamma_j[(n_j+1)(2 a_j\rho a_j^\dagger-a_j^\dagger a_j\rho-\rho a_j^\dagger a_j)\\
&&\displaystyle +n_j(2 a^\dagger_j\rho a_j-a_j a^\dagger_j\rho-\rho a_j a^\dagger_j)]\;,
\end{array}
\end{equation}
as a function of the parameters
\begin{equation}
\label{eq:underdamped_full_PM_param}
    \begin{array}{lllllllll}
    \lambda_1&=&\displaystyle\sqrt{\frac{\lambda^2}{2\Omega}}&\lambda_2&=&\displaystyle\sqrt{\frac{I_B\lambda^2}{4\Omega}}&\lambda_3&=&\displaystyle\sqrt{\frac{-I_B\lambda^2}{4\Omega}}\\
    \Omega_1&=&\Omega&\Omega_2&=&0&\Omega_3&=&0\\
    \Gamma_1&=&\Gamma&\Gamma_2&=&\Gamma-i\Omega&\Gamma_3&=&\Gamma+i\Omega\\
    n_1 &=&\displaystyle\frac{R_B-1}{2}&n_2&=&0&n_3&=&0\;,\\
    \end{array}
\end{equation}
{ characterizing the resonant modes. We note that other choices which only use two resonant modes initially prepared with complex-temperature values are possible \cite{Paul}. The parameters characterizing the Matsubara modes can be  defined by imposing $n_j=0$, $\Omega_j=0$ for $j=4,\cdots,N_\text{mats}$ and by minimizing the functional difference
\begin{equation}
    \left|\displaystyle\frac{2i}{\beta}\sum_{k>0}J(\omega^\text{M}_{k})e^{-|\omega^\text{M}_{k}||t|}-\sum_{j=4}^{N_\text{mats}}\lambda_j^2 e^{-\Gamma_j|t|}\right|\;.
\end{equation}}
{  In contrast, the stochastic method proposed in here only requires to solve the dynamics of the system coupled to a single zero-temperature resonant mode $a_0$ and driven by a complex stochastic field $\xi(t)$ as
\begin{equation}
\label{eq:dynamics_B_app_stoch}
\dot{\rho}=-i[H_B,\rho]+D_B[\rho]\;,
\end{equation}
where
\begin{equation}
\label{eq:underdamped_full_PM_param_stoch}
    \begin{array}{lll}
H_B&=&\displaystyle H_S+\lambda_0{s}(a_0+a_0^\dagger)+\Omega_0 a_0^\dagger a_0 + \xi(t){s}\\

D_B[\rho]&=&\displaystyle\Gamma_0(2 a_0\rho a_0^\dagger-a_0^\dagger a_0\rho-\rho a_0^\dagger a_0)\;.
    \end{array}
\end{equation}
In this case, the parameters for the resonant mode and the statistics of the field can be determined by the following classical-quantum decomposition of the correlation
\begin{equation}
    C(t)=C_\text{class}(t)+C_\text{quantum}(t)\;,
\end{equation}
where
\begin{equation}
\begin{array}{lll}
    C_\text{class}(t)&=&C_\text{s}(t)-f_\text{s}(t)\\
    C_\text{quantum}(t)&=&\displaystyle C_\text{as}(t)+f_\text{s}(t)=\frac{\lambda^2}{2\Omega} e^{-i\Omega t-\Gamma|t|}\;,
    \end{array}
\end{equation}
in terms of the symmetric function $f_\text{s}(t)=C_\text{quantum}(t)+C_\text{quantum}(-t)$. Specifically, the parameters in Eq.~(\ref{eq:underdamped_full_PM_param_stoch}) must then be defined as $\lambda_0=\sqrt{\lambda^2/2\Omega}$, $\Omega_0=\Omega$, $\Gamma_0=\Gamma$, while the field $\xi(t)$ must be Gaussian and such that
\begin{equation}
    \mathbb{E}[\xi(t_2)\xi(t_1)]=C_\text{class}(t_2-t_1)\;.
\end{equation}
It is actually possible to be even more explicit in the definition of the stochastic field. In fact,
}
 by using Eq.~(\ref{eq:residues_brownian}) into Eq.~(\ref{eq:c_n_analytical}), we obtain
\begin{equation}
\label{eq:c_n_brownian}
\begin{array}{lll}
   c_n
          &=&\displaystyle\frac{\lambda^2}{4\Omega}\{\coth[\beta(\Omega+i\Gamma)/2]-1\}L(i\Omega-\Gamma)\\
          &&+\displaystyle\frac{\lambda^2}{4\Omega}\{\coth[\beta(\Omega-i\Gamma)/2]-1\}L(-i\Omega-\Gamma)\\

            &+&\displaystyle\frac{2i}{\beta}\sum_{k>0}J(\omega^\text{M}_{k})L_n({-|\omega^\text{M}_{k}|})\;,

    \end{array}
\end{equation}
in terms of the function $L_n(\omega)$ defined in Eq.~(\ref{eq:L_N_app}).

\subsubsection{Over-damped limit}
The over-damped limit $\Gamma>\omega_{0}$ implies $\Omega^2<0$ so that the results in the above section can be adapted with the substitution $\Omega\mapsto i|\Omega|$, with $|\Omega|<\Gamma$. For example, the poles in the upper complex plane are now $\omega_1=i(\Gamma+\Omega)$ and $\omega_2=i(\Gamma-\Omega)$ and the residues become
\begin{equation}
    \begin{array}{lll}
    \text{Res}_{\omega_1}J_B(\omega)&=&\displaystyle-\frac{\lambda^2}{4|\Omega|}\\
    
    \text{Res}_{\omega_2}J_B(\omega)&=&\displaystyle \frac{\lambda^2}{4|\Omega|}\;.
    \end{array}
\end{equation}
Using   Eq.~(\ref{eq:C_as(t)_before_PM_3}) and noting that here the poles come in pairs (we have two such pairs, one for each of the two poles $\omega_{1,2}$ on the upper complex plane) with real residues, we can write
\begin{equation}
     \begin{array}{lll}
    C_\text{as}(t)
       &=&2\displaystyle \sum_{k_\text{i}}R^R_{k_\text{i}}\sin\left[\frac{-i(W-\omega^I_{k_\text{i}})}{2}t\right]e^{-(W+\omega^I_{k_\text{i}})|t|/2}\\

       &=&2\displaystyle \frac{\text{sg}(t)}{2i}\sum_{k_\text{i}}R^R_{k_\text{i}}e^{-\omega^I_{k_\text{i}}t}\\
       
       &=&2\displaystyle \frac{\text{sg}(t)}{2i}\frac{\lambda^2}{4|\Omega|}e^{-\Gamma|t|}\left(e^{|\Omega| |t|}-e^{-|\Omega||t|}\right)\\

              &=&i\displaystyle \frac{\lambda^2}{4|\Omega|}e^{-\Gamma |t|}\left(e^{-|\Omega| t}-e^{|\Omega|t}\right)\;.
       
    \end{array}
 \end{equation}
\begin{figure*}[t!]
\includegraphics[width = \textwidth]{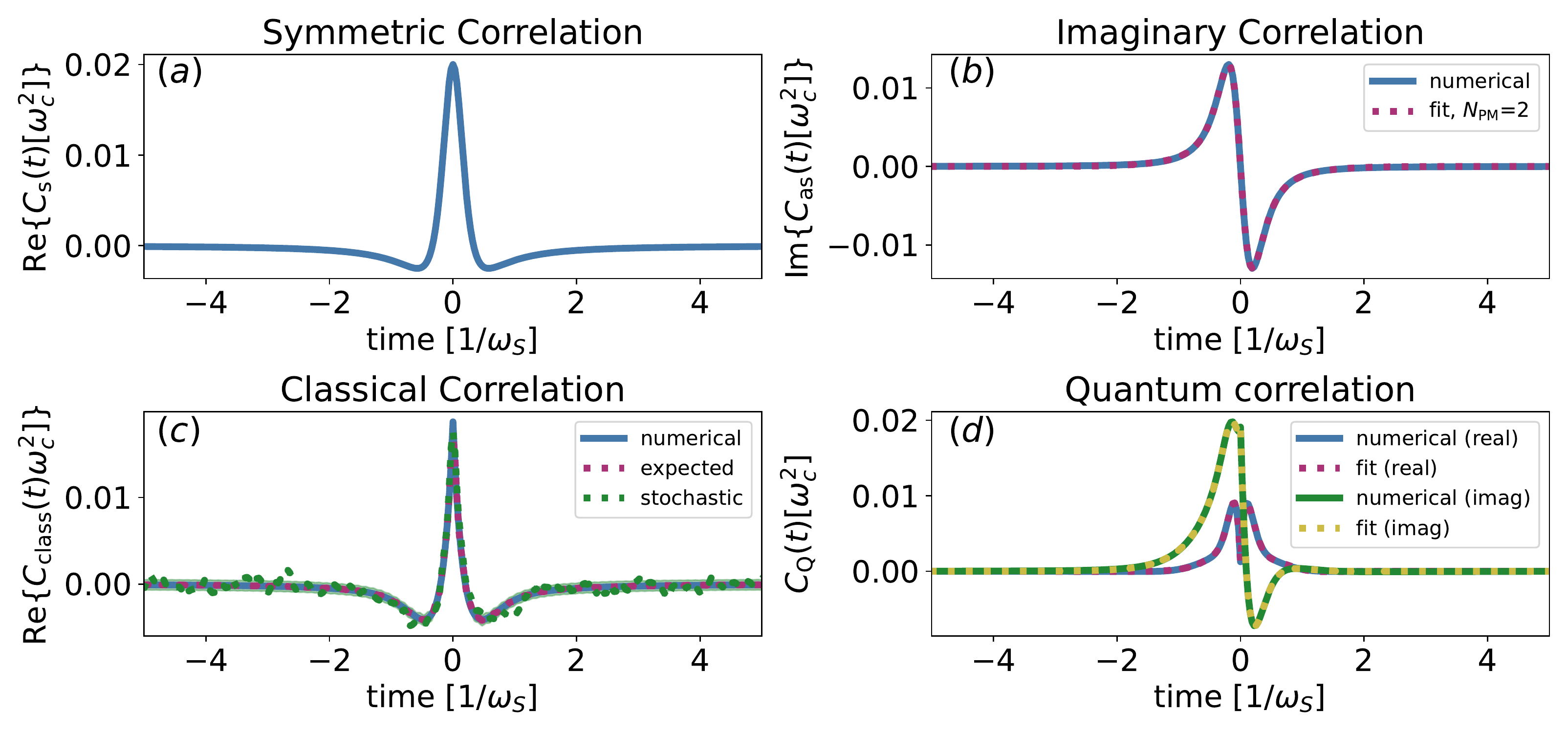} 
\caption{Correlations for the Ohmic spectral density (see \cite{BrandesDiss} for an analytical expression) at zero temperature. Here, the parameters correspond to the ones in Fig. \ref{fig:ohmic_dynamics} in the main text. The fitting of the antisymmeatric part of the correlation function is done using two sine functions (which corresponds to two pseudomodes). 
 \label{fig:ohmic_corr}}
\end{figure*}
When the poles of the spectral density $J_B(\omega)$ are not located at the poles of $\coth{{(\beta\omega)}/{2}}$, the symmetric part of the correlation reads 
\begin{equation}\begin{array}{lll}
  C_\text{s}(t)   &=&\displaystyle -\frac{\lambda^2}{4|\Omega|}\text{cot}\left[\frac{\beta(\Gamma+|\Omega|)}{2}\right]e^{-{(\Gamma+|\Omega|)}|t|}\\
     &&\displaystyle +\frac{\lambda^2}{4|\Omega|}\text{cot}\left[\frac{\beta(\Gamma-|\Omega|)}{2}\right]e^{-{(\Gamma-|\Omega|)}|t|}\\
  &&\displaystyle+\frac{2i}{\beta}\sum_{k>0}J_B(\omega^\text{M}_{k})e^{-|\omega^\text{M}_{k}||t|}\;.
\end{array}\end{equation}
When the poles $\omega_1$ and $\omega_2$ are located exactly at one of the Matsubara poles $\omega^\text{M}_k$, one can still apply the expression above after regularizing the degeneracy and then taking the limit for such regularization to be zero.

\subsubsection{Critical-damping limit} 
When $\Gamma=\omega_{0}$, we have $\Omega=0$, implying the presence of a second order pole in the upper complex plane located at $\omega_1=\omega_2=i\Gamma$. The correlation can be computed by either using the residue theorem or by simply taking the $\Omega\rightarrow 0$ limit in the corresponding expressions for the underdamped and overdamped limit to obtain
\begin{equation}
C_{as}(t)=-\frac{it\lambda^2}{2}e^{-\Gamma |t|}.
\end{equation}
Similarly, the symmetric correlation reads
\begin{equation}\begin{array}{lll}
  C_\text{s}(t)   &=&\displaystyle\frac{\lambda^2}{4}\frac{\beta+|t|\sin(\beta\Gamma)}{\sin(\beta\Gamma/2)}e^{-\Gamma|t|}\\
  &&\displaystyle +\frac{2i}{\beta}\sum_{k>0}J_B(\omega^\text{M}_{k})e^{-|\omega^\text{M}_{k}||t|}\;.
\end{array}\end{equation}

\subsection{Ohmic Spectral density}
\label{app:ohmic}
\begin{figure}[t!]
\includegraphics[width =\columnwidth]{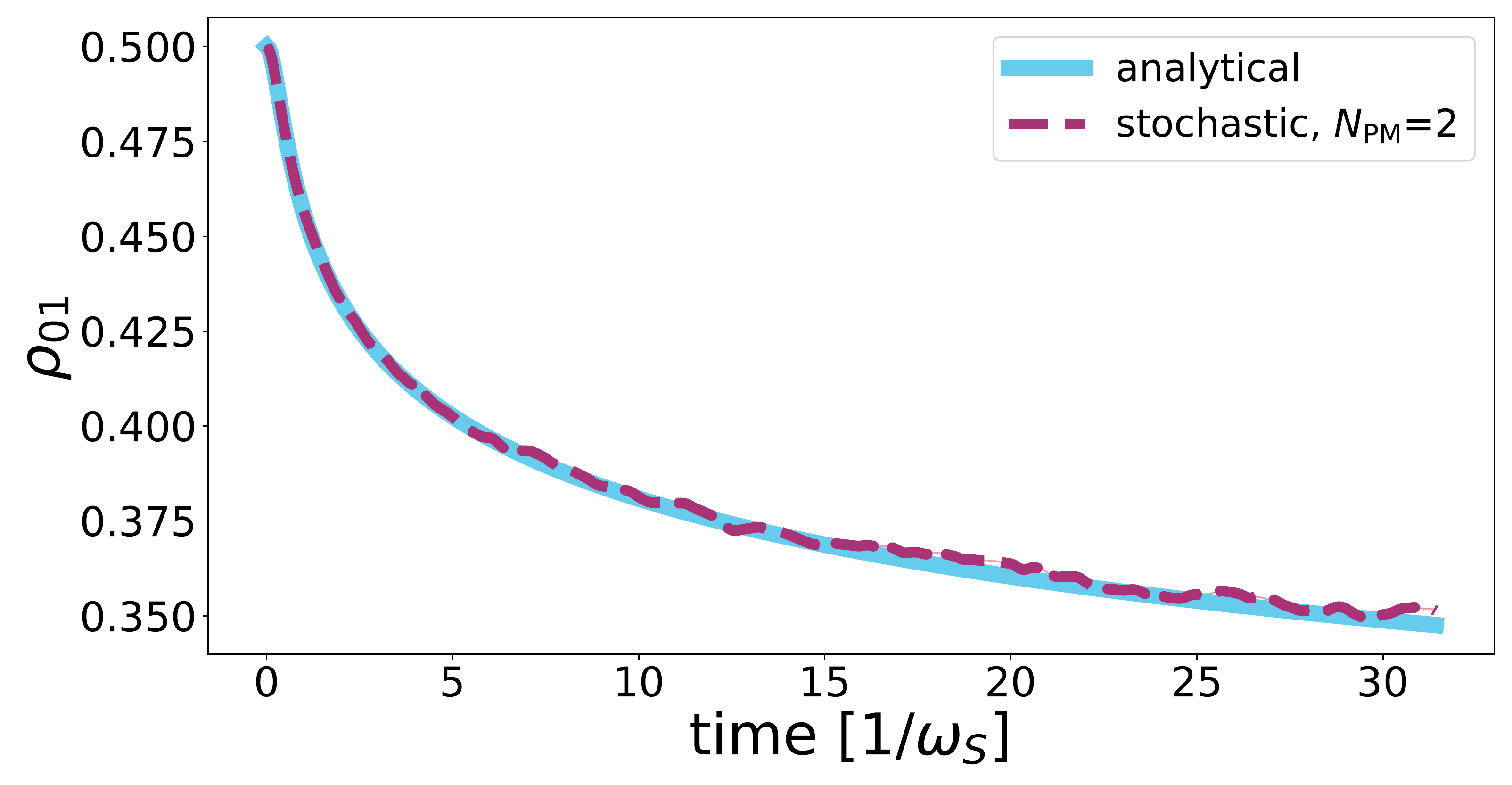} 
\caption{Dynamics of the coherences for a pure dephasing model with Ohmic spectral density at zero temperature for $N_\text{PM}=0$ (dashed-red) against the analytical result ({ solid} blue). The classical model does not have any pseudomodes and its stochastic field is defined to reproduce the \emph{symmetric} part of the correlation (see Fig.~\ref{fig:ohmic_corr}) in accordance to the pure dephasing analysis presented in section \ref{sec:pure_deph}. The initial state of the two-level system is assumed to be an equal superposition of the ground and exited state. Specifically, here we plot the off-diagonal elements of the density matrix which decay exponentially in the parameter $-4\Gamma_{\text{dep}}(t)=-2\alpha\text{Log}(1+\omega_c^2 t^2)$, see Eq.~(\ref{eq:deph_Ohm}) and \cite{Petruccione}, Eq.~(4.51). Other parameters are the same as those in Fig.~\ref{fig:ohmic_dynamics}. The averaging  is done over $10^4$ samples.
 \label{fig:ohmic_coherence_classical}}
\end{figure}

In this section, we analyze an Ohmic spectral density 
\begin{equation}
\label{eq:J_ohmic_app}
J_{\Omega}(\omega)=\pi\alpha\omega e^{-\omega/\omega_{c}}\;,
\end{equation}
characterized by a strength parameter $\alpha\in\mathbb{R}$ and a cut-off energy scale $\omega_c$. The antisymmetric contribution to the correlation takes the form
\begin{equation}
\begin{array}{lll}
C^{\Omega}_{\text{as}}(t) & =-\displaystyle\frac{\alpha}{2}\int_{0}^{\infty}\omega e^{-\omega/\omega_{c}}(e^{i\omega t}-e^{-i\omega t})d\omega\\
 & =\displaystyle -2i\alpha\omega_{c}^{2}\frac{\omega_{c}t}{(1+\omega_{c}^{2}t^{2})^{2}}\;,
 \end{array}
\end{equation}
whose behavior interpolates the following limiting regimes

\begin{equation}
\label{eq:Cas_Ohmic}
\begin{array}{lll}
C^{\Omega}_{\text{as}}(t) &\overset{t\ll1/\omega_c}{=}&\displaystyle-2i\alpha\omega_c^3 t\\
&\overset{t\gg 1/\omega_c}{=}&\displaystyle\frac{-2i\alpha\omega_c^2}{(\omega_c t)^3}\;.
 \end{array}
\end{equation}
Contrary to the rational case, this contribution does not identically follow the ansatz in Eq.~(\ref{eq:as_ansatz}). However, it is always possible to impose it, i.e. to find the optimal parameters which better approximate the expression
\begin{equation}
\label{eq:ansatz_Ohmic}
C^{\Omega}_\text{as}(t)\simeq\sum_{j=1}^{N^{\Omega}_\text{PM}}a^{\Omega}_j \sin(b^{\Omega}_j t)e^{-c^{\Omega}_j |t|}\;.
\end{equation}
It is interesting to gain some intuition on this procedure by using Eq.~(\ref{eq:Cas_Ohmic}) in Eq.~(\ref{eq:ansatz_Ohmic}) to obtain a rough estimate of the parameters. In fact, by assuming $N^{\Omega}_\text{PM}=1$, and taking the $t\ll 1/\omega_c$ limit on both sides of Eq.~(\ref{eq:ansatz_Ohmic}), we can write the following intuitive expressions
\begin{equation}
    \begin{array}{lll}
a^{\Omega}_1 \simeq-2i\alpha\omega_c^2,~~~b^{\Omega}_1 \simeq\omega_c,~~~
c^{\Omega}_1 \simeq \omega_c\;.
    \end{array}
\end{equation}
By using these expressions into Eq.~(\ref{eq:class_Q}), we can then conclude that the quantum contribution to the Ohmic spectral density can be approximated by a single pseudomode with parameters $\lambda^{\Omega}=\sqrt{2\alpha}\omega_c$, $\Omega^{\Omega}=\omega_c$, $\Gamma^{\Omega}=\omega_c$.
In Fig.~\ref{fig:ohmic_coherence_classical}, we show the results for the dynamics of the off-diagonal elements of the density matrix of a two level system interacting with an environment through a coupling operator which commutes with the system Hamiltonian. Specifically, we chose this pure dephasing model to be characterized by $s=\sigma_z$ and, for simplicity, by $H_S=0$.

As it can be seen from the analysis in section \ref{sec:pure_deph}, the dynamics of the coherences of a two level system in contact with a dephasing environment only depends on the symmetric part of the correlation. It is then interesting to note that if we were to use the protocol defined in section \ref{sec:ZeroT_main}, this model would still require a certain number of quantum pseudomodes in order to account for the quantum part of the correlation function. In turn, the classical field would then be defined in terms of the \emph{classical} part of the correlation which is defined in terms of both the symmetric and antisymmetric part. However, the procedure in section \ref{sec:ZeroT_main} is not the only possible one. In this case, we can, for example, model the full symmetric part of the correlation using a classical stochastic field (i.e., imposing $f_s(t)=0$ in the general formalism developed in section \ref{sec:Quantum-Classical}) and avoid considering any quantum degrees of freedom.

As expected from the analysis given in section \ref{sec:pure_deph}, the resulting dynamics matches the analytical one more since it is the only one which affects the system in this dephasing model [see Eq.~(\ref{eq:gamma_deph}) and the third line of Eq.~(\ref{eq:deph_parameters})].

\subsubsection{Pure Dephasing}
\label{sec:pure_deph}
Here, we consider the case in which the system-bath interaction Hamiltonian commutes with the system Hamiltonian, i.e. that $[{s},H_S]$ in the notation of App.~\ref{app:influenceSuperoperator}. In this case, the operator ${s}$ is time-independent in the interaction picture and the dynamics described by Eq.~(\ref{eq:expF_app_2}) becomes
\begin{equation}
\label{eq:app:rhoS_dephasing}
    \rho_S(t)=e^{\mathcal{F}_\text{deph}(t,{s},C(t))}\rho_S(0)\;,
\end{equation}
which, importantly, does no longer require a time-ordering. The influence superoperator for the pure dephasing case reads
 \begin{equation}
 \begin{array}{lll}
\mathcal{F}_\text{deph}[\cdot]
 &=&\displaystyle\int_{0}^{t}dt_{2}\int_{0}^{t_{2}}dt_{1}\;\{C(t_{2}-t_{1})[{s}\cdot,{s}]\\
 &&-C(t_{1}-t_{2})[\cdot{s},{s}]\}\\
 &=&\displaystyle\int_{0}^{t}dt_{2}\int_{0}^{t_{2}}dt_{1}\;\{2 C_\text{s}(t_{2}-t_{1}){s}\cdot{s}\\
 &&-C_\text{s}(t_2-t_1)[\cdot {s}^2+{s}^2\cdot]-C_\text{as}(t_2-t_1)[{s}^2,\cdot]\\
 &=&\Gamma_\text{deph}(t)D_\text{deph}[\cdot]-i\Omega_\text{deph}[{s}^2,\cdot]\;,
 \end{array}
\end{equation}
where we defined
\begin{equation}
\label{eq:deph_parameters}
    \begin{array}{lll}
\Omega_\text{deph}&=&-i\displaystyle\int_{0}^{t}dt_{2}\int_{0}^{t_{2}}dt_{1}\;C_\text{as}(t_2-t_1)\\
&=&-\displaystyle\frac{1}{\pi}\int_0^\infty d\omega J(\omega)\frac{t\omega-\sin(\omega t)}{\omega^2}\\
\Gamma_\text{deph}&=&\displaystyle\int_{0}^{t}dt_{2}\int_{0}^{t_{2}}dt_{1}\;C_\text{s}(t_2-t_1)\\
&=&\displaystyle\frac{1}{\pi}\int_0^\infty d\omega J(\omega)\coth{(\beta\omega/2)}\frac{1-\cos(\omega t)}{\omega^2}\\
D_\text{deph}[\cdot]&=&\displaystyle 2{s}\cdot{s}-{s}^2\cdot-\cdot{s}^2\;,
    \end{array}
\end{equation}
where we used
\begin{equation}
\begin{array}{lll}
\displaystyle\int_0^t d t_2\int_0^{t_2} dt_1 \cos{[\omega (t_2-t_1)]}&=&\displaystyle\int_0^t d t_2\int_0^{t_2} d u \cos{(\omega u)}\\
&=&\displaystyle\frac{1-\cos(\omega t)}{\omega^2}\\
\displaystyle\int_0^t d t_2\int_0^{t_2} dt_1 \sin{[\omega (t_2-t_1)]}
&=&\displaystyle\frac{t\omega-\sin(\omega t)}{\omega^2}\;.
\end{array}
\end{equation}
In the special case of a system made of a two-level system with ${s}=\sigma_z$, the effective Hamiltonian for the dephasing is zero because of $(\sigma_z^2=1)$ and Eq.~(\ref{eq:app:rhoS_dephasing}) becomes
\begin{equation}
\label{eq:gamma_deph}
    \rho_S(t)=e^{\Gamma_\text{deph}(t)[2\sigma_z\cdot\sigma_z-2\cdot]}\rho_S(0)\;,
\end{equation}
which leaves the diagonal elements invariant while causing a decay $\text{exp}[-4\Gamma_\text{deph}(t)]$ of the off-diagonal coherences. For the Ohmic spectral density in Eq.~(\ref{eq:J_ohmic_app})  and in the $\beta\rightarrow\infty$ limit, the decay rate can be further simplified to
\begin{equation}
\label{eq:deph_Ohm}
    \Gamma^\text{Ohmic}_\text{deph}(t)\rightarrow\frac{\alpha}{2}\log\left(1+\omega_c^2 t^2\right)\;.
\end{equation}

\nocite{apsrev41Control}
\bibliographystyle{apsrev4-2}
\bibliography{bib}

\end{document}